\documentclass[%
 reprint,
superscriptaddress,
nofootinbib,
 amsmath,amssymb,
 prd,
floatfix,
showkeys]{revtex4-2}

\usepackage[margin=5pt,caption=false]{subfig}
\usepackage{slantsc}
\usepackage{comment}
\usepackage{longtable} 
\usepackage[online]{threeparttablex}	
\usepackage{threeparttable}
\usepackage{booktabs}
\usepackage[figuresright]{rotating}
\usepackage{textcase}
\usepackage{multirow}
\newcommand{\be}{\begin{equation}}
\newcommand{\ee}{\end{equation}}
\newcommand{\bea}{\begin{eqnarray}}
\newcommand{\eea}{\end{eqnarray}}
\newcommand{\hunit}{$\rm{km \ s^{-1} \ Mpc^{-1}}$}
\newcommand{\lcdm}{$\Lambda$CDM}
\newcommand{\pcdm}{$\phi$CDM}

\usepackage{array}

\newcommand{\hii}{H\,\textsc{ii}}
\newcommand{\Om}{\Omega_{m0}}
\newcommand{\Ok}{\Omega_{k0}}
\newcommand{\om}{$\Omega_{m0}$}
\newcommand{\ok}{$\Omega_{k0}$}
\newcommand{\wx}{$w_{\rm X}$}
\newcommand{\wX}{w_{\rm X}}

\newcommand{\mii}{Mg\,\textsc{ii}}

\newcommand{\civ}{C\,\textsc{iv}}
\newcommand{\OIII}{[O\,\textsc{iii}]}

\newcommand{\obh}{\Omega_{b}h^2}
\newcommand{\och}{\Omega_{c}h^2}
\newcommand{\onh}{\Omega_{\nu}h^2}
\newcommand{\obhs}{$\Omega_{b}h^2$}
\newcommand{\ochs}{$\Omega_{c}h^2$}

\usepackage[T1]{fontenc}
\usepackage{scalerel}
\usepackage{tikz}
\usetikzlibrary{svg.path}

\definecolor{orcidlogocol}{HTML}{A6CE39}
\tikzset{
  orcidlogo/.pic={
    \fill[orcidlogocol] svg{M256,128c0,70.7-57.3,128-128,128C57.3,256,0,198.7,0,128C0,57.3,57.3,0,128,0C198.7,0,256,57.3,256,128z};
    \fill[white] svg{M86.3,186.2H70.9V79.1h15.4v48.4V186.2z}
                 svg{M108.9,79.1h41.6c39.6,0,57,28.3,57,53.6c0,27.5-21.5,53.6-56.8,53.6h-41.8V79.1z M124.3,172.4h24.5c34.9,0,42.9-26.5,42.9-39.7c0-21.5-13.7-39.7-43.7-39.7h-23.7V172.4z}
                 svg{M88.7,56.8c0,5.5-4.5,10.1-10.1,10.1c-5.6,0-10.1-4.6-10.1-10.1c0-5.6,4.5-10.1,10.1-10.1C84.2,46.7,88.7,51.3,88.7,56.8z};
  }
}

\newcommand\orcidicon[1]{\href{https://orcid.org/#1}{\mbox{\scalerel*{
\begin{tikzpicture}[yscale=-1,transform shape]
\pic{orcidlogo};
\end{tikzpicture}
}{|}}}}

\usepackage{graphicx}	
\usepackage{amsmath,bm}	
\usepackage{amssymb}	
\usepackage{fnpct}

\usepackage[colorlinks=true,allcolors=blue]{hyperref}
\begin{document}

\preprint{APS/123-QED}

\title{Standardizing reverberation-mapped H$\beta$ active galactic nuclei using time-averaged radius-luminosity relations with 5100\,\AA\,, broad H$\beta$, and narrow \OIII\ luminosities}

\author{Shulei Cao$^{\orcidicon{0000-0003-2421-7071}}$}
\email{shuleic@mail.smu.edu}
\affiliation{Department of Physics, Kansas State University, 116 Cardwell Hall, Manhattan, KS 66506, USA}
\affiliation{Department of Physics, Southern Methodist University, Dallas, TX 75205, USA}%
\author{Amit Kumar Mandal$^{\orcidicon{0000-0001-9957-6349}}$}
\email{amitastro.am@gmail.com}
\affiliation{Center for Theoretical Physics, Polish Academy of Sciences, Al. Lotnik\'{o}w 32/46, 02-668 Warsaw, Poland}%
\author{Michal Zaja\v{c}ek$^{\orcidicon{0000-0001-6450-1187}}$}
\email{zajacek@physics.muni.cz}
\affiliation{Department of Theoretical Physics and Astrophysics, Faculty of Science, Masaryk University, Kotlá\v{r}ská 2, 611 37 Brno, Czech Republic}%
\author{Bharat Ratra$^{\orcidicon{0000-0002-7307-0726}}$}%
\email{ratra@phys.ksu.edu}
\affiliation{Department of Physics, Kansas State University, 116 Cardwell Hall, Manhattan, KS 66506, USA}%

\date{\today}

\begin{abstract}

Active galactic nuclei (AGN) have been studied as alternate probes in cosmology due to their large and stable luminosities and broad redshift range. Previously it was shown that higher-redshift AGN that were reverberation-mapped (RM) using broad Mg\,\textsc{ii} and C\,\textsc{iv} lines are standardizable and yield weak cosmological constraints that are consistent with those from better-established probes. In contrast, lower-redshift AGN that were reverberation-mapped using the broad H$\beta$ line exhibited tensions with the standard cosmological model, in particular they preferred currently decelerating cosmological expansion. Here we study the standardizability of a homogeneous RM H$\beta$ sample of $\sim 100$ AGN (over redshifts $0.00308 \leq z \leq 0.8429$), whose H$\beta$ time delays and three luminosity tracers (at 5100\,\AA\,, broad H$\beta$, and narrow [O\,\textsc{iii}]) are averaged over several epochs. We find that this averaged sample is standardizable using three $R-L$ relations. While for luminosities corresponding to 5100\,\AA\, and the broad H$\beta$ line the cosmological constraints prefer currently decelerating cosmological expansion, the cosmological parameters for the narrow [O\,\textsc{iii}] luminosity are more consistent with those from better-established probes and they are in agreement with currently accelerating cosmological expansion. This demonstrates for the first time that narrow-line region [O\,\textsc{iii}] can be utilized for AGN standardization and cosmological constraints. Selecting proper photoionizing flux proxies for the broad-line region is thus crucial in studies of RM AGN standardizability. 

\end{abstract}


\maketitle

\section{Introduction}

The current standard model of cosmology -- the spatially flat $\Lambda$CDM model \citep{peeb84} -- can successfully accommodate most observed characteristics concerning the structure and evolution of the expanding Universe, at both lower and higher cosmological redshifts. These include spatial homogeneity and isotropy on large length scales ($>100\,{\rm Mpc}$), flat spatial geometry, and the cosmic microwave background (CMB), a fingerprint of the earlier hot and dense state of the expanding Universe.  

The current cosmological energy budget is dominated by dark energy (with a current fractional contribution of $\Omega_{\rm DE0}\sim 70\%$), which is responsible for the current accelerated expansion of the Universe that was first discovered using Type Ia supernovae \citep{Riess_1998, Perlmutter_1999,2024SSRv..220...24K}. Matter, dominated by cold dark matter (CDM), currently contributes $\Om\sim 30\%$ of the current total energy budget. The properties of nonrelativistic (cold) and weakly interacting dark matter determined how structure formed, hierarchically (bottom-up) and induced by primordial fluctuations in the energy density that grew under gravity.

The standard flat $\Lambda$CDM model certainly does not fit all observations. There are some observational discrepancies, the most well-known one being the difference between some low-redshift measurements of the Hubble constant $H_0$ and the CMB anisotropy measurements \citep{Hu:2023jqc, 2025PDU....4901965D}.\footnote{We note that the differences in measurements of the matter density fluctuations ($\sigma_8$ or $S_8$) now appears to have been resolved, through a better accounting of the errors in one of the probes, \citep{DESY32025,KiDS2025}.} Another well-known discrepancy is the indication that dark energy might be dynamical, \citep[][and references therein]{CaoRatra2023, deCruzPerez:2024shj, DESI_2025_2025JCAP...02..021A, Parketal2024, Parketal2025, ParkRatra2025a, 2025PhRvD.112h3515A}. These discrepancies could be caused by unknown systematic effects in at least one of the cosmological probes or because the actual cosmological model deviates from the standard flat $\Lambda$CDM model. Different probes can possess various and different biases, for instance due to their intrinsic change caused by their evolution from higher to lower redshifts (e.g., due to changing metallicity content in the interstellar medium). 

It is therefore essential to develop and apply alternate cosmological probes, in addition to the better-established now-standard ones, so that any biases and hidden systematic effects can be identified and eventually accounted for. Especially valuable are new probes that cover a wide redshift range and/or partially bridge between the better-studied low-redshift $(z \leq 2.3)$ and higher-redshift $(z \sim 1100)$ regimes. For each probe it is necessary to test whether the corresponding data sample is standardizable, i.e., whether the parameters of the relation used for the standardization of the probe are independent of the cosmological model. To accomplish this while avoiding the circularity problem both the standardization relation parameters and the cosmological model parameters need to be determined simultaneously, \cite{KhadkaRatra2020c, Caoetal_2021}.

Several alternate probes have been tested. An example is \hii\ starburst galaxies to $z = 2.5$, using a claimed relation between the H$\beta$ absolute luminosity and the ionized gas velocity dispersion, $L_{\rm H\beta}$-$\sigma_{\rm g}$, \cite{Mania_2012, GonzalezMoran2019, GonzalezMoranetal2021, CaoRyanRatra2020, CaoRyanRatra2022, Johnsonetal2022, Mehrabietal2022}, which actually is not standardizable because for current data the $L_{\rm H\beta}$-$\sigma_{\rm g}$ relation evolves with redshift \cite{CaoRatra2024a, Melnick:2024ywi}. Another alternate probe tested is $\gamma$-ray bursts (GRBs) to $z = 8.2$ (using, for example, the Amati relation between the rest-frame photon peak energy and the rest-frame isotropic radiated energy, $E_{\rm p}-E_{\rm iso}$, \cite{Amatietal2002}, or other relations), \citep{Huetal_2021, Khadkaetal_2021b, Luongoetal2021, CaoRatra2022, CaoKhadkaRatra2022, CaoDainottiRatra2022, CaoDainottiRatra2022b, Favaleetal2024, Lietal2024, CaoRatra2024b, Bargiacchietal2025, Dengetal2025, CaoRatra2025, Huangetal2025, OkazakiDesai2025, Xieetal2025}, with a number of GRB relations being standardizable and with the currently most-restrictive constraints coming from the currently largest-available standardizable sample of Amati-correlated GRBs \cite{Khadkaetal_2021b}.

Active galactic nuclei (AGN \cite{2021bhns.confE...1K,2024SSRv..220...29Z,Zajacek25}), including quasars (QSOs) that are the most luminous AGN, have been considered as especially appealing alternate probes due to their large and stable luminosities and because they cover a broad redshift range (from $z=0.00106$ for NGC4395 \cite{2019MNRAS.486..691B} up to $z=7.642$ \cite{2021ApJ...907L...1W} for J0313-1806). Several methods have been developed to standardize AGN and apply them in cosmology, specifically (i) the AGN angular size -- redshift relation for compact radio sources up to $z=2.7$ \cite[e.g.,][]{gurvits_kellermann_frey_1999, ChenRatra2003, Cao_et_al2017b}; (ii) a power-law $L_{X}-L_{UV}$ relation between X-ray and UV luminosities \cite[e.g.,][]{RisalitiandLusso_2015, Lussoetal2020}; (iii) the $R-L$ relation between the broad-line region (BLR) radius $R$ and the photoionizing absolute luminosity $L$ for reverberation-mapped (RM) AGN (the BLR radius is determined from a time delay, $R=c\tau$, where $c$ is the speed of light) \cite[e.g.][]{2011ApJ...740L..49W,2011A&A...535A..73H,2013A&A...556A..97C}; (iv) QSOs accreting near the Eddington limit, for which the bolometric luminosity is proportional to the supermassive black hole (SMBH) mass, which allows using them as Eddington standardizable candles \cite[e.g.,][]{2014ApJ...793..108W}; (v) time-delay distances between strongly lensed QSO images \cite[e.g.,][]{2017MNRAS.468.2590S}; (vi) continuum reverberation mapping of accretion discs allows determination of the Hubble constant \cite[e.g.,][]{1999MNRAS.302L..24C,2007MNRAS.380..669C, 2025A&A...702A..92J}; and (vii) angular-diameter measurement of the BLR in combination with its inferred radius from reverberation mapping can also be used to constrain the Hubble constant \cite[e.g.,][]{2002ApJ...581L..67E,2020NatAs...4..517W}.  

Cosmological constraints based on $L_{X}-L_{UV}$ relation AGN data initially were largely consistent with the standard flat $\Lambda$CDM model, \cite{RisalitiandLusso_2015, KhadkaRatra2020a}. Updated $L_{X}-L_{UV}$ data seemed to indicate a significant deviation from the standard model \citep{RisalitiandLusso_2019, Lussoetal2020}. However, from a proper analysis in which the $L_{X}-L_{UV}$ relation parameters and the cosmological parameters for several cosmological models were determined simultaneously \cite{Khadka_2020b, KhadkaRatra2021, KhadkaRatra2022}, it was found that the $L_{X}-L_{UV}$ QSO sample of \cite{Lussoetal2020} is not standardizable. More specifically, for the sample of \cite{Lussoetal2020} the $L_{X}-L_{UV}$ relation parameter values depend on both the cosmological model and the redshift range \citep{KhadkaRatra2021, KhadkaRatra2022, Petrosian_2022, Khadka:2022aeg, Lietal2025, Wuetal25, Montieletal2025, Chiraetal2025}. \citet{Khadka:2022aeg} found, for a sample of 58 \mii\ RM AGN with UV and X-ray flux measurements, that the luminosity distance based on the $L_{X}-L_{UV}$ relation is shorter than the one inferred from the $R-L$ relation, which is consistent with the $L_{X}-L_{UV}$-based cosmological constraints favoring a larger nonrelativistic matter density parameter $\Om$. A contributing factor that can explain the differences in luminosity distances based on $L_{X}-L_{UV}$ and $R-L$ relations is differential extinction between X-ray and UV domains due to dust in host galaxies, which is difficult to mitigate \cite{Zajaceketal2024}. 

In contrast to the X-ray/UV-flux sample, RM AGN/QSOs are standardizable when using the $R-L$ relation. In addition, the RM AGN samples yield cosmological parameter values that are consistent with the flat $\Lambda$CDM model with a few exceptions, though these constraints are weak and therefore mild dark energy dynamics and a small amount of spatial curvature cannot be excluded \cite{Khadkaetal_2021a, Khadkaetal2021c, Khadkaetal2022a, 2022MNRAS.516.1721C, Caoetal2024, cao_2025a, cao25}. For higher-redshift \mii\ and \civ\ RM AGN \cite{Khadkaetal_2021a,2022MNRAS.516.1721C}, with redshift ranges $0.0033 \leq z \leq 1.89$ and $0.001064 \leq z \leq 3.368$, respectively, the samples are standardizable and the weak cosmological constraints are consistent with better-established probes, such as a joint sample of Hubble parameter [$H(z)$] and baryon acoustic oscillation (BAO) measurements, $H(z)$ + BAO. In contrast, although the sample of 118 lower-redshift H$\beta$ RM AGN ($0.0023 \leq z \leq 0.89$) was found to be standardizable as well, the inferred cosmological constraints were in $\gtrsim 2\sigma$ tension with those from better-established $H(z)$ + BAO data and favored a currently decelerating cosmological expansion \cite{Khadkaetal2021c}. A contributing factor to this tension was a large degree of heterogeneity of the H$\beta$ data sample, with several methodologies applied to infer rest-frame BLR time delays, which resulted in systematic effects that were difficult to account for.

To mitigate the effect of different methodologies used to infer the time delay, we investigated a homogeneous sample of 41 AGN with H$\alpha$ and H$\beta$ time delays \citep{Choetal2023}. This sample was found to be standardizable using four $R-L$ relations involving H$\alpha$ and H$\beta$ time delays and monochromatic and broad H$\alpha$ luminosities \cite{cao_2025a}. The sample consisted of low-redshift sources ($0.00415 \leq z \leq 0.474$) with an absence of highly-accreting sources, which can explain the inferred steeper $R-L$ relations than the slope of 0.5 expected from a simple photoionization argument. Cosmological parameters were only weakly constrained, however, in contrast to the larger sample investigated by \citet{Khadkaetal2021c}, they were found to be consistent with those from better-established $H(z)$ + BAO data. A larger homogeneous sample of 157 AGN ($0.00308 \leq z \leq 0.8429$) that were reverberation-mapped using the broad H$\beta$ line was also found to be standardizable using the $R-L$ relation with the monochromatic luminosity at 5100\,\AA\, \cite{cao25}. For this sample, which contained sources with a broader range of Eddington ratios, the $R-L$ relation slope was found to be slightly smaller than the expected slope of 0.5, which can be explained by the presence of high-accreting sources. Cosmological parameter constraints were inferred to be consistent within $\lesssim 2\sigma$ with those from better-established $H(z)$ + BAO data, except for the nonflat $\Lambda$CDM model and the nonflat XCDM parametrization, which are, however, disfavored based on other cosmological probes, \citep{deCruzPerez:2024shj}. Overall, this shows that sample homogeneity helps to decrease systematic effects and can partially address the tension with respect to better-established probes as well as with higher-redshift AGN.

However, some tension persisted for the H$\beta$ RM sources since the inferred cosmological parameters for $R-L$ relations involving broad $H\alpha$ and 5100\,\AA\, luminosities preferred currently decelerating cosmological expansion and larger $\Om$ values in contrast to the joint $H(z)$ + BAO data predictions and those from most higher-redshift RM AGN data compilations \cite{Khadkaetal_2021a,2022MNRAS.516.1721C}. Improvements could come from a unified time-delay determination methodology, as well as from a better proxy for the BLR photoionizing flux. The traditionally used monochromatic luminosity at 5100\,\AA\, for the broad H$\beta$ line might not be the best representation since the AGN spectral energy distribution and hence the fraction of photoionizing photons depends on the relative accretion rate (Eddington ratio) as well as the SMBH mass. Moreover, the radius (rest-frame time delay) of the H$\beta$ line-emitting material changes with the luminosity state of the AGN (the so-called BLR ``breathing'' \cite{Barth_2015, 2006MNRAS.365.1180C}) and the adopted time delays and luminosities for different epochs do not have to exactly correspond to each other since the luminosity is variable. 

Therefore, in this work, we again consider the homogeneous sample used in \cite{cao25}, but more specifically we consider subsamples of it, where several measurements at different epochs were averaged, both in terms of the H$\beta$ time delay and the corresponding luminosity. The redshift ranges of the subsamples are the same as those of the larger sample of 157 measurements \cite{2024ApJS..275...13W,cao25}, i.e., $0.00308\leq z \leq 0.8429$. Importantly we also consider three photoionizing luminosity proxies: 5100\,\AA\, luminosity (113 AGN), broad H$\beta$ luminosity (100 AGN), and narrow \OIII\ luminosity (100 AGN), as they exhibit significant mutual correlations \citep{2024ApJS..275...13W} and can therefore be used as proxies for the underlying photoionizing luminosity. However, it is important to note that these luminosities originate from different regions of the AGN, with $L_{5100}$ arising from the accretion disc, whereas $L_{\rm H\beta}$ and $L_{\mathrm{{\OIII}}}$ originate from the BLR and the narrow-line region (NLR), respectively. We especially stress that emission regions related to $L_{5100}$ and $L_{\rm H\beta}$ on the one hand and $L_{\mathrm{\OIII}}$ on the other hand are spatially separated by $\sim 100\text{--}1000\,{\rm pc}$. Also, in the unified model of AGN \citep{1985ApJ...297..621A,1993ARA&A..31..473A,1995PASP..107..803U} the BLR and the NLR are found in different geometrical configurations: while the BLR lies close to the accretion disc plane, the NLR is found in bipolar ionization cones above the accretion disc plane. Despite considerable spatial separation, these three subsamples with three different $R-L$ relations are found to be standardizable, with inferred $R-L$ relation slopes that are closer to the simple photoionization prediction of 0.5 than for the non-averaged larger sample.

Our most interesting finding is that while cosmological constraints obtained using $L_{5100}$ and $L_{\rm H\beta}$ more favored currently decelerating cosmological expansion, the constraints for $L_{\mathrm{{\OIII}}}$ are more consistent with currently accelerating cosmological expansion and overall more consistent with those obtained using better-established $H(z)$ + BAO data. This is likely related to the longer timescales associated with the NLR \OIII\ emission in contrast to the emission at 5100\,\AA\, and the broad H$\beta$ line emission, which originate within or close to the accretion disc. Therefore, the \OIII\ luminosity is insensitive to the actual short-term accretion state and hence the current Eddington ratio, unlike $L_{5100}$ and $L_{\rm H\beta}$. Although the intrinsic scatter is the largest for the $R-L$ relation involving $L_{\mathrm{{\OIII}}}$, which is related to its own intrinsic scatter \citep{2005ApJ...634..161H}, it suffers from smaller systematic effects related to the Eddington ratio, host contamination correction, and extinction. 

In principle, since the narrow \OIII\, emission is approximately orientation independent, it can serve as a proxy for the total AGN power, which motivates its application as a luminosity-distance indicator. However, previous attempts for AGN standardization using \OIII\, have failed because of obscuration effects (type II AGN) for the hard X-ray/\OIII\, relation \citep{2005ApJ...634..161H} as well as a large intrinsic scatter and systematic effects \cite{2017MNRAS.468.1433P,2014ARA&A..52..589H}. The radius-luminosity relation in this work thus demonstrates the first successful application of the NLR \OIII\, luminosity to AGN standardization and cosmological constraints.

Our paper is structured as follows. In Sec.~\ref{sec:model} we introduce cosmological models that we use to test the standardizability of the three time-averaged AGN subsamples. In Sec.~\ref{sec:data} we introduce the three H$\beta$ AGN subsamples as well as the joint $H(z)$ + BAO sample used for comparison. Subsequently, in Sec.~\ref{sec:analysis} we describe the methodology to simultaneously infer $R-L$ relation parameters and cosmological model parameters, thus avoiding the circularity problem. We present the main results in Sec.~\ref{sec:results} and we discuss the differences with respect to our previous analyses in Sec.~\ref{sec:discussion}. We summarize the main conclusions in Sec.~\ref{sec:conclusion}.

\section{Cosmological models}
\label{sec:model}

To assess whether H$\beta$ RM AGN can be standardized through the $R-L$ relation, we simultaneously fit the $R-L$ relation parameters and the cosmological parameters of six spatially flat and nonflat relativistic dark energy models. If the derived $R-L$ relation parameters are consistent across different cosmologies, the AGN dataset can be treated as standardizable. This method \cite{KhadkaRatra2020c, Caoetal_2021} avoids the circularity problem common in such analyses. 

For each of the three types of source luminosities we consider, source absolute luminosities are determined from corresponding source fluxes and the luminosity distance $D_L(z)$ in each cosmological model we consider. The luminosity distance
\begin{equation}
  \label{eq:DL}
\resizebox{0.48\textwidth}{!}{%
    $D_L(z) = 
    \begin{cases}
    \frac{c(1+z)}{H_0\sqrt{\Omega_{\rm k0}}}\sinh\left[\frac{H_0\sqrt{\Omega_{\rm k0}}}{c}D_C(z)\right] & \text{if}\ \Omega_{\rm k0} > 0, \\
    \vspace{1mm}
    (1+z)D_C(z) & \text{if}\ \Omega_{\rm k0} = 0,\\
    \vspace{1mm}
    \frac{c(1+z)}{H_0\sqrt{|\Omega_{\rm k0}|}}\sin\left[\frac{H_0\sqrt{|\Omega_{\rm k0}|}}{c}D_C(z)\right] & \text{if}\ \Omega_{\rm k0} < 0,
    \end{cases}$%
    }
\end{equation}
where the comoving distance is given by
\begin{equation}
\label{eq:gz}
   D_C(z) = \frac{c}{H_0}\int^z_0 \frac{dz'}{E(z')}.
\end{equation}
In these expressions $H_0$ is the Hubble constant and $\Ok$ the current spatial curvature density.\footnote{For recent discussions on the implications of, and constraints on, spatial curvature refer to Refs.\ \cite{Oobaetal2018b, ParkRatra2019b, DiValentinoetal2021a, ArjonaNesseris2021, Dhawanetal2021, Renzietal2021, Gengetal2022, MukherjeeBanerjee2022, Glanvilleetal2022, Wuetal2023, deCruzPerezetal2023, DahiyaJain2022, Stevensetal2023,Favaleetal2023, Qietal2023, Shimon:2024mbm, Wu:2024faw}.} We adopt a standard neutrino setup with one massive and two massless species, giving $N_{\rm eff} = 3.046$ and total mass $\sum m_\nu = 0.06$ eV. The present neutrino energy density is $\onh=\sum m_{\nu}/(93.14\ \rm eV)$, where $h = H_0/100$~\hunit. The total nonrelativistic matter density is $\Om = (\onh + \obh + \och)/{h^2}$ including neutrinos, baryons, and cold dark matter. Radiation is neglected at late times.

We explore six cosmological models/parameterizations in this work. In the \lcdm\ model dark energy is a cosmological constant $\Lambda$ with equation of state parameter $w_{\rm DE}=-1$. The XCDM parametrization treats dark energy as a homogeneous fluid with constant $w_{\rm DE}\neq -1$, providing a phenomenological description of dynamical dark energy. For both \lcdm\ and XCDM, the expansion rate is given by
\be
\label{eq:EzL}
\resizebox{0.48\textwidth}{!}{%
    $E(z) = \sqrt{\Om\left(1 + z\right)^3 + \Ok\left(1 + z\right)^2 + \Omega_{\rm DE0}\left(1+z\right)^{3(1+w_{\rm DE})}},$%
    }
\ee
with $\Omega_{\rm DE0} = 1 - \Om - \Ok$. In AGN-only analyses, we fix $h=0.7$ and $\Omega_b = 0.05$; free parameters are $\{\Om,\Ok\}$ for \lcdm\ and $\{\Om,\Ok,\wX\}$ for XCDM (with $\Ok=0$ for flat models). When including $H(z)$ + BAO data, the fundamental parameters $\{H_0, \obh, \och\}$ replace $\Om$.

For \pcdm\ models \citep{peebrat88, ratpeeb88, pavlov13}\footnote{For recent studies constraining \pcdm, see Refs.\ \cite{ooba_etal_2018b, ooba_etal_2019, park_ratra_2018, park_ratra_2019b, park_ratra_2020, Singhetal2019, UrenaLopezRoy2020, SinhaBanerjee2021, deCruzetal2021, Xuetal2022, Jesusetal2022, Adiletal2023, Dongetal2023, VanRaamsdonkWaddell2023, Avsajanishvilietal2024, VanRaamsdonkWaddell2024a, Thompson2024, ParkRatra2025b, Zhangetal2025}.}, dark energy arises from a scalar field $\phi$ with an inverse power-law potential energy density
\be
\label{PE}
V(\phi)=\frac{1}{2}\kappa m_p^2\phi^{-\alpha}.
\ee
The expansion rate evolves as
\be
\label{eq:Ezp}
    E(z) = \sqrt{\Om\left(1 + z\right)^3 + \Ok\left(1 + z\right)^2 + \Omega_{\phi}(z,\alpha)},
\ee
with
\be
\label{Op}
\Omega_{\phi}(z,\alpha)=\frac{1}{6H_0^2}\bigg[\frac{1}{2}\dot{\phi}^2+V(\phi)\bigg],
\ee
and $\phi$ obeying
\be
\label{em}
\ddot{\phi}+3H\dot{\phi}+V'(\phi)=0.
\ee
In these expressions $\alpha$ is the potential exponent ($\alpha = 0$ recovers \lcdm), $m_p$ is the Planck mass, and overdots and a prime denote derivatives with respect to time and $\phi$. The normalization $\kappa$ is fixed using the shooting method in \textsc{class} \citep{class}. In AGN-only \pcdm\ fits, $\{\Om,\Ok,\alpha\}$ are free ($\Ok=0$ for flat models), whereas in the fits including $H(z)$ + BAO data, $\Om$ is replaced by $\{H_0, \obh, \och\}$. Note that we have improved our \textsc{class} implementation of \pcdm\ by adopting radiation-dominated initial conditions aligned with the scale factor, a more robust initial-guess and root-finding procedure, and an explicit constraint enforcing $\Omega_{\phi}>0$.

\section{Sample and Data}
\label{sec:data}

\subsection{AGN Sample and Data}

We employ the largest and most homogeneous compilation of H$\beta$ RM AGN data currently available, derived from the uniform lag analysis conducted by \citet{2024ApJS..275...13W}. Their compilation incorporates 312 H$\beta$ RM measurements, each comprising both H$\beta$ and continuum light curves collected from the literature. This comprehensive dataset provides an exceptional opportunity to refine the calibration of the radius-luminosity ($R-L$) relation, where the H$\beta$ BLR radius $R=c\tau_{\rm H\beta}$ ($\tau_{\rm H\beta}$ is the rest-frame time delay of the broad $H{\beta}$ line with respect to the photoionizing continuum) and $L$ is the AGN absolute luminosity.

Within this compilation, \citet{2024ApJS..275...13W} identified 157 high-quality RM measurements through multiple lag-quality assessments  \citep[see][for details]{2024ApJS..275...13W, cao25}. This subset, designated as the ``Best sample'', serves as the foundation for testing standardizability of the AGN sample and for constraining cosmological parameters. The corresponding cosmological results derived from this Best sample, which spans a broad luminosity range of $10^{41.73} \, \mathrm{erg \, s^{-1}} < L_{5100} < 10^{45.9} \, \mathrm{erg \, s^{-1}}$, were presented in our previous work \citep{cao25}.

\begin{figure*}[t]
\centering
\includegraphics[width=\textwidth]{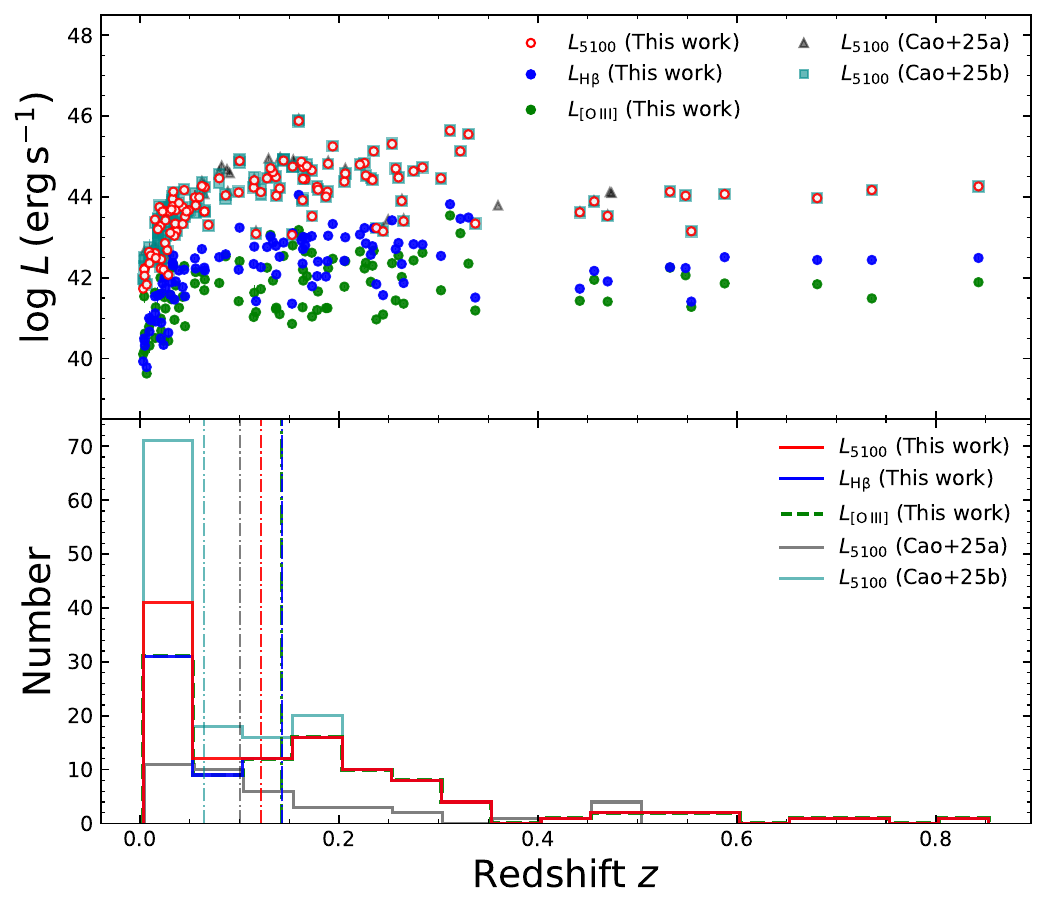}
\caption{ Comparison of the properties of the selected Average Scheme sample of H$\beta$ RM AGN, characterized by AGN monochromatic luminosity $L_{5100}$ (red), broad H$\beta$ luminosity $L_{\text{H}\beta}$ (blue), and optical narrow {\OIII} luminosity $L_{\mathrm{{\OIII}}}$ (green), with the earlier samples of both H$\alpha$ and H$\beta$ RM AGN from Cao+25a \cite{cao_2025a} (black) and the H$\beta$-only sample from Cao+25b \cite{cao25} (cyan). The top panel presents AGN luminosity as a function of redshift, while the bottom panel shows the corresponding redshift distributions in histogram form, constructed using a uniform bin size for all samples. Dot-dashed vertical lines of different colors mark the median redshift of each sample. Note that the H$\beta$ and {\OIII} samples analyzed in this work originate from the same AGN dataset; therefore, the median redshift values indicated by the blue and green dot-dashed lines overlap. Additionally, the AGN sample with $L_{5100}$ analyzed in this work and that of Cao+25b \cite{cao25} span the same redshift range.  The luminosities are calculated from the observed fluxes assuming a flat $\Lambda$CDM cosmology with $H_0 = 72 \, {\rm km \, s^{-1} \, Mpc^{-1}}$ and $\Omega_{m0} = 0.3$. }
\label{fig:dist}
\end{figure*}

\begin{table*}
    \centering
    \caption{Comparison of different low-redshift AGN samples studied in this work as well as in Cao+25a \cite{cao_2025a} and Cao+25b \cite{cao25}. From the left to the right columns, we include sample designation, its size, redshift range, median redshift, luminosity range, and median luminosity.}
    \begin{tabular}{c|c|c|c|c|c}
    \hline
    \hline 
    Sample  & Size & redshift range & median redshift & $\log{(L\,[{\rm erg\,s^{-1}}])}$ range & $\log{(L\,[{\rm erg\,s^{-1}}])}$ median  \\
    \hline
    $L_{5100}$ (this work) & 113  & $(0.00308, 0.8429)$ & $0.12141$ & $(41.73, 45.88)$ & 43.97
 \\
    $L_{H\beta}$ (this work) & 100 & (0.00308, 0.8429)  &  0.142255 & (39.79, 44.05) & 42.325  \\
    $L_{{\rm \OIII}}$ (this work) & 100 & (0.00308, 0.8429)  &  0.142255 & (39.63, 43.54) & 41.69 \\
    $L_{5100}$ (Cao+25a) & 41  & (0.0041488, 0.4741278) &  0.1000625 & (41.972, 45.937) &  44.016 \\
    $L_{5100}$ (Cao+25b) & 157  & (0.00308, 0.8429) & 0.06458 & (41.73, 45.9) & 43.79 \\    
    \hline
    \end{tabular}
    \label{tab_sample_overview}
\end{table*}

However, several AGN in the Best sample have been monitored repeatedly through independent RM campaigns conducted at different epochs. Numerous RM studies have demonstrated that, for many AGN, the measured H$\beta$ time delay can vary substantially between campaigns, often corresponding to different luminosity states. Such variations suggest that the BLR structure may evolve dynamically on timescales of only a few years \citep[e.g.,][]{2014Sci...345...64K, 2018ApJ...856..108P}, potentially in response to changes in the accretion state \citep{2006MNRAS.365.1180C, 2014MNRAS.438.3340E}, inhomogeneities within the BLR gas distribution \citep{1995A&A...296..332W}, or the influence of radiation pressure \citep{2023MNRAS.520.1807C}. The precise origin of this variability, however, remains uncertain, since the observed luminosity changes across epochs are often modest relative to their measurement uncertainties. The effect of the increase/decrease of the BLR radius with the rise/fall in the source luminosity is often referred to as BLR breathing \citep{1990ApJ...365L...5N,2004ApJ...606..749K}. For individual sources, low-ionization recombination emission lines (H$\alpha$ and H$\beta$) exhibit the breathing effect while the \mii\ line seems to have a weaker response, potentially due to the line-emitting region being located towards the outer BLR radius \citep{2020ApJ...888...58G}. This results in the existence of the H$\alpha$ and H$\beta$ $R-L$ correlations even for individual AGN while \mii\ may lack such a correlation for single sources. However, most studied broad lines, including \mii, exhibit the $R-L$ relation for a sample of sources \citep[see e.g.][]{Czerny:2022xfj}.  

A typical example of such BLR dynamical behavior was detected in NGC~5548, one of the most extensively monitored AGN, with more than two decades of reverberation data. This source exhibits H$\beta$ delays ranging from 2.3 days at $L_{5100} \sim 10^{42.7} \, \mathrm{erg \, s^{-1}}$ \citep{2010ApJ...721..715D} to 21.5 days at $L_{5100} \sim 10^{43.4}  \, \mathrm{erg \, s^{-1}}$ \citep{2002ApJ...581..197P, 2006MNRAS.365.1180C, 2004ApJ...613..682P}, clearly illustrating the BLR breathing effect. Similar multi-epoch variations have been reported for NGC~4151 \citep{2023MNRAS.520.1807C}, Arp~151 \citep{2018ApJ...856..108P}, Mrk~335 \citep{1998PASP..110..660P, 2004ApJ...613..682P, 2012ApJ...755...60G, 2014ApJ...782...45D, 2015ApJ...804..138H, 2021ApJS..253...20H}, and several other AGN.

Because these epoch-dependent variations can introduce additional scatter into the $R-L$ relation, it is useful to adopt an approach that minimizes this effect. Therefore, we use the ``Average Scheme'' sample compiled by \citet{2024ApJS..275...13W}, in which multiple RM measurements for the same AGN are combined through uncertainty-weighted averages of both the H$\beta$ lag and the corresponding luminosities. Hence, this approach provides a representative, time-averaged characterization of each AGN’s BLR radius and luminosity state, thereby reducing the impact of short-term BLR variability on the $R-L$ relation.

Applying this scheme to the 157 high-quality H$\beta$ lags in the Best sample, \citet{2024ApJS..275...13W} identified 113 distinct AGN with reliable measurements of the monochromatic luminosity at 5100 {\AA} ($L_{5100}$). Among these, 100 AGN also have available measurements for all three luminosity indicators: $L_{5100}$, broad H$\beta$ luminosity ($L_{\text{H}\beta}$), and narrow {\OIII} luminosity ($L_{\mathrm{{\OIII}}}$). In the present study, we adopt this Average Scheme sample, comprising 113 AGN with $L_{5100}$ measurements and a subsample of 100 AGN with $L_{\text{H}\beta}$ and $L_{\mathrm{{\OIII}}}$ data.

In Fig.~\ref{fig:dist} we present the luminosity distributions and number counts as a function of redshift for these samples, characterized by the different luminosity indicators ($L_{5100}$, $L_{\text{H}\beta}$, and $L_{\mathrm{{\OIII}}}$). For context, we also overlay the samples analyzed in our previous works \citep[i.e.,][]{cao_2025a, cao25} for a direct comparison. The sample with the luminosity measurements at 5100\,\AA\, spans a redshift range from 0.00308 to 0.8429 and covers a wide dynamic range in luminosity: $10^{41.73}  \, \mathrm{erg \, s^{-1}} \leq L_{5100} \leq 10^{45.88}  \, \mathrm{erg \, s^{-1}}$. The subsamples with reliable broad H$\beta$ and \OIII\, luminosities span the same redshift range with corresponding luminosity ranges of $10^{39.79}  \, \mathrm{erg \, s^{-1}} \leq L_{\text{H}\beta} \leq 10^{44.05}  \, \mathrm{erg \, s^{-1}}$ and $10^{39.63}  \, \mathrm{erg \, s^{-1}} \leq L_{\mathrm{{\OIII}}} \leq 10^{43.54}  \, \mathrm{erg \, s^{-1}}$, respectively. Therefore, these samples provide comprehensive coverage of $z$ and $L$ for exploring radius-luminosity correlations across different AGN luminosity indicators. In Table~\ref{tab_sample_overview}, we provide an overview of the five samples corresponding to low-redshift sources with BLR radius measurements based on broad H$\beta$ and H$\alpha$ lines: three samples studied here with different luminosities and two samples studied in \cite{cao_2025a, cao25}. In Table~\ref{tab:sam_avg} we list the measurements of the three samples we study here. 

Since our sample covers a wide redshift range, we apply corrections to account for the Milky Way’s peculiar velocity, which has a particularly strong impact at low redshifts \citep[see, e.g.,][]{Khadkaetal2021c,Khadka:2022aeg}. These corrections were calculated using the NED Velocity Correction Calculator \footnote{\url{https://ned.ipac.caltech.edu/help/velc_help.html}}, which incorporates multiple components of local motion, including Galactic rotation, the Milky Way’s motion within the Local Group, the Local Group’s infall toward the center of the Local Supercluster, and the motion relative to the CMB reference frame. We apply these corrections consistently to all AGN in our sample, ensuring that the resulting redshifts more accurately represent their true cosmological values by incorporating the effects of local peculiar velocities. The final, corrected redshifts are presented in Table~\ref{tab:sam_avg}.

\subsection{$H(z)$ and BAO sample and data}

In addition, we incorporate 32 $H(z)$ measurements, derived from cosmic chronometers, along with 12 BAO measurements. Together, these datasets constitute the combined $H(z)$ + BAO sample used in our analyses. The $H(z)$ data span the redshift range $0.070 \leq z \leq 1.965$, while the BAO measurements extend from $0.122 \leq z \leq 2.334$. A comprehensive description of these datasets can be found in tables 1 and 2 and in Sec.~III of \citet{CaoRatra2023}.

\begin{turnpage}
\begin{table*}
\centering
\resizebox{2.73\columnwidth}{!}{%
\begin{threeparttable}
\caption{Properties of the Average Scheme sample.}
\label{tab:sam_avg}
\setlength{\tabcolsep}{18pt}
\begin{tabular}{lcccccccr}
\toprule\toprule
Object & $z$ & $\log F_{5100}$ ($\mathrm{erg \, s^{-1} \, cm^{-2}}$) & $\log L_{5100}$ ($\mathrm{erg \, s^{-1}}$) & $\log F_{\mathrm{H\beta}}$ ($\mathrm{erg \, s^{-1} \, cm^{-2}}$) & $\log L_{\mathrm{H\beta}}$ ($\mathrm{erg \, s^{-1}}$) & $\log F_{\mathrm{{\OIII}}}$ ($\mathrm{erg \, s^{-1} \, cm^{-2}}$) & $\log L_{\mathrm{{\OIII}}}$ ($\mathrm{erg \, s^{-1}}$) & $\tau_{\mathrm{H\beta, ICCF}}$ (days) \\
\midrule

3C120 & 0.03279 & $ -10.426 \pm 0.09 $ & $ 43.944 \pm 0.09 $ & $ -12.066 \pm 0.06 $ & $ 42.304 \pm 0.06 $ & $ -12.116 \pm 0.07 $ & $ 42.254 \pm 0.07 $ & $ 33.9 _ {-14.8} ^ {+22.6} $ \\
3C382 & 0.05512 & $ -10.89 \pm 0.19 $ & $ 43.946 \pm 0.19 $ & $ -12.4 \pm 0.03 $ & $ 42.436 \pm 0.03 $ & $ -13.02 \pm 0.03 $ & $ 41.816 \pm 0.03 $ & $ 24.1 _ {-15.5} ^ {+15.5} $ \\
3C390.3 & 0.0559 & $ -11.061 \pm 0.15 $ & $ 43.787 \pm 0.15 $ & $ -12.621 \pm 0.05 $ & $ 42.227 \pm 0.05 $ & $ -12.711 \pm 0.03 $ & $ 42.137 \pm 0.03 $ & $ 77.6 _ {-69.7} ^ {+80.4} $ \\
Ark120 & 0.03274 & $ -10.597 \pm 0.23 $ & $ 43.771 \pm 0.23 $ & $ -11.997 \pm 0.07 $ & $ 42.371 \pm 0.07 $ & $ -12.887 \pm 0.05 $ & $ 41.481 \pm 0.05 $ & $ 26.3 _ {-9.1} ^ {+9.1} $ \\
IRASF12397+3333 & 0.04439 & $ -10.452 \pm 0.04 $ & $ 44.188 \pm 0.04 $ & $ -12.422 \pm 0.04 $ & $ 42.218 \pm 0.04 $ & $ -12.332 \pm 0.04 $ & $ 42.308 \pm 0.04 $ & $ 11.5 _ {-1.8} ^ {+2.9} $ \\
J074352.02+271239.5 & 0.25334 & $ -10.952 \pm 0.02 $ & $ 45.315 \pm 0.02 $ & $ -12.842 \pm 0.01 $ & $ 43.425 \pm 0.01 $ & $ -13.902 \pm 0.05 $ & $ 42.365 \pm 0.05 $ & $ 66.0 _ {-18.2} ^ {+7.6} $ \\
J075101.42+291419.1 & 0.12141 & $ -11.436 \pm 0.07 $ & $ 44.124 \pm 0.07 $ & $ -13.296 \pm 0.07 $ & $ 42.264 \pm 0.07 $ & $ -13.836 \pm 0.04 $ & $ 41.724 \pm 0.04 $ & $ 24.0 _ {-5.5} ^ {+6.6} $ \\
J075949.54+320023.8 & 0.18836 & $ -11.846 \pm 0.06 $ & $ 44.132 \pm 0.06 $ & $ -13.566 \pm 0.04 $ & $ 42.412 \pm 0.04 $ & $ -14.726 \pm 0.07 $ & $ 41.252 \pm 0.07 $ & $ 18.6 _ {-5.6} ^ {+11.1} $ \\
J080101.41+184840.7 & 0.14027 & $ -11.482 \pm 0.03 $ & $ 44.215 \pm 0.03 $ & $ -13.172 \pm 0.02 $ & $ 42.525 \pm 0.02 $ & $ -14.592 \pm 0.05 $ & $ 41.105 \pm 0.05 $ & $ 9.1 _ {-4.0} ^ {+7.3} $ \\
J081441.91+212918.5 & 0.16338 & $ -11.917 \pm 0.07 $ & $ 43.925 \pm 0.07 $ & $ -13.487 \pm 0.03 $ & $ 42.355 \pm 0.03 $ & $ -14.567 \pm 0.07 $ & $ 41.275 \pm 0.07 $ & $ 21.4 _ {-7.9} ^ {+8.9} $ \\
J083553.46+055317.1 & 0.2054 & $ -11.68 \pm 0.02 $ & $ 44.381 \pm 0.02 $ & $ -13.64 \pm 0.02 $ & $ 42.421 \pm 0.02 $ & $ -14.28 \pm 0.03 $ & $ 41.781 \pm 0.03 $ & $ 28.8 _ {-4.6} ^ {+5.3} $ \\
J084533.28+474934.5 & 0.30251 & $ -11.984 \pm 0.02 $ & $ 44.46 \pm 0.02 $ & $ -13.904 \pm 0.03 $ & $ 42.54 \pm 0.03 $ & $ -14.754 \pm 0.03 $ & $ 41.69 \pm 0.03 $ & $ 21.9 _ {-4.0} ^ {+9.1} $ \\
J093302.68+385228.0 & 0.17802 & $ -11.669 \pm 0.13 $ & $ 44.254 \pm 0.13 $ & $ -13.879 \pm 0.05 $ & $ 42.044 \pm 0.05 $ & $ -14.719 \pm 0.06 $ & $ 41.204 \pm 0.06 $ & $ 20.9 _ {-4.8} ^ {+4.3} $ \\
J093922.89+370943.9 & 0.18688 & $ -11.955 \pm 0.04 $ & $ 44.015 \pm 0.04 $ & $ -13.935 \pm 0.04 $ & $ 42.035 \pm 0.04 $ & $ -14.715 \pm 0.13 $ & $ 41.255 \pm 0.13 $ & $ 10.7 _ {-3.2} ^ {+3.2} $ \\
J101000.68+300321.5 & 0.25702 & $ -11.579 \pm 0.02 $ & $ 44.702 \pm 0.02 $ & $ -13.569 \pm 0.02 $ & $ 42.712 \pm 0.02 $ & $ -14.839 \pm 0.11 $ & $ 41.442 \pm 0.11 $ & $ 38.0 _ {-6.1} ^ {+11.4} $ \\
J102339.64+523349.6 & 0.13695 & $ -11.64 \pm 0.03 $ & $ 44.034 \pm 0.03 $ & $ -13.59 \pm 0.03 $ & $ 42.084 \pm 0.03 $ & $ -14.41 \pm 0.04 $ & $ 41.264 \pm 0.04 $ & $ 26.3 _ {-7.9} ^ {+5.4} $ \\
MCG+04-22-042 & 0.03405 & $ -11.207 \pm 0.11 $ & $ 43.197 \pm 0.11 $ & $\cdots$ & $\cdots$ & $\cdots$ & $\cdots$ & $ 13.5 _ {-1.9} ^ {+3.1} $ \\
MCG+08-11-011 & 0.02051 & $ -10.674 \pm 0.04 $ & $ 43.28 \pm 0.04 $ & $ -12.354 \pm 0.09 $ & $ 41.6 \pm 0.09 $ & $ -11.954 \pm 0.07 $ & $ 42.0 \pm 0.07 $ & $ 15.5 _ {-0.4} ^ {+0.7} $ \\
Mrk110 & 0.03584 & $ -10.766 \pm 0.19 $ & $ 43.684 \pm 0.19 $ & $ -12.526 \pm 0.04 $ & $ 41.924 \pm 0.04 $ & $ -12.626 \pm 0.05 $ & $ 41.824 \pm 0.05 $ & $ 27.5 _ {-6.3} ^ {+4.4} $ \\
Mrk1310 & 0.02077 & $ -11.684 \pm 0.17 $ & $ 42.281 \pm 0.17 $ & $ -13.414 \pm 0.1 $ & $ 40.551 \pm 0.1 $ & $ -12.924 \pm 0.08 $ & $ 41.041 \pm 0.08 $ & $ 3.7 _ {-0.5} ^ {+0.5} $ \\
Mrk1392 & 0.03652 & $ -11.306 \pm 0.16 $ & $ 43.16 \pm 0.16 $ & $\cdots$ & $\cdots$ & $\cdots$ & $\cdots$ & $ 28.2 _ {-4.5} ^ {+3.2} $ \\
Mrk142 & 0.04521 & $ -11.141 \pm 0.05 $ & $ 43.516 \pm 0.05 $ & $ -13.111 \pm 0.04 $ & $ 41.546 \pm 0.04 $ & $ -13.851 \pm 0.04 $ & $ 40.806 \pm 0.04 $ & $ 5.6 _ {-1.4} ^ {+1.6} $ \\
Mrk1511 & 0.03458 & $ -11.36 \pm 0.06 $ & $ 43.058 \pm 0.06 $ & $ -12.94 \pm 0.06 $ & $ 41.478 \pm 0.06 $ & $ -13.44 \pm 0.05 $ & $ 40.978 \pm 0.05 $ & $ 6.5 _ {-0.9} ^ {+1.0} $ \\
Mrk202 & 0.02359 & $ -11.775 \pm 0.18 $ & $ 42.303 \pm 0.18 $ & $ -13.635 \pm 0.09 $ & $ 40.443 \pm 0.09 $ & $ -13.535 \pm 0.07 $ & $ 40.543 \pm 0.07 $ & $ 5.2 _ {-3.3} ^ {+7.4} $ \\
Mrk290 & 0.03037 & $ -11.169 \pm 0.06 $ & $ 43.133 \pm 0.06 $ & $ -12.699 \pm 0.06 $ & $ 41.603 \pm 0.06 $ & $ -12.699 \pm 0.06 $ & $ 41.603 \pm 0.06 $ & $ 9.5 _ {-1.1} ^ {+1.3} $ \\
Mrk335 & 0.02461 & $ -10.497 \pm 0.12 $ & $ 43.618 \pm 0.12 $ & $ -12.157 \pm 0.11 $ & $ 41.958 \pm 0.11 $ & $ -12.637 \pm 0.08 $ & $ 41.478 \pm 0.08 $ & $ 13.2 _ {-4.6} ^ {+5.2} $ \\
Mrk374 & 0.04283 & $ -10.854 \pm 0.04 $ & $ 43.755 \pm 0.04 $ & $ -12.834 \pm 0.04 $ & $ 41.775 \pm 0.04 $ & $ -13.074 \pm 0.04 $ & $ 41.535 \pm 0.04 $ & $ 14.8 _ {-5.8} ^ {+3.4} $ \\
Mrk40 & 0.02128 & $ -11.519 \pm 0.1 $ & $ 42.467 \pm 0.1 $ & $ -13.089 \pm 0.11 $ & $ 40.897 \pm 0.11 $ & $ -13.289 \pm 0.07 $ & $ 40.697 \pm 0.07 $ & $ 4.8 _ {-0.9} ^ {+0.7} $ \\
Mrk50 & 0.02502 & $ -11.211 \pm 0.09 $ & $ 42.919 \pm 0.09 $ & $\cdots$ & $\cdots$ & $\cdots$ & $\cdots$ & $ 7.8 _ {-0.9} ^ {+1.2} $ \\
Mrk509 & 0.03346 & $ -10.283 \pm 0.05 $ & $ 44.105 \pm 0.05 $ & $ -11.863 \pm 0.04 $ & $ 42.525 \pm 0.04 $ & $ -12.093 \pm 0.05 $ & $ 42.295 \pm 0.05 $ & $ 77.7 _ {-5.4} ^ {+5.4} $ \\
Mrk590 & 0.02553 & $ -10.758 \pm 0.19 $ & $ 43.39 \pm 0.19 $ & $ -12.428 \pm 0.15 $ & $ 41.72 \pm 0.15 $ & $ -12.938 \pm 0.06 $ & $ 41.21 \pm 0.06 $ & $ 22.9 _ {-8.4} ^ {+7.9} $ \\
Mrk6 & 0.01951 & $ -10.128 \pm 0.06 $ & $ 43.783 \pm 0.06 $ & $\cdots$ & $\cdots$ & $\cdots$ & $\cdots$ & $ 9.8 _ {-7.2} ^ {+3.6} $ \\
Mrk704 & 0.02962 & $ -10.627 \pm 0.11 $ & $ 43.653 \pm 0.11 $ & $ --$ & $\cdots$ & $\cdots$ & $\cdots$ & $ 19.9 _ {-7.3} ^ {+8.3} $ \\
Mrk79 & 0.02258 & $ -10.384 \pm 0.08 $ & $ 43.655 \pm 0.08 $ & $ -12.164 \pm 0.08 $ & $ 41.875 \pm 0.08 $ & $ -12.414 \pm 0.07 $ & $ 41.625 \pm 0.07 $ & $ 16.2 _ {-6.7} ^ {+6.7} $ \\
Mrk817 & 0.0316 & $ -10.644 \pm 0.09 $ & $ 43.693 \pm 0.09 $ & $ -12.444 \pm 0.12 $ & $ 41.893 \pm 0.12 $ & $ -12.864 \pm 0.05 $ & $ 41.473 \pm 0.05 $ & $ 19.1 _ {-9.7} ^ {+9.2} $ \\
NGC2617 & 0.01517 & $ -11.021 \pm 0.1 $ & $ 42.668 \pm 0.1 $ & $ -12.511 \pm 0.12 $ & $ 41.178 \pm 0.12 $ & $ -13.121 \pm 0.1 $ & $ 40.568 \pm 0.1 $ & $ 4.5 _ {-1.4} ^ {+1.0} $ \\
NGC3227 & 0.00483 & $ -10.262 \pm 0.11 $ & $ 42.426 \pm 0.11 $ & $ -12.182 \pm 0.1 $ & $ 40.506 \pm 0.1 $ & $ -11.882 \pm 0.1 $ & $ 40.806 \pm 0.1 $ & $ 4.3 _ {-1.0} ^ {+1.0} $ \\
NGC3516 & 0.00906 & $ -10.572 \pm 0.23 $ & $ 42.665 \pm 0.23 $ & $ -12.212 \pm 0.18 $ & $ 41.025 \pm 0.18 $ & $ -12.422 \pm 0.17 $ & $ 40.815 \pm 0.17 $ & $ 9.5 _ {-2.6} ^ {+2.4} $ \\
NGC3783 & 0.01079 & $ -10.747 \pm 0.18 $ & $ 42.643 \pm 0.18 $ & $ -12.347 \pm 0.18 $ & $ 41.043 \pm 0.18 $ & $ -12.407 \pm 0.17 $ & $ 40.983 \pm 0.17 $ & $ 3.8 _ {-3.5} ^ {+3.0} $ \\
NGC4051 & 0.00308 & $ -10.312 \pm 0.11 $ & $ 41.984 \pm 0.11 $ & $ -12.112 \pm 0.18 $ & $ 40.184 \pm 0.18 $ & $ -11.932 \pm 0.17 $ & $ 40.364 \pm 0.17 $ & $ 5.7 _ {-1.5} ^ {+1.9} $ \\

NGC4151 & 0.00415 & $ -10.146 \pm 0.12 $ & $ 42.41 \pm 0.12 $ & $ -11.856 \pm 0.2 $ & $ 40.7 \pm 0.2 $ & $ -10.816 \pm 0.17 $ & $ 41.74 \pm 0.17 $ & $ 6.8 _ {-0.5} ^ {+0.5} $ \\
NGC4593 & 0.00946 & $ -10.871 \pm 0.07 $ & $ 42.404 \pm 0.07 $ & $ -12.561 \pm 0.18 $ & $ 40.714 \pm 0.18 $ & $ -12.911 \pm 0.18 $ & $ 40.364 \pm 0.18 $ & $ 3.2 _ {-2.3} ^ {+1.5} $ \\
NGC4748 & 0.01575 & $ -11.165 \pm 0.13 $ & $ 42.557 \pm 0.13 $ & $ -12.735 \pm 0.1 $ & $ 40.987 \pm 0.1 $ & $ -12.385 \pm 0.1 $ & $ 41.337 \pm 0.1 $ & $ 5.7 _ {-1.6} ^ {+1.9} $ \\
NGC5548 & 0.01788 & $ -10.599 \pm 0.22 $ & $ 43.234 \pm 0.22 $ & $ -12.239 \pm 0.21 $ & $ 41.594 \pm 0.21 $ & $ -12.229 \pm 0.09 $ & $ 41.604 \pm 0.09 $ & $ 11.7 _ {-6.8} ^ {+6.8} $ \\
NGC6814 & 0.00451 & $ -10.702 \pm 0.29 $ & $ 41.926 \pm 0.29 $ & $ -12.312 \pm 0.28 $ & $ 40.316 \pm 0.28 $ & $ -12.552 \pm 0.28 $ & $ 40.076 \pm 0.28 $ & $ 6.5 _ {-1.0} ^ {+0.9} $ \\
NGC7469 & 0.01503 & $ -10.312 \pm 0.05 $ & $ 43.369 \pm 0.05 $ & $ -12.212 \pm 0.1 $ & $ 41.469 \pm 0.1 $ & $ -12.122 \pm 0.09 $ & $ 41.559 \pm 0.09 $ & $ 16.6 _ {-2.3} ^ {+2.3} $ \\
NPM1G+27.0587 & 0.05946 & $ -10.952 \pm 0.04 $ & $ 43.952 \pm 0.04 $ & $\cdots$ & $\cdots$ & $\cdots$ & $\cdots$ & $ 13.5 _ {-4.4} ^ {+3.7} $ \\
PG0026+129 & 0.14424 & $ -10.808 \pm 0.03 $ & $ 44.915 \pm 0.03 $ & $ -12.838 \pm 0.04 $ & $ 42.885 \pm 0.04 $ & $ -13.168 \pm 0.02 $ & $ 42.555 \pm 0.02 $ & $ 50.0 _ {-47.2} ^ {+46.1} $ \\
PG0804+761 & 0.10006 & $ -10.49 \pm 0.05 $ & $ 44.891 \pm 0.05 $ & $ -12.14 \pm 0.04 $ & $ 43.241 \pm 0.04 $ & $ -12.97 \pm 0.06 $ & $ 42.411 \pm 0.06 $ & $ 104.7 _ {-38.6} ^ {+38.6} $ \\
PG0923+201 & 0.19368 & $ -10.747 \pm 0.06 $ & $ 45.258 \pm 0.06 $ & $ -12.667 \pm 0.04 $ & $ 43.338 \pm 0.04 $ & $ -13.527 \pm 0.04 $ & $ 42.478 \pm 0.04 $ & $ 93.2 _ {-10.7} ^ {+10.7} $ \\
PG0953+414 & 0.23487 & $ -11.06 \pm 0.01 $ & $ 45.133 \pm 0.01 $ & $ -12.96 \pm 0.04 $ & $ 43.233 \pm 0.04 $ & $ -13.52 \pm 0.02 $ & $ 42.673 \pm 0.02 $ & $ 154.8 _ {-17.8} ^ {+17.8} $ \\
PG1001+291 & 0.3301 & $ -10.973 \pm 0.03 $ & $ 45.559 \pm 0.03 $ & $ -13.033 \pm 0.04 $ & $ 43.499 \pm 0.04 $ & $ -14.173 \pm 0.04 $ & $ 42.359 \pm 0.04 $ & $ 42.6 _ {-6.9} ^ {+5.9} $ \\
PG1226+023 & 0.15949 & $ -9.931 \pm 0.04 $ & $ 45.887 \pm 0.04 $ & $ -11.761 \pm 0.03 $ & $ 44.057 \pm 0.03 $ & $ -12.631 \pm 0.02 $ & $ 43.187 \pm 0.02 $ & $ 234.2 _ {-102.5} ^ {+91.7} $ \\
PG1229+204 & 0.06458 & $ -11.316 \pm 0.06 $ & $ 43.662 \pm 0.06 $ & $ -12.706 \pm 0.06 $ & $ 42.272 \pm 0.06 $ & $ -13.266 \pm 0.03 $ & $ 41.712 \pm 0.03 $ & $ 37.1 _ {-15.4} ^ {+28.2} $ \\
PG1310-108 & 0.03534 & $ -11.07 \pm 0.06 $ & $ 43.367 \pm 0.06 $ & $ -11.207 \pm 0.11 $ & $ 43.23 \pm 0.11 $ & $ -11.207 \pm 0.11 $ & $ 43.23 \pm 0.11 $ & $ 6.6 _ {-2.9} ^ {+2.1} $ \\
PG1402+261 & 0.16404 & $ -11.107 \pm 0.05 $ & $ 44.738 \pm 0.05 $ & $\cdots$ & $\cdots$ & $\cdots$ & $\cdots$ & $ 66.1 _ {-6.1} ^ {+4.6} $ \\
PG1404+226 & 0.09908 & $ -11.243 \pm 0.03 $ & $ 44.128 \pm 0.03 $ & $ -13.153 \pm 0.04 $ & $ 42.218 \pm 0.04 $ & $ -13.933 \pm 0.04 $ & $ 41.438 \pm 0.04 $ & $ 19.9 _ {-3.7} ^ {+3.7} $ \\
PG1415+451 & 0.1142 & $ -11.265 \pm 0.04 $ & $ 44.238 \pm 0.04 $ & $ -13.155 \pm 0.04 $ & $ 42.348 \pm 0.04 $ & $ -14.465 \pm 0.04 $ & $ 41.038 \pm 0.04 $ & $ 30.2 _ {-4.2} ^ {+4.2} $ \\
PG1448+273 & 0.06506 & $ -10.739 \pm 0.03 $ & $ 44.246 \pm 0.03 $ & $ -12.799 \pm 0.04 $ & $ 42.186 \pm 0.04 $ & $ -13.019 \pm 0.04 $ & $ 41.966 \pm 0.04 $ & $ 31.6 _ {-4.4} ^ {+5.1} $ \\
PG1519+226 & 0.13656 & $ -11.171 \pm 0.06 $ & $ 44.5 \pm 0.06 $ & $ -12.871 \pm 0.04 $ & $ 42.8 \pm 0.04 $ & $ -14.431 \pm 0.04 $ & $ 41.24 \pm 0.04 $ & $ 70.7 _ {-6.5} ^ {+4.9} $ \\
PG1535+547 & 0.0391 & $ -10.663 \pm 0.02 $ & $ 43.864 \pm 0.02 $ & $ -12.333 \pm 0.04 $ & $ 42.194 \pm 0.04 $ & $ -13.253 \pm 0.04 $ & $ 41.274 \pm 0.04 $ & $ 25.7 _ {-3.0} ^ {+3.5} $ \\
PG1617+175 & 0.11482 & $ -11.078 \pm 0.07 $ & $ 44.43 \pm 0.07 $ & $ -12.718 \pm 0.09 $ & $ 42.79 \pm 0.09 $ & $ -13.858 \pm 0.18 $ & $ 41.65 \pm 0.18 $ & $ 48.9 _ {-40.5} ^ {+29.3} $ \\
PG1626+554 & 0.13361 & $ -11.026 \pm 0.06 $ & $ 44.624 \pm 0.06 $ & $ -12.756 \pm 0.04 $ & $ 42.894 \pm 0.04 $ & $ -13.706 \pm 0.04 $ & $ 41.944 \pm 0.04 $ & $ 77.5 _ {-3.6} ^ {+3.6} $ \\

\bottomrule\bottomrule
\end{tabular}
\end{threeparttable}%
}
\end{table*}
\end{turnpage}

\begin{turnpage}
\begin{table*}
\centering
\resizebox{2.73\columnwidth}{!}{%
\begin{threeparttable}
\setlength{\tabcolsep}{18pt}
\begin{tabular}{lcccccccr}
\toprule\toprule
Object & $z$ & $\log F_{5100}$ ($\mathrm{erg \, s^{-1} \, cm^{-2}}$) & $\log L_{5100}$ ($\mathrm{erg \, s^{-1}}$) & $\log F_{\mathrm{H\beta}}$ ($\mathrm{erg \, s^{-1} \, cm^{-2}}$) & $\log L_{\mathrm{H\beta}}$ ($\mathrm{erg \, s^{-1}}$) & $\log F_{\mathrm{{\OIII}}}$ ($\mathrm{erg \, s^{-1} \, cm^{-2}}$) & $\log L_{\mathrm{{\OIII}}}$ ($\mathrm{erg \, s^{-1}}$) & $\tau_{\mathrm{H\beta, ICCF}}$ (days) \\
(1) & (2) & (3) & (4) & (5) & (6) & (7) & (8) & (9) \\
\midrule
PG2130+099 & 0.06217 & $ -10.686 \pm 0.12 $ & $ 44.258 \pm 0.12 $ & $ -12.246 \pm 0.04 $ & $ 42.698 \pm 0.04 $ & $ -12.816 \pm 0.04 $ & $ 42.128 \pm 0.04 $ & $ 18.6 _ {-4.7} ^ {+3.0} $ \\
PG2209+184 & 0.06873 & $ -11.742 \pm 0.11 $ & $ 43.293 \pm 0.11 $ & $\cdots$ & $\cdots$ & $\cdots$ & $\cdots$ & $ 13.5 _ {-2.5} ^ {+2.8} $ \\
RBS1303 & 0.04277 & $ -11.257 \pm 0.07 $ & $ 43.351 \pm 0.07 $ & $\cdots$ & $\cdots$ & $\cdots$ & $\cdots$ & $ 18.6 _ {-3.9} ^ {+3.0} $ \\
RBS1917 & 0.06477 & $ -11.358 \pm 0.04 $ & $ 43.623 \pm 0.04 $ & $\cdots$ & $\cdots$ & $\cdots$ & $\cdots$ & $ 12.0 _ {-4.4} ^ {+4.2} $ \\
RM017 & 0.45657 & $ -12.972 \pm 0.01 $ & $ 43.891 \pm 0.01 $ & $ -14.692 \pm 0.01 $ & $ 42.171 \pm 0.01 $ & $ -14.912 \pm 0.01 $ & $ 41.951 \pm 0.01 $ & $ 21.9 _ {-7.0} ^ {+4.5} $ \\
RM191 & 0.44206 & $ -13.21 \pm 0.01 $ & $ 43.62 \pm 0.01 $ & $ -15.1 \pm 0.01 $ & $ 41.73 \pm 0.01 $ & $ -15.4 \pm 0.01 $ & $ 41.43 \pm 0.01 $ & $ 8.7 _ {-1.6} ^ {+1.8} $ \\
RM229 & 0.47008 & $ -13.364 \pm 0.01 $ & $ 43.53 \pm 0.01 $ & $ -14.984 \pm 0.02 $ & $ 41.91 \pm 0.02 $ & $ -15.484 \pm 0.01 $ & $ 41.41 \pm 0.01 $ & $ 14.5 _ {-4.0} ^ {+3.0} $ \\
RM265 & 0.7357 & $ -13.193 \pm 0.01 $ & $ 44.172 \pm 0.01 $ & $ -14.923 \pm 0.03 $ & $ 42.442 \pm 0.03 $ & $ -15.873 \pm 0.03 $ & $ 41.492 \pm 0.03 $ & $ 9.1 _ {-7.1} ^ {+4.6} $ \\
RM267 & 0.5877 & $ -13.056 \pm 0.01 $ & $ 44.071 \pm 0.01 $ & $ -14.616 \pm 0.02 $ & $ 42.511 \pm 0.02 $ & $ -15.266 \pm 0.01 $ & $ 41.861 \pm 0.01 $ & $ 20.4 _ {-2.3} ^ {+2.3} $ \\
RM272 & 0.26303 & $ -12.404 \pm 0.02 $ & $ 43.9 \pm 0.02 $ & $ -13.964 \pm 0.01 $ & $ 42.34 \pm 0.01 $ & $ -14.284 \pm 0.01 $ & $ 42.02 \pm 0.01 $ & $ 17.0 _ {-5.1} ^ {+4.7} $ \\
RM301 & 0.54846 & $ -13.024 \pm 0.01 $ & $ 44.031 \pm 0.01 $ & $ -14.814 \pm 0.01 $ & $ 42.241 \pm 0.01 $ & $ -14.994 \pm 0.01 $ & $ 42.061 \pm 0.01 $ & $ 12.0 _ {-4.7} ^ {+3.6} $ \\
RM320 & 0.26493 & $ -12.912 \pm 0.01 $ & $ 43.4 \pm 0.01 $ & $ -14.442 \pm 0.01 $ & $ 41.87 \pm 0.01 $ & $ -14.952 \pm 0.01 $ & $ 41.36 \pm 0.01 $ & $ 23.4 _ {-10.8} ^ {+5.9} $ \\
RM377 & 0.33701 & $ -13.213 \pm 0.01 $ & $ 43.34 \pm 0.01 $ & $ -15.043 \pm 0.02 $ & $ 41.51 \pm 0.02 $ & $ -15.363 \pm 0.01 $ & $ 41.19 \pm 0.01 $ & $ 7.1 _ {-2.9} ^ {+10.9} $ \\
RM392 & 0.8429 & $ -13.25 \pm 0.01 $ & $ 44.26 \pm 0.01 $ & $ -15.02 \pm 0.04 $ & $ 42.49 \pm 0.04 $ & $ -15.62 \pm 0.07 $ & $ 41.89 \pm 0.07 $ & $ 13.5 _ {-5.0} ^ {+3.1} $ \\
RM519 & 0.55421 & $ -13.915 \pm 0.01 $ & $ 43.15 \pm 0.01 $ & $ -15.655 \pm 0.03 $ & $ 41.41 \pm 0.03 $ & $ -15.785 \pm 0.01 $ & $ 41.28 \pm 0.01 $ & $ 11.0 _ {-3.8} ^ {+4.5} $ \\
RM551 & 0.68079 & $ -13.311 \pm 0.01 $ & $ 43.971 \pm 0.01 $ & $ -14.841 \pm 0.08 $ & $ 42.441 \pm 0.08 $ & $ -15.441 \pm 0.01 $ & $ 41.841 \pm 0.01 $ & $ 6.5 _ {-3.1} ^ {+7.4} $ \\
RM694 & 0.53277 & $ -12.893 \pm 0.01 $ & $ 44.132 \pm 0.01 $ & $ -14.763 \pm 0.02 $ & $ 42.262 \pm 0.02 $ & $ -14.773 \pm 0.01 $ & $ 42.252 \pm 0.01 $ & $ 14.8 _ {-5.4} ^ {+5.4} $ \\

RM775 & 0.17293 & $ -12.37 \pm 0.01 $ & $ 43.525 \pm 0.01 $ & $ -14.1 \pm 0.01 $ & $ 41.795 \pm 0.01 $ & $ -14.85 \pm 0.01 $ & $ 41.045 \pm 0.01 $ & $ 19.0 _ {-10.1} ^ {+6.6} $ \\
RM776 & 0.11655 & $ -12.428 \pm 0.01 $ & $ 43.094 \pm 0.01 $ & $ -14.098 \pm 0.02 $ & $ 41.424 \pm 0.02 $ & $ -14.368 \pm 0.01 $ & $ 41.154 \pm 0.01 $ & $ 10.0 _ {-2.8} ^ {+2.8} $ \\
RM779 & 0.15281 & $ -12.712 \pm 0.01 $ & $ 43.065 \pm 0.01 $ & $ -14.412 \pm 0.01 $ & $ 41.365 \pm 0.01 $ & $ -14.912 \pm 0.01 $ & $ 40.865 \pm 0.01 $ & $ 11.2 _ {-7.2} ^ {+4.9} $ \\
RM790 & 0.23754 & $ -12.972 \pm 0.01 $ & $ 43.232 \pm 0.01 $ & $ -14.362 \pm 0.03 $ & $ 41.842 \pm 0.03 $ & $ -15.232 \pm 0.01 $ & $ 40.972 \pm 0.01 $ & $ 8.7 _ {-2.8} ^ {+4.0} $ \\
RM840 & 0.2443 & $ -13.08 \pm 0.01 $ & $ 43.151 \pm 0.01 $ & $ -14.66 \pm 0.01 $ & $ 41.571 \pm 0.01 $ & $ -15.14 \pm 0.01 $ & $ 41.091 \pm 0.01 $ & $ 5.8 _ {-2.0} ^ {+2.1} $ \\
RXJ2044.0+2833 & 0.04907 & $ -11.077 \pm 0.04 $ & $ 43.653 \pm 0.04 $ & $\cdots$ & $\cdots$ & $\cdots$ & $\cdots$ & $ 14.1 _ {-2.0} ^ {+2.0} $ \\
SBS1116+583A & 0.02836 & $ -12.156 \pm 0.28 $ & $ 42.085 \pm 0.28 $ & $ -13.586 \pm 0.07 $ & $ 40.655 \pm 0.07 $ & $ -13.786 \pm 0.06 $ & $ 40.455 \pm 0.06 $ & $ 2.2 _ {-0.5} ^ {+0.9} $ \\
UGC06728 & 0.00659 & $ -11.117 \pm 0.24 $ & $ 41.842 \pm 0.24 $ & $ -13.157 \pm 0.05 $ & $ 39.802 \pm 0.05 $ & $ -13.317 \pm 0.02 $ & $ 39.642 \pm 0.02 $ & $ 1.1 _ {-0.9} ^ {+0.5} $ \\
Zw229-015 & 0.02735 & $ -11.546 \pm 0.06 $ & $ 42.662 \pm 0.06 $ & $\cdots$ & $ --$ & $\cdots$ & $\cdots$ & $ 4.1 _ {-1.0} ^ {+1.4} $ \\
Zw535-012 & 0.04673 & $ -11.063 \pm 0.06 $ & $ 43.623 \pm 0.06 $ & $\cdots$ & $\cdots$ & $\cdots$ & $\cdots$ & $ 19.1 _ {-4.0} ^ {+7.0} $ \\
J0101+422 & 0.18909 & $ -11.166 \pm 0.01 $ & $ 44.815 \pm 0.01 $ & $ -12.946 \pm 0.01 $ & $ 43.035 \pm 0.01 $ & $ -13.746 \pm 0.02 $ & $ 42.235 \pm 0.02 $ & $ 75.9 _ {-12.2} ^ {+14.0} $ \\
J0140+234 & 0.32204 & $ -11.37 \pm 0.01 $ & $ 45.136 \pm 0.01 $ & $ -13.04 \pm 0.01 $ & $ 43.466 \pm 0.01 $ & $ -13.4 \pm 0.02 $ & $ 43.106 \pm 0.02 $ & $ 114.6 _ {-10.6} ^ {+10.6} $ \\
J0939+375 & 0.23194 & $ -11.717 \pm 0.01 $ & $ 44.464 \pm 0.01 $ & $ -13.607 \pm 0.01 $ & $ 42.574 \pm 0.01 $ & $ -14.277 \pm 0.02 $ & $ 41.904 \pm 0.02 $ & $ 19.5 _ {-13.9} ^ {+9.0} $ \\
PG0947+396 & 0.20633 & $ -11.482 \pm 0.01 $ & $ 44.584 \pm 0.01 $ & $ -13.062 \pm 0.01 $ & $ 43.004 \pm 0.01 $ & $ -13.652 \pm 0.02 $ & $ 42.414 \pm 0.02 $ & $ 36.3 _ {-10.9} ^ {+9.2} $ \\
J1026+523 & 0.25979 & $ -11.81 \pm 0.01 $ & $ 44.483 \pm 0.01 $ & $ -13.43 \pm 0.01 $ & $ 42.863 \pm 0.01 $ & $ -13.74 \pm 0.02 $ & $ 42.553 \pm 0.02 $ & $ 33.9 _ {-3.9} ^ {+3.9} $ \\
PG1100+772 & 0.31163 & $ -10.833 \pm 0.01 $ & $ 45.64 \pm 0.01 $ & $ -12.653 \pm 0.02 $ & $ 43.82 \pm 0.02 $ & $ -12.933 \pm 0.03 $ & $ 43.54 \pm 0.03 $ & $ 53.7 _ {-22.3} ^ {+14.8} $ \\
J1120+423 & 0.22702 & $ -11.626 \pm 0.01 $ & $ 44.534 \pm 0.01 $ & $ -13.306 \pm 0.01 $ & $ 42.854 \pm 0.01 $ & $ -13.786 \pm 0.02 $ & $ 42.374 \pm 0.02 $ & $ 44.6 _ {-15.4} ^ {+15.4} $ \\
PG1121+422 & 0.22568 & $ -11.31 \pm 0.01 $ & $ 44.843 \pm 0.01 $ & $ -12.92 \pm 0.01 $ & $ 43.233 \pm 0.01 $ & $ -13.92 \pm 0.01 $ & $ 42.233 \pm 0.01 $ & $ 114.7 _ {-21.1} ^ {+23.8} $ \\
PG1202+281 & 0.166 & $ -11.411 \pm 0.01 $ & $ 44.446 \pm 0.01 $ & $ -13.051 \pm 0.01 $ & $ 42.806 \pm 0.01 $ & $ -13.201 \pm 0.02 $ & $ 42.656 \pm 0.02 $ & $ 38.9 _ {-9.0} ^ {+9.0} $ \\
J1217+333 & 0.17888 & $ -11.743 \pm 0.02 $ & $ 44.185 \pm 0.02 $ & $ -13.523 \pm 0.01 $ & $ 42.405 \pm 0.01 $ & $ -14.263 \pm 0.02 $ & $ 41.665 \pm 0.02 $ & $ 26.3 _ {-20.6} ^ {+21.2} $ \\
VIII\_Zw218 & 0.12763 & $ -11.14 \pm 0.01 $ & $ 44.467 \pm 0.01 $ & $ -12.84 \pm 0.01 $ & $ 42.767 \pm 0.01 $ & $ -13.35 \pm 0.02 $ & $ 42.257 \pm 0.02 $ & $ 63.0 _ {-16.0} ^ {+16.0} $ \\
J1415+483 & 0.27495 & $ -11.707 \pm 0.01 $ & $ 44.642 \pm 0.01 $ & $ -13.517 \pm 0.01 $ & $ 42.832 \pm 0.01 $ & $ -13.917 \pm 0.02 $ & $ 42.432 \pm 0.02 $ & $ 25.1 _ {-11.0} ^ {+11.6} $ \\
J1456+380 & 0.28392 & $ -11.649 \pm 0.01 $ & $ 44.732 \pm 0.01 $ & $ -13.559 \pm 0.01 $ & $ 42.822 \pm 0.01 $ & $ -13.759 \pm 0.02 $ & $ 42.622 \pm 0.02 $ & $ 77.6 _ {-8.9} ^ {+8.9} $ \\
J1540+355 & 0.16379 & $ -11.392 \pm 0.01 $ & $ 44.452 \pm 0.01 $ & $ -13.132 \pm 0.01 $ & $ 42.712 \pm 0.01 $ & $ -13.642 \pm 0.01 $ & $ 42.202 \pm 0.01 $ & $ 57.5 _ {-14.6} ^ {+18.5} $ \\
J1619+501 & 0.23366 & $ -11.768 \pm 0.01 $ & $ 44.42 \pm 0.01 $ & $ -13.408 \pm 0.01 $ & $ 42.78 \pm 0.01 $ & $ -13.888 \pm 0.01 $ & $ 42.3 \pm 0.01 $ & $ 32.4 _ {-6.7} ^ {+6.7} $ \\
PG1427+480 & 0.22105 & $ -11.331 \pm 0.01 $ & $ 44.802 \pm 0.01 $ & $ -13.061 \pm 0.01 $ & $ 43.072 \pm 0.01 $ & $ -13.481 \pm 0.02 $ & $ 42.652 \pm 0.02 $ & $ 33.1 _ {-19.1} ^ {+20.6} $ \\
PG1612+261 & 0.13116 & $ -10.921 \pm 0.02 $ & $ 44.712 \pm 0.02 $ & $ -12.611 \pm 0.01 $ & $ 43.022 \pm 0.01 $ & $ -12.571 \pm 0.02 $ & $ 43.062 \pm 0.02 $ & $ 63.1 _ {-14.5} ^ {+13.1} $ \\
Mrk1014 & 0.1622 & $ -10.97 \pm 0.02 $ & $ 44.865 \pm 0.02 $ & $ -12.92 \pm 0.02 $ & $ 42.915 \pm 0.02 $ & $ -12.81 \pm 0.03 $ & $ 43.025 \pm 0.03 $ & $ 107.2 _ {-24.7} ^ {+22.2} $ \\
PG2349-014 & 0.17261 & $ -11.24 \pm 0.02 $ & $ 44.653 \pm 0.02 $ & $ -12.87 \pm 0.02 $ & $ 43.023 \pm 0.02 $ & $ -13.28 \pm 0.03 $ & $ 42.613 \pm 0.03 $ & $ 56.3 _ {-10.4} ^ {+11.7} $ \\
PG0052+251 & 0.15339 & $ -11.038 \pm 0.02 $ & $ 44.743 \pm 0.02 $ & $ -12.688 \pm 0.04 $ & $ 43.093 \pm 0.04 $ & $ -12.988 \pm 0.07 $ & $ 42.793 \pm 0.07 $ & $ 77.7 _ {-23.3} ^ {+23.3} $ \\
Mrk1501 & 0.08603 & $ -11.213 \pm 0.06 $ & $ 44.028 \pm 0.06 $ & $ -12.673 \pm 0.02 $ & $ 42.568 \pm 0.02 $ & $ -12.723 \pm 0.02 $ & $ 42.518 \pm 0.02 $ & $ 12.0 _ {-6.7} ^ {+5.5} $ \\
PG1322+659 & 0.16769 & $ -11.108 \pm 0.05 $ & $ 44.758 \pm 0.05 $ & $ -12.888 \pm 0.04 $ & $ 42.978 \pm 0.04 $ & $ -13.828 \pm 0.05 $ & $ 42.038 \pm 0.05 $ & $ 40.7 _ {-15.0} ^ {+17.8} $ \\
PG1440+356 & 0.07958 & $ -10.704 \pm 0.12 $ & $ 44.466 \pm 0.12 $ & $ -12.654 \pm 0.07 $ & $ 42.516 \pm 0.07 $ & $ -13.294 \pm 0.09 $ & $ 41.876 \pm 0.09 $ & $ 41.7 _ {-18.2} ^ {+16.3} $ \\

\bottomrule\bottomrule
\end{tabular}
\begin{tablenotes}
\item Columns are: (1) object name, (2) redshift corrected for peculiar velocity, (3) monochromatic flux at 5100 {\AA} in log scale, (4) monochromatic luminosity at 5100 {\AA} in log scale,  (5) H$\beta$ flux in log scale, (6) H$\beta$ luminosity in log scale, (7) {\OIII} flux in log scale, (8) {\OIII} luminosity in log scale, and (9) rest-frame average H$\beta$ time lag from ICCF. Note that luminosities are derived assuming a flat $\Lambda$CDM cosmology with $H_0 = 72$~\hunit and $\Omega_{m0} = 0.3$, using redshifts corrected for peculiar velocities. The luminosity values and rest-frame H$\beta$ lags correspond to the Average Scheme sample as reported by \citet{2024ApJS..275...13W}.
\end{tablenotes}
\end{threeparttable}%
}
\end{table*}
\end{turnpage}

\section{Data Analysis Methodology}
\label{sec:analysis}

The correlation between the H$\beta$ rest-frame time lag ($\tau_{\mathrm{H}\beta}=R_{\mathrm{H}\beta}/c$) in days and the monochromatic luminosity at 5100\ \AA\ ($L_{5100}$), or the broad H$\beta$ or narrow {\OIII} luminosity ($L_{\rm H\beta/{\OIII}}$) in units of $\rm erg\ s^{-1}$ can be expressed as
\begin{equation}
\begin{aligned}
    \log{\left(\frac{\tau_{\mathrm{H}\beta}}{\rm day}\right)} &= 
        \beta + \gamma \log{\left(\frac{L_{5100}}{10^{44}\,{\rm erg\ s^{-1}}}\right)}, \\
    \log{\left(\frac{\tau_{\mathrm{H}\beta}}{\rm day}\right)} &= 
        \beta + \gamma \log{\left(\frac{L_{\mathrm{H\beta/{\OIII}}}}{10^{42}\,{\rm erg\ s^{-1}}}\right)},
\end{aligned}
\label{eq:R-L}
\end{equation}
where $\beta$ and $\gamma$ are the intercept and slope parameters, respectively. The luminosity $L_{\mathrm{5100/H\beta/{\OIII}}}$ is derived from the measured AGN flux $F_{\mathrm{5100/H\beta/{\OIII}}}$ (in $\rm erg\ s^{-1}\ cm^{-2}$) and the luminosity distance $D_L$ is computed from Eq.\ \eqref{eq:DL}
\be
\label{eq:L5100}
    L_{\mathrm{5100/H\beta/{\OIII}}}=4\pi D_L^2F_{\mathrm{5100/H\beta/{\OIII}}}.
\ee

The natural logarithm of the likelihood function \citep{D'Agostini_2005} is
\be
\label{eq:LH}
    \ln\mathcal{L}= -\frac{1}{2}\Bigg[\chi^2+\sum^{N}_{i=1}\ln\left(2\pi\sigma^2_{\mathrm{tot},i}\right)\Bigg],
\ee
where
\be
\label{eq:chi2}
    \chi^2 = \sum^{N}_{i=1}\bigg[\frac{(\log \tau_{\mathrm{H}\beta,i} - \beta - \gamma\log L_{\mathrm{5100/H\beta/{\OIII}},i})^2}{\sigma^2_{\mathrm{tot},i}}\bigg].
\ee
and the total variance for each of the $N$ measurements is
\be
\label{eq:sigma_Hbeta}
\sigma^2_{\mathrm{tot},i}=\sigma_{\rm int}^2+\sigma_{{\log \tau_{\mathrm{H}\beta,i}}}^2+\gamma^2\sigma_{\log F_{\mathrm{5100/H\beta/{\OIII}},i}}^2.
\ee
Here $\sigma_{\rm int}$ represents the intrinsic scatter of the H$\beta$ AGN ensemble, accounting for unmodeled or systematic contributions. The quantities $\sigma_{{\log \tau_{\mathrm{H}\beta,i}}}$ and $\sigma_{\log F_{5100,i}}$ denote the measurement uncertainties in the logarithm of the time delay and the 5100\,\AA\ flux density, respectively. To properly treat the asymmetric uncertainties in $\tau_{\mathrm{H}\beta}$, we adopt the upper error ($\sigma_{\tau_{\mathrm{H}\beta},+}$) when the model prediction for $\log{\tau_{\mathrm{H}\beta}}$ exceeds the observed value, and the lower error ($\sigma_{\tau_{\mathrm{H}\beta},-}$) otherwise.

Descriptions of the likelihood expressions for the $H(z)$ and BAO datasets, as well as their covariance matrices, are provided in Secs.~IV and III of Ref.~\cite{CaoRatra2023}. 

To compare the performance of different cosmological models, we use the Akaike information criterion (AIC), the Bayesian information criterion (BIC), and the deviance information criterion (DIC). Further discussion of these statistical measures is available in Sec.~IV of Ref.~\cite{CaoRatra2023}.

We carry out the Bayesian inference analyses using the MCMC code \textsc{MontePython} \cite{Brinckmann2019}, adopting flat (uniform) prior probability distributions for all free parameters, as specified in Table~\ref{tab:priors}. Posterior summaries and visualizations are produced with the \textsc{GetDist} package \cite{Lewis_2019}.

\begin{table}[htbp]
\centering
\setlength\tabcolsep{3.3pt}
\begin{threeparttable}
\caption{Flat (uniform) priors of the constrained parameters.}
\label{tab:priors}
\begin{tabular}{lcc}
\toprule\toprule
Parameter & & Prior\\
\midrule
 & Cosmological parameters & \\
\midrule
$H_0$\,\tnote{a} &  & [None, None]\\
\obhs\,\tnote{b} &  & [0, 1]\\
\ochs\,\tnote{b} &  & [0, 1]\\
\ok &  & [$-2$, 2]\\
$\alpha$ &  & [0, 10]\\
\wx &  & [$-5$, 0.33]\\
\om\,\tnote{c} &  & [0.051314766115, 1]\\
\\
\midrule
 & $R-L$ relation parameters & \\
\midrule
$\beta$ &  & [0, 10]\\
$\gamma$ &  & [0, 5]\\
$\sigma_{\mathrm{int}}$ &  & [0, 5]\\
\bottomrule\bottomrule
\end{tabular}
\begin{tablenotes}
\item [a] \hunit. In the AGN-only cases, $H_0=70$ \hunit.
\item [b] Analyses involving $H(z)$ + BAO data. In the AGN-only cases $\Omega_{b}=0.05$.
\item [c] RM AGN only, to ensure that $\Omega_{c}$ remains positive.
\end{tablenotes}
\end{threeparttable}%
\end{table}

\section{Results}
\label{sec:results}

\begin{turnpage}
\begin{table*}
\centering
\resizebox{2.4\columnwidth}{!}{%
\begin{threeparttable}
\caption{Unmarginalized best-fitting parameter values for all models from various combinations of data.}\label{tab:BFP}
\begin{tabular}{lccccccccccccccccc}
\toprule\toprule
Model & dataset & $\Omega_{b}h^2$ & $\Omega_{c}h^2$ & $\Omega_{m0}$ & $\Omega_{k0}$ & $w_{\mathrm{X}}$/$\alpha$\tnote{a} & $H_0$\tnote{b} & $\gamma$ & $\beta$ & $\sigma_{\mathrm{int}}$ & $-2\ln\mathcal{L}_{\mathrm{max}}$ & AIC & BIC & DIC & $\Delta \mathrm{AIC}$ & $\Delta \mathrm{BIC}$ & $\Delta \mathrm{DIC}$ \\
\midrule
 & $\tau_{\mathrm{H}\beta}\text{-}L_{5100}$ & $\cdots$ & $\cdots$ & 0.998 & $\cdots$ & $\cdots$ & $\cdots$ & 0.453 & 1.383 & 0.166 & $-28.51$ & $-22.51$ & $-14.33$ & $-18.45$ & 0.00 & 0.00 & 0.00\\
 & $\tau_{\mathrm{H}\beta}\text{-}L_{\mathrm{H}\beta}$ & $\cdots$ & $\cdots$ & 0.996 & $\cdots$ & $\cdots$ & $\cdots$ & 0.475 & 1.275 & 0.152 & $-32.12$ & $-26.12$ & $-18.30$ & $-21.19$ & 0.00 & 0.00 & 0.00\\
 & $\tau_{\mathrm{H}\beta}\text{-}L_{\mathrm{{\OIII}}}$ & $\cdots$ & $\cdots$ & 0.559 & $\cdots$ & $\cdots$ & $\cdots$ & 0.519 & 1.462 & 0.245 & 28.14 & 34.14 & 41.95 & 35.88 & 0.00 & 0.00 & 0.00\\
Flat \lcdm & $H(z)$ + BAO & 0.0254 & 0.1200 & 0.297 & $\cdots$ & $\cdots$ & 70.12 & $\cdots$ & $\cdots$ & $\cdots$ & 30.56 & 36.56 & 41.91 & 37.32 & 0.00 & 0.00 & 0.00\\
 & $H(z)$ + BAO + $\tau_{\mathrm{H}\beta}\text{-}L_{5100}$ & 0.0256 & 0.1200 & 0.297 & $\cdots$ & $\cdots$ & 70.13 & 0.437 & 1.363 & 0.170 & 7.54 & 19.54 & 37.88 & 20.54 & 0.00 & 0.00 & 0.00\\
 & $H(z)$ + BAO + $\tau_{\mathrm{H}\beta}\text{-}L_{\mathrm{H}\beta}$ & 0.0249 & 0.1204 & 0.299 & $\cdots$ & $\cdots$ & 69.83 & 0.465 & 1.235 & 0.166 & 5.44 & 17.44 & 35.26 & 18.46 & 0.00 & 0.00 & 0.00\\
 & $H(z)$ + BAO + $\tau_{\mathrm{H}\beta}\text{-}L_{\mathrm{{\OIII}}}$ & 0.0257 & 0.1208 & 0.298 & $\cdots$ & $\cdots$ & 70.27 & 0.513 & 1.452 & 0.249 & 59.08 & 71.08 & 88.90 & 72.78 & 0.00 & 0.00 & 0.00\\
[6pt]
 & $\tau_{\mathrm{H}\beta}\text{-}L_{5100}$ & $\cdots$ & $\cdots$ & 0.998 & 1.999 & $\cdots$ & $\cdots$ & 0.472 & 1.426 & 0.162 & $-31.21$ & $-23.21$ & $-12.30$ & $-19.56$ & $-0.70$ & 2.03 & $-1.11$\\
 & $\tau_{\mathrm{H}\beta}\text{-}L_{\mathrm{H}\beta}$ & $\cdots$ & $\cdots$ & 0.993 & 1.993 & $\cdots$ & $\cdots$ & 0.502 & 1.300 & 0.149 & $-35.50$ & $-27.50$ & $-17.08$ & $-22.81$ & $-1.38$ & 1.22 & $-1.62$\\
 & $\tau_{\mathrm{H}\beta}\text{-}L_{\mathrm{{\OIII}}}$ & $\cdots$ & $\cdots$ & 0.998 & $-1.365$ & $\cdots$ & $\cdots$ & 0.509 & 1.443 & 0.243 & 26.86 & 34.86 & 45.28 & 37.43 & 0.72 & 3.33 & 1.55\\
Nonflat \lcdm & $H(z)$ + BAO & 0.0269 & 0.1128 & 0.289 & 0.041 & $\cdots$ & 69.61 & $\cdots$ & $\cdots$ & $\cdots$ & 30.34 & 38.34 & 45.48 & 38.80 & 1.78 & 3.56 & 1.48\\
 & $H(z)$ + BAO + $\tau_{\mathrm{H}\beta}\text{-}L_{5100}$ & 0.0273 & 0.1113 & 0.288 & 0.051 & $\cdots$ & 69.46 & 0.435 & 1.361 & 0.173 & 7.21 & 21.21 & 42.61 & 22.01 & 1.67 & 4.73 & 1.47\\
 & $H(z)$ + BAO + $\tau_{\mathrm{H}\beta}\text{-}L_{\mathrm{H}\beta}$ & 0.0280 & 0.1104 & 0.286 & 0.057 & $\cdots$ & 69.72 & 0.463 & 1.237 & 0.161 & 5.12 & 19.12 & 39.91 & 19.86 & 1.68 & 4.65 & 1.41\\
 & $H(z)$ + BAO + $\tau_{\mathrm{H}\beta}\text{-}L_{\mathrm{{\OIII}}}$ & 0.0265 & 0.1107 & 0.290 & 0.054 & $\cdots$ & 68.99 & 0.517 & 1.449 & 0.248 & 58.90 & 72.90 & 93.69 & 74.36 & 1.82 & 4.79 & 1.57\\
[6pt]
 & $\tau_{\mathrm{H}\beta}\text{-}L_{5100}$ & $\cdots$ & $\cdots$ & 0.082 & $\cdots$ & 0.142 & $\cdots$ & 0.457 & 1.389 & 0.160 & $-29.52$ & $-21.52$ & $-10.61$ & $-18.26$ & 0.99 & 3.72 & 0.19\\
 & $\tau_{\mathrm{H}\beta}\text{-}L_{\mathrm{H}\beta}$ & $\cdots$ & $\cdots$ & 0.064 & $\cdots$ & 0.126 & $\cdots$ & 0.480 & 1.280 & 0.154 & $-33.37$ & $-25.37$ & $-14.95$ & $-20.82$ & 0.75 & 3.35 & 0.37\\
 & $\tau_{\mathrm{H}\beta}\text{-}L_{\mathrm{{\OIII}}}$ & $\cdots$ & $\cdots$ & 0.361 & $\cdots$ & $-4.978$ & $\cdots$ & 0.473 & 1.378 & 0.242 & 24.37 & 32.37 & 42.79 & 37.35 & $-1.77$ & 0.84 & 1.47\\
Flat XCDM & $H(z)$ + BAO & 0.0320 & 0.0932 & 0.283 & $\cdots$ & $-0.731$ & 66.69 & $\cdots$ & $\cdots$ & $\cdots$ & 26.57 & 34.57 & 41.71 & 34.52 & $-1.98$ & $-0.20$ & $-2.80$\\
 & $H(z)$ + BAO + $\tau_{\mathrm{H}\beta}\text{-}L_{5100}$ & 0.0327 & 0.0879 & 0.279 & $\cdots$ & $-0.692$ & 65.94 & 0.439 & 1.353 & 0.170 & 2.33 & 16.33 & 37.72 & 16.50 & $-3.21$ & $-0.15$ & $-4.04$\\
 & $H(z)$ + BAO + $\tau_{\mathrm{H}\beta}\text{-}L_{\mathrm{H}\beta}$ & 0.0348 & 0.0845 & 0.275 & $\cdots$ & $-0.702$ & 66.07 & 0.468 & 1.223 & 0.161 & $-0.19$ & 13.81 & 34.60 & 14.23 & $-3.63$ & $-0.66$ & $-4.23$\\
 & $H(z)$ + BAO + $\tau_{\mathrm{H}\beta}\text{-}L_{\mathrm{{\OIII}}}$ & 0.0316 & 0.0936 & 0.282 & $\cdots$ & $-0.739$ & 66.76 & 0.516 & 1.435 & 0.245 & 55.09 & 69.09 & 89.88 & 70.01 & $-2.00$ & 0.97 & $-2.78$\\
[6pt]
 & $\tau_{\mathrm{H}\beta}\text{-}L_{5100}$ & $\cdots$ & $\cdots$ & 0.493 & $-1.992$ & 0.125 & $\cdots$ & 0.481 & 1.426 & 0.158 & $-33.48$ & $-23.48$ & $-9.85$ & $-19.98$ & $-0.98$ & 4.48 & $-1.53$\\
 & $\tau_{\mathrm{H}\beta}\text{-}L_{\mathrm{H}\beta}$ & $\cdots$ & $\cdots$ & 0.068 & $-1.983$ & 0.128 & $\cdots$ & 0.506 & 1.302 & 0.144 & $-38.22$ & $-28.22$ & $-15.20$ & $-24.01$ & $-2.10$ & 3.11 & $-2.83$\\
 & $\tau_{\mathrm{H}\beta}\text{-}L_{\mathrm{{\OIII}}}$ & $\cdots$ & $\cdots$ & 0.993 & $-1.078$ & $-4.987$ & $\cdots$ & 0.451 & 1.333 & 0.227 & 19.22 & 29.22 & 42.24 & 41.72 & $-4.92$ & 0.29 & 5.84\\
Nonflat XCDM & $H(z)$ + BAO & 0.0312 & 0.0990 & 0.293 & $-0.085$ & $-0.693$ & 66.84 & $\cdots$ & $\cdots$ & $\cdots$ & 26.00 & 36.00 & 44.92 & 36.17 & $-0.56$ & 3.01 & $-1.15$\\
 & $H(z)$ + BAO + $\tau_{\mathrm{H}\beta}\text{-}L_{5100}$ & 0.0332 & 0.0927 & 0.287 & $-0.083$ & $-0.658$ & 66.42 & 0.439 & 1.352 & 0.168 & 1.73 & 17.73 & 42.18 & 17.76 & $-1.81$ & 4.30 & $-2.78$\\
 & $H(z)$ + BAO + $\tau_{\mathrm{H}\beta}\text{-}L_{\mathrm{H}\beta}$ & 0.0325 & 0.0916 & 0.287 & $-0.093$ & $-0.651$ & 65.94 & 0.471 & 1.220 & 0.163 & $-0.90$ & 15.10 & 38.85 & 15.59 & $-2.35$ & 3.59 & $-2.86$\\
 & $H(z)$ + BAO + $\tau_{\mathrm{H}\beta}\text{-}L_{\mathrm{{\OIII}}}$ & 0.0311 & 0.0961 & 0.290 & $-0.053$ & $-0.709$ & 66.42 & 0.516 & 1.433 & 0.248 & 54.66 & 70.66 & 94.42 & 71.25 & $-0.43$ & 5.51 & $-1.53$\\
[6pt]
 & $\tau_{\mathrm{H}\beta}\text{-}L_{5100}$ & $\cdots$ & $\cdots$ & 0.998 & $\cdots$ & 5.155 & $\cdots$ & 0.455 & 1.389 & 0.165 & $-28.50$ & $-20.50$ & $-9.59$ & $-19.11$ & 2.01 & 4.73 & $-0.66$\\
 & $\tau_{\mathrm{H}\beta}\text{-}L_{\mathrm{H}\beta}$ & $\cdots$ & $\cdots$ & 0.998 & $\cdots$ & 0.732 & $\cdots$ & 0.475 & 1.276 & 0.155 & $-32.12$ & $-24.12$ & $-13.70$ & $-21.73$ & 2.00 & 4.60 & $-0.54$\\
 & $\tau_{\mathrm{H}\beta}\text{-}L_{\mathrm{{\OIII}}}$ & $\cdots$ & $\cdots$ & 0.454 & $\cdots$ & 0.046 & $\cdots$ & 0.518 & 1.459 & 0.243 & 28.21 & 36.21 & 46.63 & 35.68 & 2.07 & 4.68 & $-0.20$\\
Flat \pcdm & $H(z)$ + BAO & 0.0337 & 0.0864 & 0.271 & $\cdots$ & 1.169 & 66.78 & $\cdots$ & $\cdots$ & $\cdots$ & 26.50 & 34.50 & 41.64 & 34.01 & $-2.05$ & $-0.27$ & $-3.31$\\
 & $H(z)$ + BAO + $\tau_{\mathrm{H}\beta}\text{-}L_{5100}$ & 0.0354 & 0.0806 & 0.264 & $\cdots$ & 1.391 & 66.45 & 0.435 & 1.352 & 0.168 & 2.27 & 16.27 & 37.66 & 15.91 & $-3.27$ & $-0.21$ & $-4.62$\\
 & $H(z)$ + BAO + $\tau_{\mathrm{H}\beta}\text{-}L_{\mathrm{H}\beta}$ & 0.0389 & 0.0664 & 0.248 & $\cdots$ & 2.013 & 65.40 & 0.470 & 1.219 & 0.160 & $-0.13$ & 13.87 & 34.66 & 13.62 & $-3.57$ & $-0.60$ & $-4.84$\\
 & $H(z)$ + BAO + $\tau_{\mathrm{H}\beta}\text{-}L_{\mathrm{{\OIII}}}$ & 0.0330 & 0.0858 & 0.269 & $\cdots$ & 1.140 & 66.61 & 0.516 & 1.433 & 0.246 & 55.15 & 69.15 & 89.94 & 69.02 & $-1.93$ & 1.04 & $-3.76$\\
[6pt]
 & $\tau_{\mathrm{H}\beta}\text{-}L_{5100}$ & $\cdots$ & $\cdots$ & 0.995 & $-1.702$ & 4.884 & $\cdots$ & 0.457 & 1.399 & 0.159 & $-30.20$ & $-20.20$ & $-6.56$ & $-18.07$ & 2.31 & 7.76 & 0.37\\
 & $\tau_{\mathrm{H}\beta}\text{-}L_{\mathrm{H}\beta}$ & $\cdots$ & $\cdots$ & 0.970 & $-1.597$ & 4.956 & $\cdots$ & 0.483 & 1.279 & 0.155 & $-34.03$ & $-24.03$ & $-11.01$ & $-20.67$ & 2.09 & 7.30 & 0.52\\
 & $\tau_{\mathrm{H}\beta}\text{-}L_{\mathrm{{\OIII}}}$ & $\cdots$ & $\cdots$ & 0.815 & $-1.949$ & 1.380 & $\cdots$ & 0.515 & 1.450 & 0.239 & 26.83 & 36.83 & 49.86 & 36.46 & 2.70 & 7.91 & 0.58\\
Nonflat \pcdm & $H(z)$ + BAO & 0.0338 & 0.0878 & 0.273 & $-0.077$ & 1.441 & 66.86 & $\cdots$ & $\cdots$ & $\cdots$ & 25.92 & 35.92 & 44.84 & 35.12 & $-0.64$ & 2.93 & $-2.20$\\
 & $H(z)$ + BAO + $\tau_{\mathrm{H}\beta}\text{-}L_{5100}$ & 0.0345 & 0.0873 & 0.274 & $-0.112$ & 1.656 & 66.77 & 0.437 & 1.354 & 0.174 & 1.74 & 17.74 & 42.19 & 17.19 & $-1.79$ & 4.32 & $-3.34$\\
 & $H(z)$ + BAO + $\tau_{\mathrm{H}\beta}\text{-}L_{\mathrm{H}\beta}$ & 0.0358 & 0.0812 & 0.266 & $-0.076$ & 1.720 & 66.49 & 0.468 & 1.222 & 0.168 & $-0.70$ & 15.30 & 39.06 & 14.88 & $-2.14$ & 3.80 & $-3.58$\\
 & $H(z)$ + BAO + $\tau_{\mathrm{H}\beta}\text{-}L_{\mathrm{{\OIII}}}$ & 0.0344 & 0.0882 & 0.273 & $-0.095$ & 1.513 & 67.16 & 0.517 & 1.439 & 0.245 & 54.38 & 70.38 & 94.13 & 70.52 & $-0.71$ & 5.23 & $-2.26$\\
\bottomrule\bottomrule
\end{tabular}
\begin{tablenotes}
\item [a] \wx\ corresponds to flat/nonflat XCDM and $\alpha$ corresponds to flat/nonflat \pcdm.
\item [b] \hunit.
\end{tablenotes}
\end{threeparttable}%
}
\end{table*}
\end{turnpage}

\begin{turnpage}
\begin{table*}
\centering
\resizebox{2.4\columnwidth}{!}{%
\begin{threeparttable}
\caption{One-dimensional marginalized posterior mean values and uncertainties ($\pm 1\sigma$ error bars or $2\sigma$ limits) of the parameters for all models from various combinations of data.}\label{tab:1d_BFP}
\begin{tabular}{lcccccccccc}
\toprule\toprule
Model & dataset & $\Omega_{b}h^2$ & $\Omega_{c}h^2$ & $\Omega_{m0}$ & $\Omega_{k0}$ & $w_{\mathrm{X}}$/$\alpha$\tnote{a} & $H_0$\tnote{b} & $\gamma$ & $\beta$ & $\sigma_{\mathrm{int}}$\\
\midrule
 & $\tau_{\mathrm{H}\beta}\text{-}L_{5100}$ & $\cdots$ & $\cdots$ & $>0.348$ & $\cdots$ & $\cdots$ & $\cdots$ & $0.450\pm0.026$ & $1.380\pm0.024$ & $0.172^{+0.016}_{-0.020}$ \\
 & $\tau_{\mathrm{H}\beta}\text{-}L_{\mathrm{H}\beta}$ & $\cdots$ & $\cdots$ & $>0.400$ & $\cdots$ & $\cdots$ & $\cdots$ & $0.479\pm0.029$ & $1.262\pm0.025$ & $0.163^{+0.017}_{-0.020}$ \\
 & $\tau_{\mathrm{H}\beta}\text{-}L_{\mathrm{{\OIII}}}$ & $\cdots$ & $\cdots$ & $0.570^{+0.412}_{-0.160}$ & $\cdots$ & $\cdots$ & $\cdots$ & $0.494\pm0.047$ & $1.466\pm0.036$ & $0.252^{+0.021}_{-0.026}$ \\
Flat \lcdm & $H(z)$ + BAO & $0.0260\pm0.0040$ & $0.1213^{+0.0091}_{-0.0103}$ & $0.298^{+0.015}_{-0.018}$ & $\cdots$ & $\cdots$ & $70.51\pm2.72$ & $\cdots$ & $\cdots$ & $\cdots$ \\
 & $H(z)$ + BAO + $\tau_{\mathrm{H}\beta}\text{-}L_{5100}$ & $0.0258^{+0.0036}_{-0.0040}$ & $0.1217^{+0.0088}_{-0.0099}$ & $0.299^{+0.015}_{-0.017}$ & $\cdots$ & $\cdots$ & $70.39\pm2.61$ & $0.438\pm0.025$ & $1.364\pm0.026$ & $0.176^{+0.016}_{-0.019}$ \\
 & $H(z)$ + BAO + $\tau_{\mathrm{H}\beta}\text{-}L_{\mathrm{H}\beta}$ & $0.0257^{+0.0037}_{-0.0040}$ & $0.1217^{+0.0088}_{-0.0099}$ & $0.299^{+0.015}_{-0.017}$ & $\cdots$ & $\cdots$ & $70.37\pm2.62$ & $0.466\pm0.028$ & $1.245\pm0.029$ & $0.169^{+0.016}_{-0.020}$ \\
 & $H(z)$ + BAO + $\tau_{\mathrm{H}\beta}\text{-}L_{\mathrm{{\OIII}}}$ & $0.0259\pm0.0038$ & $0.1213^{+0.0087}_{-0.0099}$ & $0.298^{+0.015}_{-0.017}$ & $\cdots$ & $\cdots$ & $70.45\pm2.63$ & $0.489\pm0.046$ & $1.454^{+0.035}_{-0.036}$ & $0.252^{+0.020}_{-0.026}$ \\
[6pt]
 & $\tau_{\mathrm{H}\beta}\text{-}L_{5100}$ & $\cdots$ & $\cdots$ & $>0.210$ & $>-0.359$ & $\cdots$ & $\cdots$ & $0.459\pm0.027$ & $1.398\pm0.026$ & $0.170^{+0.016}_{-0.019}$ \\
 & $\tau_{\mathrm{H}\beta}\text{-}L_{\mathrm{H}\beta}$ & $\cdots$ & $\cdots$ & $>0.237$ & $>-0.150$ & $\cdots$ & $\cdots$ & $0.490\pm0.029$ & $1.280^{+0.028}_{-0.025}$ & $0.160^{+0.016}_{-0.020}$ \\
 & $\tau_{\mathrm{H}\beta}\text{-}L_{\mathrm{{\OIII}}}$ & $\cdots$ & $\cdots$ & $0.603^{+0.397}_{-0.129}$ & $0.028^{+1.125}_{-1.103}$ & $\cdots$ & $\cdots$ & $0.494\pm0.047$ & $1.466\pm0.042$ & $0.252^{+0.021}_{-0.026}$ \\
Nonflat \lcdm & $H(z)$ + BAO & $0.0275^{+0.0047}_{-0.0053}$ & $0.1132\pm0.0183$ & $0.289\pm0.023$ & $0.047^{+0.083}_{-0.091}$ & $\cdots$ & $69.81\pm2.87$ & $\cdots$ & $\cdots$ & $\cdots$ \\
 & $H(z)$ + BAO + $\tau_{\mathrm{H}\beta}\text{-}L_{5100}$ & $0.0276^{+0.0046}_{-0.0052}$ & $0.1120\pm0.0178$ & $0.289\pm0.023$ & $0.055^{+0.082}_{-0.089}$ & $\cdots$ & $69.59\pm2.80$ & $0.439\pm0.025$ & $1.361\pm0.027$ & $0.176^{+0.016}_{-0.019}$ \\
 & $H(z)$ + BAO + $\tau_{\mathrm{H}\beta}\text{-}L_{\mathrm{H}\beta}$ & $0.0277^{+0.0046}_{-0.0052}$ & $0.1117\pm0.0179$ & $0.289\pm0.023$ & $0.057^{+0.082}_{-0.090}$ & $\cdots$ & $69.55^{+2.79}_{-2.80}$ & $0.466\pm0.028$ & $1.241\pm0.029$ & $0.169^{+0.016}_{-0.020}$ \\
 & $H(z)$ + BAO + $\tau_{\mathrm{H}\beta}\text{-}L_{\mathrm{{\OIII}}}$ & $0.0274^{+0.0046}_{-0.0052}$ & $0.1133\pm0.0179$ & $0.290\pm0.023$ & $0.046^{+0.081}_{-0.089}$ & $\cdots$ & $69.78\pm2.80$ & $0.489\pm0.045$ & $1.451^{+0.035}_{-0.036}$ & $0.252^{+0.020}_{-0.025}$ \\
[6pt]
 & $\tau_{\mathrm{H}\beta}\text{-}L_{5100}$ & $\cdots$ & $\cdots$ & $>0.229$ & $\cdots$ & $-1.784^{+1.943}_{-1.169}$ & $\cdots$ & $0.449\pm0.027$ & $1.380\pm0.025$ & $0.172^{+0.016}_{-0.020}$ \\
 & $\tau_{\mathrm{H}\beta}\text{-}L_{\mathrm{H}\beta}$ & $\cdots$ & $\cdots$ & $>0.213$ & $\cdots$ & $-1.679^{+1.841}_{-1.189}$ & $\cdots$ & $0.479^{+0.029}_{-0.028}$ & $1.262\pm0.025$ & $0.163^{+0.016}_{-0.020}$ \\
 & $\tau_{\mathrm{H}\beta}\text{-}L_{\mathrm{{\OIII}}}$ & $\cdots$ & $\cdots$ & $0.569^{+0.245}_{-0.259}$ & $\cdots$ & $<-0.187$ & $\cdots$ & $0.487\pm0.047$ & $1.444^{+0.050}_{-0.042}$ & $0.250^{+0.021}_{-0.026}$ \\
Flat XCDM & $H(z)$ + BAO & $0.0308^{+0.0053}_{-0.0046}$ & $0.0980^{+0.0182}_{-0.0161}$ & $0.286\pm0.019$ & $\cdots$ & $-0.778^{+0.132}_{-0.104}$ & $67.20^{+3.05}_{-3.06}$ & $\cdots$ & $\cdots$ & $\cdots$ \\
 & $H(z)$ + BAO + $\tau_{\mathrm{H}\beta}\text{-}L_{5100}$ & $0.0315^{+0.0055}_{-0.0043}$ & $0.0942^{+0.0183}_{-0.0164}$ & $0.284\pm0.020$ & $\cdots$ & $-0.745^{+0.125}_{-0.095}$ & $66.62\pm2.94$ & $0.442\pm0.025$ & $1.349\pm0.027$ & $0.174^{+0.016}_{-0.019}$ \\
 & $H(z)$ + BAO + $\tau_{\mathrm{H}\beta}\text{-}L_{\mathrm{H}\beta}$ & $0.0316^{+0.0057}_{-0.0042}$ & $0.0936^{+0.0183}_{-0.0164}$ & $0.283\pm0.020$ & $\cdots$ & $-0.740^{+0.125}_{-0.094}$ & $66.51\pm2.93$ & $0.470\pm0.028$ & $1.228\pm0.030$ & $0.167^{+0.016}_{-0.020}$ \\
 & $H(z)$ + BAO + $\tau_{\mathrm{H}\beta}\text{-}L_{\mathrm{{\OIII}}}$ & $0.0310^{+0.0053}_{-0.0047}$ & $0.0971^{+0.0183}_{-0.0160}$ & $0.285\pm0.019$ & $\cdots$ & $-0.771^{+0.132}_{-0.101}$ & $67.10^{+2.97}_{-2.99}$ & $0.492\pm0.046$ & $1.440\pm0.036$ & $0.252^{+0.020}_{-0.026}$ \\
[6pt]
 & $\tau_{\mathrm{H}\beta}\text{-}L_{5100}$ & $\cdots$ & $\cdots$ & $>0.551$\tnote{c} & $>0.792$\tnote{c} & $<0.043$ & $\cdots$ & $0.470\pm0.030$ & $1.423^{+0.035}_{-0.039}$ & $0.168^{+0.016}_{-0.019}$ \\
 & $\tau_{\mathrm{H}\beta}\text{-}L_{\mathrm{H}\beta}$ & $\cdots$ & $\cdots$ & $>0.196$ & $>1.015$\tnote{c} & $<0.040$ & $\cdots$ & $0.505\pm0.033$ & $1.309^{+0.037}_{-0.036}$ & $0.158^{+0.016}_{-0.020}$ \\
 & $\tau_{\mathrm{H}\beta}\text{-}L_{\mathrm{{\OIII}}}$ & $\cdots$ & $\cdots$ & $>0.511$\tnote{c} & $-0.169^{+0.589}_{-0.881}$ & $<-0.148$ & $\cdots$ & $0.482\pm0.049$ & $1.437^{+0.070}_{-0.052}$ & $0.250^{+0.020}_{-0.026}$ \\
Nonflat XCDM & $H(z)$ + BAO & $0.0305^{+0.0055}_{-0.0047}$ & $0.1011\pm0.0196$ & $0.292\pm0.024$ & $-0.059\pm0.106$ & $-0.746^{+0.135}_{-0.090}$ & $67.17^{+2.96}_{-2.97}$ & $\cdots$ & $\cdots$ & $\cdots$ \\
 & $H(z)$ + BAO + $\tau_{\mathrm{H}\beta}\text{-}L_{5100}$ & $0.0312^{+0.0057}_{-0.0046}$ & $0.0978^{+0.0197}_{-0.0198}$ & $0.290^{+0.025}_{-0.024}$ & $-0.069\pm0.108$ & $-0.714^{+0.129}_{-0.082}$ & $66.66\pm2.90$ & $0.442\pm0.025$ & $1.350\pm0.027$ & $0.174^{+0.016}_{-0.019}$ \\
 & $H(z)$ + BAO + $\tau_{\mathrm{H}\beta}\text{-}L_{\mathrm{H}\beta}$ & $0.0313^{+0.0059}_{-0.0044}$ & $0.0972^{+0.0196}_{-0.0198}$ & $0.290\pm0.024$ & $-0.070\pm0.109$ & $-0.708^{+0.127}_{-0.082}$ & $66.56^{+2.89}_{-2.90}$ & $0.470\pm0.028$ & $1.229\pm0.030$ & $0.167^{+0.016}_{-0.020}$ \\
 & $H(z)$ + BAO + $\tau_{\mathrm{H}\beta}\text{-}L_{\mathrm{{\OIII}}}$ & $0.0306^{+0.0057}_{-0.0048}$ & $0.1005\pm0.0196$ & $0.292\pm0.024$ & $-0.063^{+0.107}_{-0.108}$ & $-0.739^{+0.137}_{-0.089}$ & $67.09^{+2.93}_{-2.94}$ & $0.492\pm0.045$ & $1.440^{+0.036}_{-0.035}$ & $0.252^{+0.020}_{-0.026}$ \\
[6pt]
 & $\tau_{\mathrm{H}\beta}\text{-}L_{5100}$ & $\cdots$ & $\cdots$ & $>0.237$ & $\cdots$ & $2.747^{+2.571}_{-1.319}$ & $\cdots$ & $0.451\pm0.026$ & $1.382\pm0.023$ & $0.171^{+0.016}_{-0.019}$ \\
 & $\tau_{\mathrm{H}\beta}\text{-}L_{\mathrm{H}\beta}$ & $\cdots$ & $\cdots$ & $>0.269$ & $\cdots$ & $2.779^{+2.562}_{-1.400}$ & $\cdots$ & $0.480\pm0.028$ & $1.263\pm0.025$ & $0.163^{+0.016}_{-0.020}$ \\
 & $\tau_{\mathrm{H}\beta}\text{-}L_{\mathrm{{\OIII}}}$ & $\cdots$ & $\cdots$ & $>0.382$\tnote{c} & $\cdots$ & $2.486^{+2.547}_{-1.767}$ & $\cdots$ & $0.497\pm0.047$ & $1.472\pm0.034$ & $0.252^{+0.020}_{-0.026}$ \\%
Flat \pcdm & $H(z)$ + BAO & $0.0327^{+0.0060}_{-0.0031}$ & $0.0867^{+0.0184}_{-0.0185}$ & $0.272\pm0.022$ & $\cdots$ & $1.260^{+0.497}_{-0.804}$ & $66.24\pm2.86$ & $\cdots$ & $\cdots$ & $\cdots$ \\
 & $H(z)$ + BAO + $\tau_{\mathrm{H}\beta}\text{-}L_{5100}$ & $0.0332^{+0.0065}_{-0.0020}$ & $0.0840\pm0.0186$ & $0.270^{+0.023}_{-0.022}$ & $\cdots$ & $1.390^{+0.528}_{-0.833}$ & $65.90\pm2.82$ & $0.442\pm0.025$ & $1.345\pm0.027$ & $0.174^{+0.016}_{-0.019}$ \\
 & $H(z)$ + BAO + $\tau_{\mathrm{H}\beta}\text{-}L_{\mathrm{H}\beta}$ & $0.0333^{+0.0065}_{-0.0020}$ & $0.0835\pm0.0187$ & $0.270^{+0.022}_{-0.023}$ & $\cdots$ & $1.411^{+0.533}_{-0.839}$ & $65.83\pm2.83$ & $0.470\pm0.028$ & $1.224\pm0.030$ & $0.167^{+0.016}_{-0.020}$ \\
 & $H(z)$ + BAO + $\tau_{\mathrm{H}\beta}\text{-}L_{\mathrm{{\OIII}}}$ & $0.0329^{+0.0068}_{-0.0023}$ & $0.0864^{+0.0194}_{-0.0178}$ & $0.272\pm0.022$ & $\cdots$ & $1.271^{+0.498}_{-0.821}$ & $66.28\pm2.84$ & $0.492\pm0.046$ & $1.435\pm0.035$ & $0.252^{+0.020}_{-0.026}$ \\
[6pt]
 & $\tau_{\mathrm{H}\beta}\text{-}L_{5100}$ & $\cdots$ & $\cdots$ & $>0.240$ & $-0.492^{+0.653}_{-0.502}$ & $2.933^{+2.285}_{-0.916}$ & $\cdots$ & $0.451\pm0.026$ & $1.382\pm0.023$ & $0.171^{+0.016}_{-0.019}$ \\
 & $\tau_{\mathrm{H}\beta}\text{-}L_{\mathrm{H}\beta}$ & $\cdots$ & $\cdots$ & $>0.276$ & $-0.522^{+0.659}_{-0.498}$ & $3.014^{+2.222}_{-0.890}$ & $\cdots$ & $0.480\pm0.028$ & $1.263\pm0.025$ & $0.162^{+0.016}_{-0.020}$ \\
 & $\tau_{\mathrm{H}\beta}\text{-}L_{\mathrm{{\OIII}}}$ & $\cdots$ & $\cdots$ & $>0.435$\tnote{c} & $-0.396^{+0.655}_{-0.501}$ & $2.537^{+2.442}_{-1.297}$ & $\cdots$ & $0.497\pm0.046$ & $1.472\pm0.034$ & $0.251^{+0.021}_{-0.025}$ \\
Nonflat \pcdm & $H(z)$ + BAO & $0.0324^{+0.0062}_{-0.0031}$ & $0.0901^{+0.0200}_{-0.0201}$ & $0.277^{+0.025}_{-0.024}$ & $-0.076^{+0.097}_{-0.107}$ & $1.447^{+0.586}_{-0.794}$ & $66.51\pm2.86$ & $\cdots$ & $\cdots$ & $\cdots$ \\
 & $H(z)$ + BAO + $\tau_{\mathrm{H}\beta}\text{-}L_{5100}$ & $0.0330^{+0.0068}_{-0.0021}$ & $0.0872\pm0.0202$ & $0.275\pm0.025$ & $-0.082\pm0.105$ & $1.593^{+0.621}_{-0.833}$ & $66.18\pm2.84$ & $0.442\pm0.025$ & $1.347\pm0.027$ & $0.174^{+0.016}_{-0.019}$ \\
 & $H(z)$ + BAO + $\tau_{\mathrm{H}\beta}\text{-}L_{\mathrm{H}\beta}$ & $0.0331^{+0.0067}_{-0.0020}$ & $0.0867^{+0.0192}_{-0.0210}$ & $0.274\pm0.025$ & $-0.082\pm0.105$ & $1.612^{+0.629}_{-0.827}$ & $66.11^{+2.83}_{-2.84}$ & $0.470\pm0.028$ & $1.225\pm0.030$ & $0.167^{+0.016}_{-0.020}$ \\
 & $H(z)$ + BAO + $\tau_{\mathrm{H}\beta}\text{-}L_{\mathrm{{\OIII}}}$ & $0.0326^{+0.0070}_{-0.0023}$ & $0.0897\pm0.0203$ & $0.277^{+0.024}_{-0.025}$ & $-0.083\pm0.104$ & $1.487^{+0.598}_{-0.828}$ & $66.53^{+2.85}_{-2.86}$ & $0.492\pm0.045$ & $1.437\pm0.035$ & $0.251^{+0.020}_{-0.025}$ \\
\bottomrule\bottomrule
\end{tabular}
\begin{tablenotes}
\item [a] \wx\ corresponds to flat/nonflat XCDM and $\alpha$ corresponds to flat/nonflat \pcdm.
\item [b] \hunit. For AGN-only cases, $\Omega_b=0.05$ and $H_0=70$ \hunit.
\item [c] This is the 1$\sigma$ limit. The 2$\sigma$ limit is set by the prior and not shown here.
\end{tablenotes}
\end{threeparttable}%
}
\end{table*}
\end{turnpage}

\begin{table}[htbp]
\centering
\setlength\tabcolsep{17pt}
\begin{threeparttable}
\caption{The largest differences between results for considered cosmological models from different datasets with $1\sigma$ being the quadrature sum of the two corresponding $1\sigma$ error bars.}\label{tab:diff1}
\begin{tabular}{lccc}
\toprule\toprule
 Dataset & $\Delta\gamma$ & $\Delta\beta$ & $\Delta\sigma_{\mathrm{int}}$ \\
\midrule
$\tau_{\mathrm{H}\beta}\text{-}L_{5100}$\tnote{a} & $0.05\sigma$ & $0.06\sigma$ & $0.04\sigma$\\[2pt]
$\tau_{\mathrm{H}\beta}\text{-}L_{\mathrm{H}\beta}$\tnote{a} & $0.03\sigma$ & $0.03\sigma$ & $0.04\sigma$\\[2pt]
$\tau_{\mathrm{H}\beta}\text{-}L_{\mathrm{{\OIII}}}$\tnote{a} & $0.15\sigma$ & $0.46\sigma$ & $0.06\sigma$\\[2pt]
$\tau_{\mathrm{H}\beta}\text{-}L_{5100}$\tnote{b} & $0.52\sigma$ & $0.93\sigma$ & $0.16\sigma$\\[2pt]
$\tau_{\mathrm{H}\beta}\text{-}L_{\mathrm{H}\beta}$\tnote{b} & $0.59\sigma$ & $1.07\sigma$ & $0.20\sigma$\\[2pt]
$\tau_{\mathrm{H}\beta}\text{-}L_{\mathrm{{\OIII}}}$\tnote{b} & $0.22\sigma$ & $0.45\sigma$ & $0.06\sigma$\\
\bottomrule\bottomrule
\end{tabular}
\begin{tablenotes}[flushleft]
\item [a] Flat \lcdm\ and XCDM, and flat and nonflat \pcdm\ models.
\item [b] All six flat and nonflat \lcdm, XCDM, and \pcdm\ models.
\end{tablenotes}
\end{threeparttable}%
\end{table}

\begin{figure*}[htbp]
\centering
 \subfloat[]{%
    \includegraphics[width=0.35\textwidth,height=0.33\textwidth]{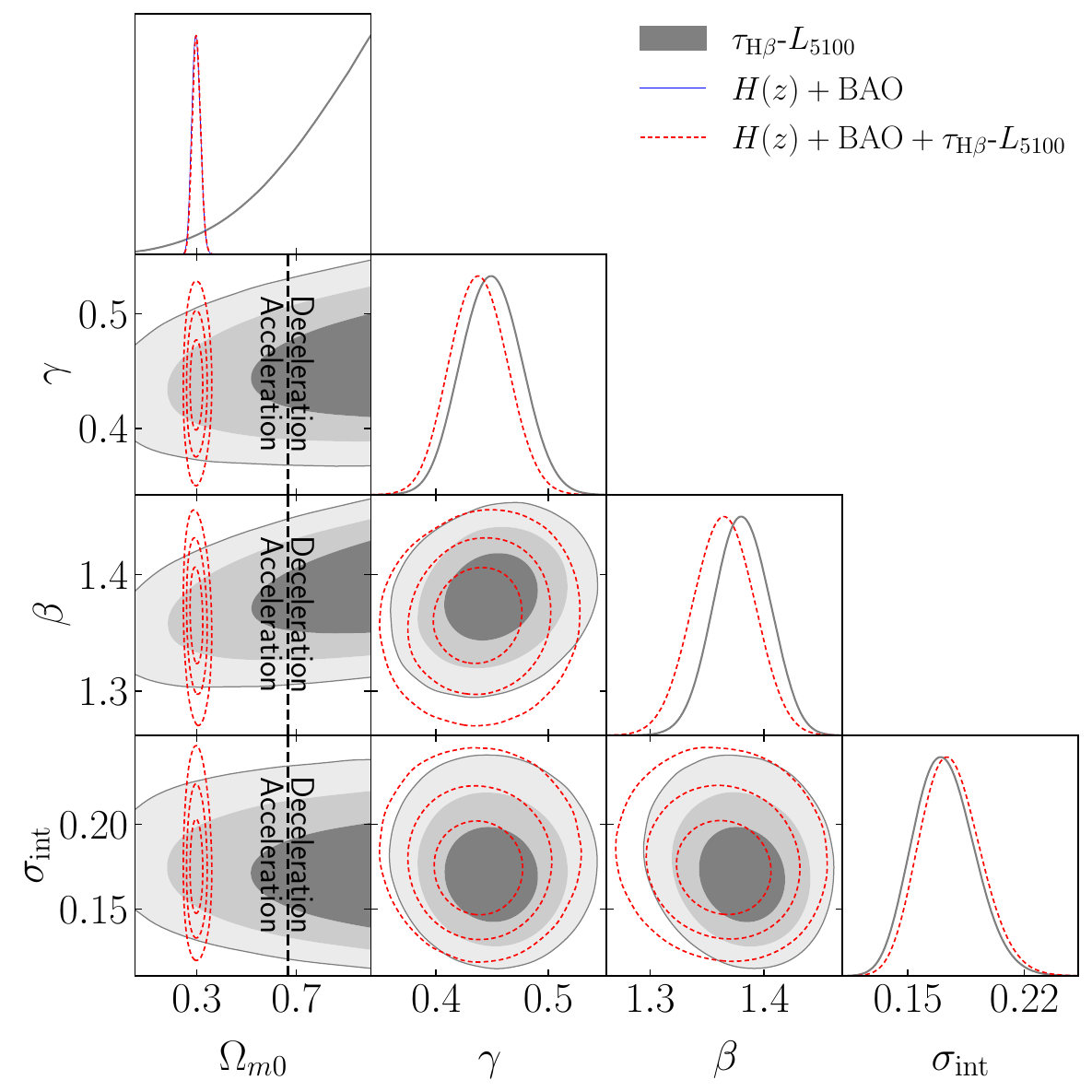}}
 \hspace{0.1\textwidth}
 \subfloat[]{%
    \includegraphics[width=0.35\textwidth,height=0.33\textwidth]{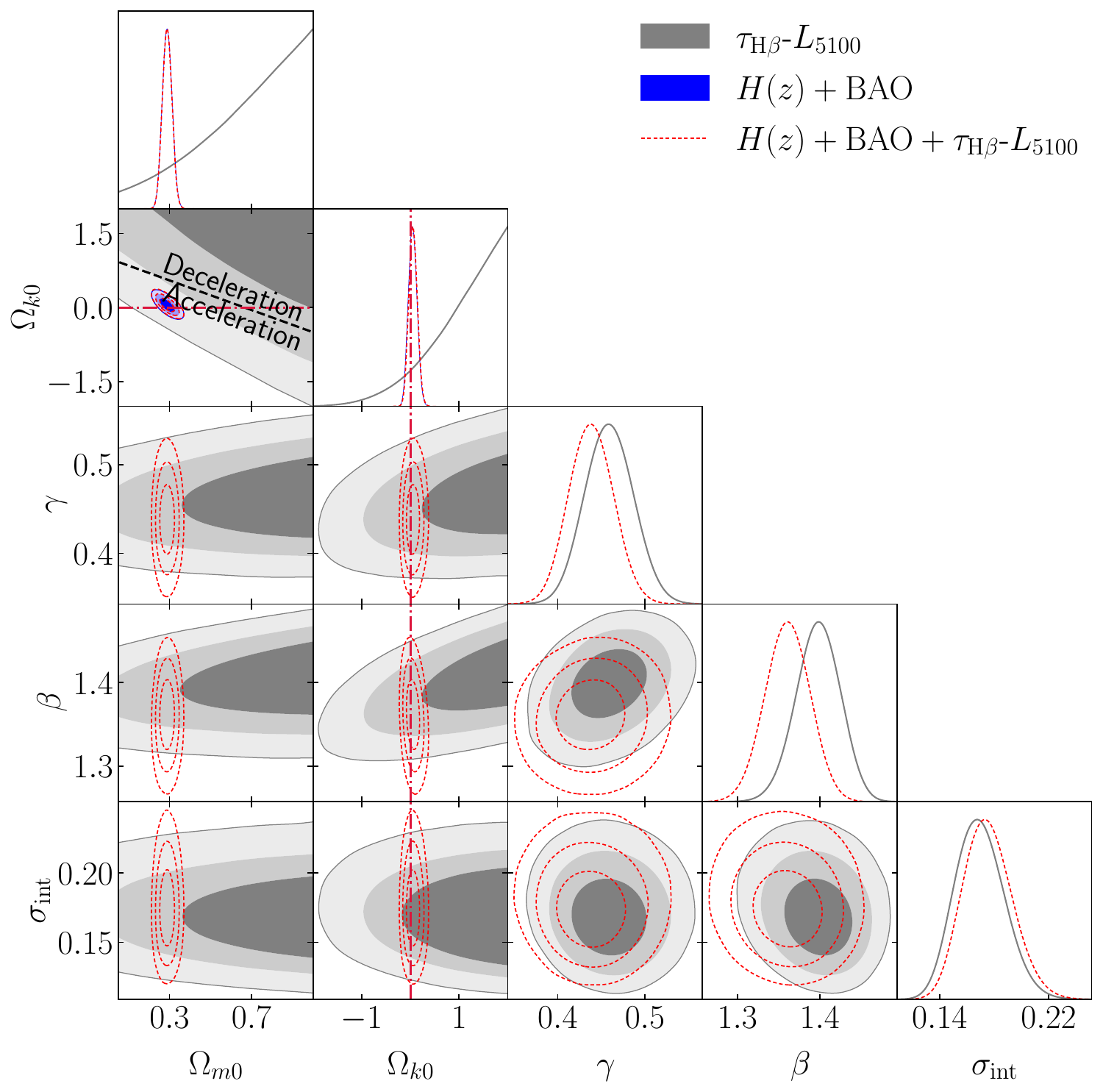}}\\
 \subfloat[]{%
    \includegraphics[width=0.35\textwidth,height=0.33\textwidth]{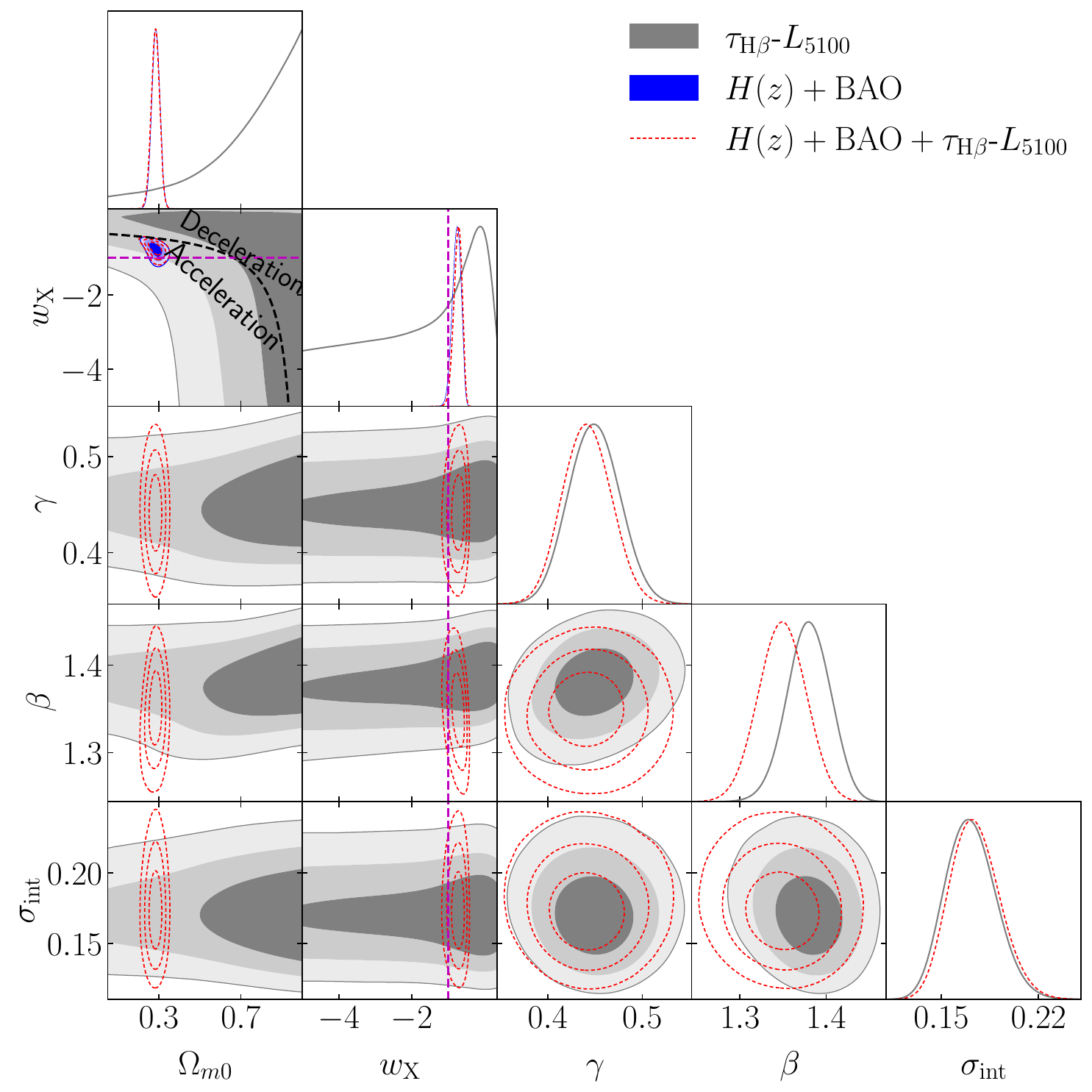}}
 \hspace{0.1\textwidth}
 \subfloat[]{%
    \includegraphics[width=0.35\textwidth,height=0.33\textwidth]{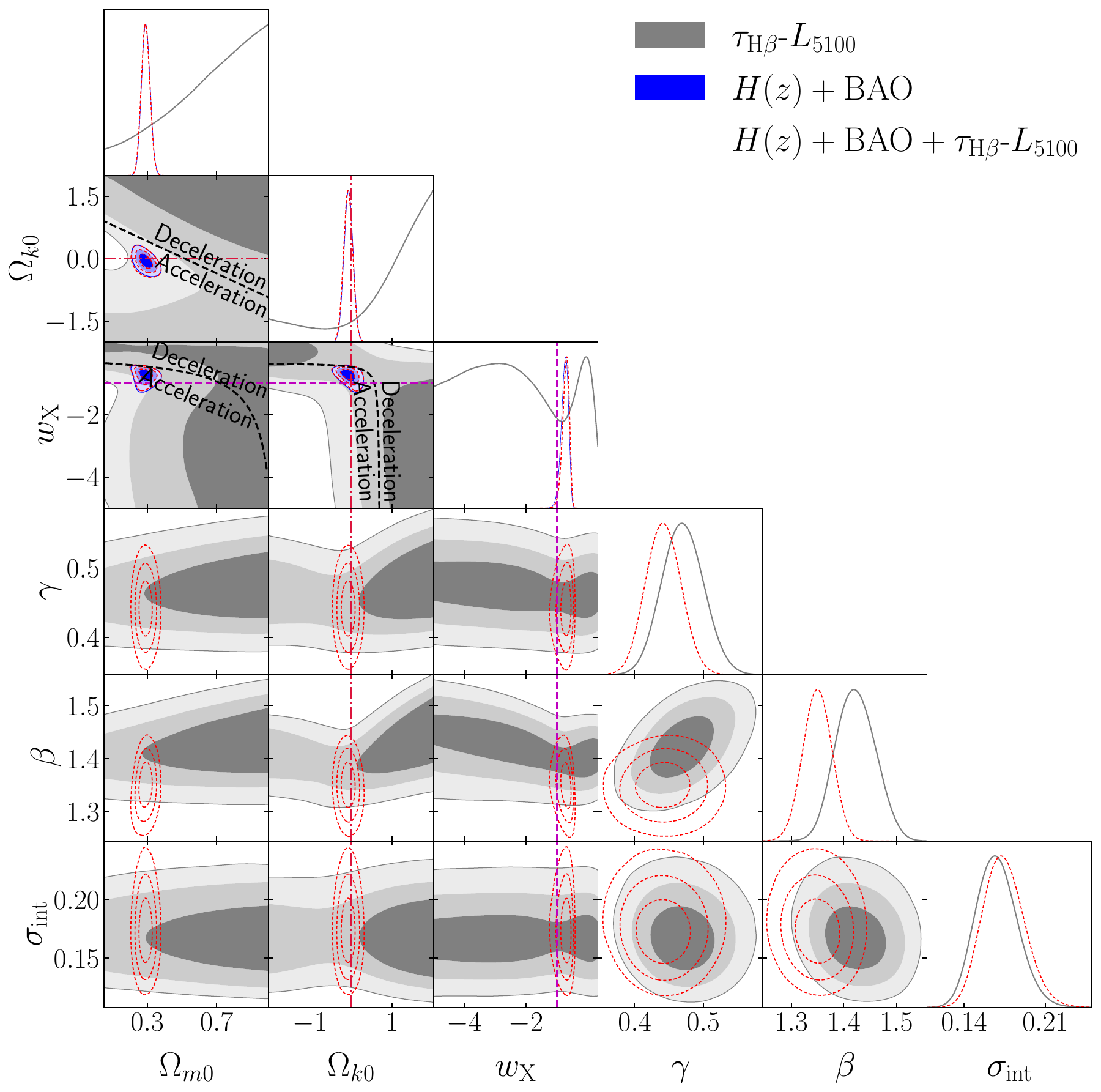}}\\
 \subfloat[]{%
    \includegraphics[width=0.35\textwidth,height=0.33\textwidth]{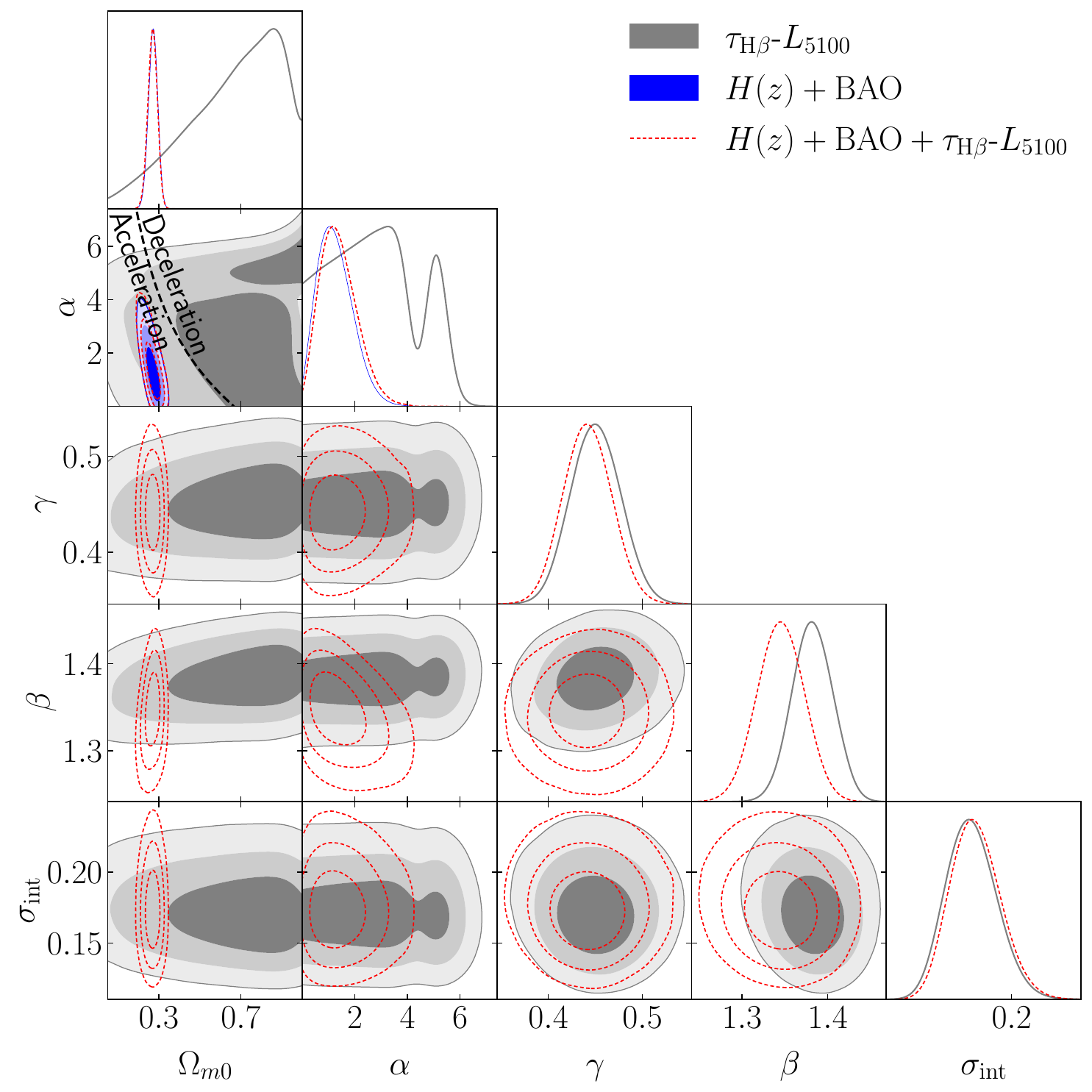}}
 \hspace{0.1\textwidth}
 \subfloat[]{%
    \includegraphics[width=0.35\textwidth,height=0.33\textwidth]{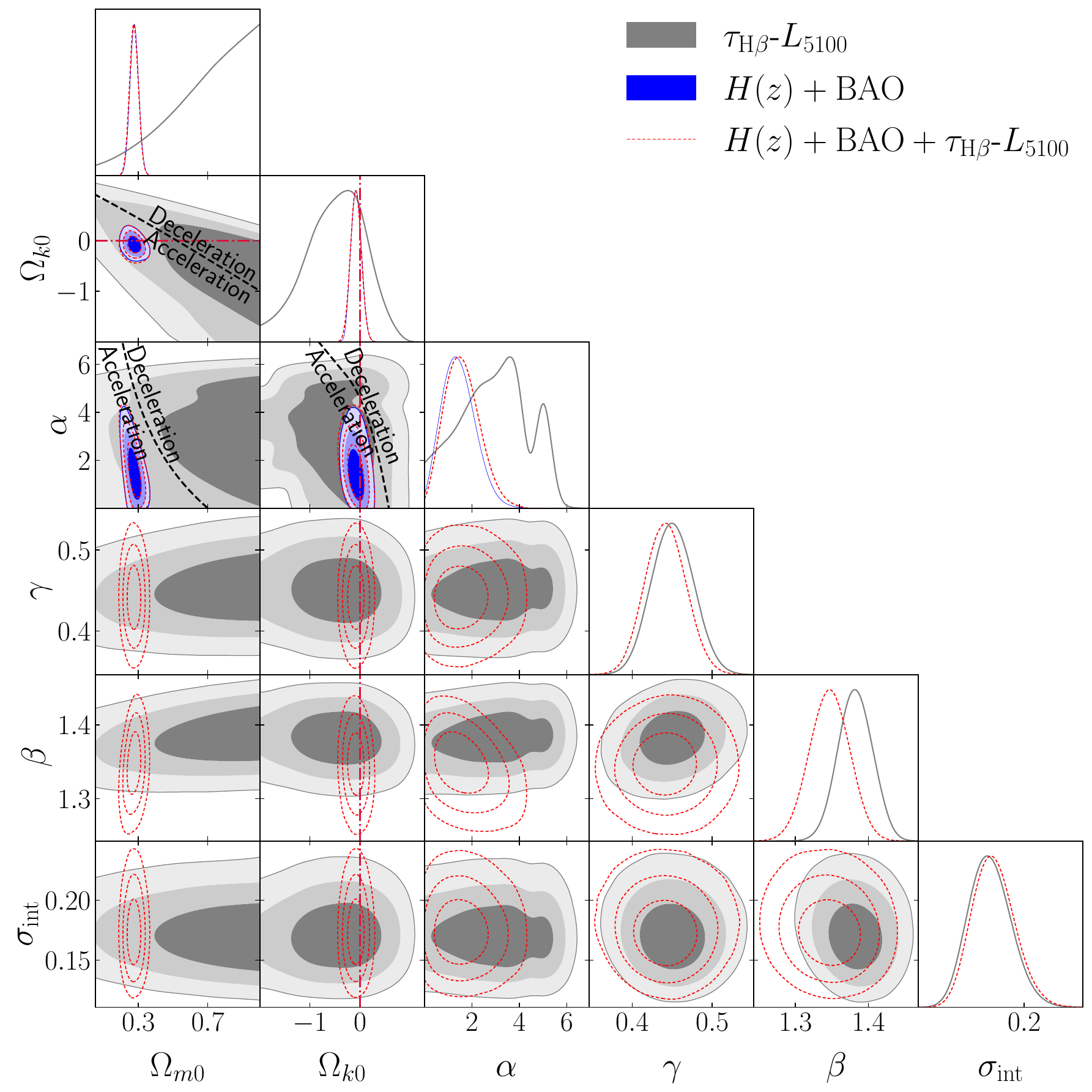}}\\
\caption{One-dimensional likelihoods and 1$\sigma$, 2$\sigma$, and 3$\sigma$ two-dimensional likelihood confidence contours from $\tau_{\mathrm{H}\beta}\text{-}L_{5100}$ (gray), $H(z)$ + BAO (blue), and $H(z)$ + BAO + $\tau_{\mathrm{H}\beta}\text{-}L_{5100}$ (dashed red) data for six different models, with \lcdm, XCDM, and \pcdm\ in the top, middle, and bottom rows, and flat (nonflat) models in the left (right) column. The black dashed zero-acceleration lines, computed for the third cosmological parameter set to the $H(z)$ + BAO data best-fitting values listed in Table \ref{tab:BFP} in (d), (f), divide the parameter space into regions associated with currently accelerating (below or below left) and currently decelerating (above or above right) cosmological expansion. The crimson dash-dot lines represent flat hypersurfaces, with closed spatial hypersurfaces either below or to the left. The magenta lines represent $w_{\rm X}=-1$, i.e.\ flat or nonflat \lcdm\ models. The $\alpha = 0$ axes correspond to flat and nonflat \lcdm\ models in (e), (f), respectively. (a) Flat \lcdm. (b) Nonflat \lcdm. (c) Flat XCDM. (d) Nonflat XCDM. (e) Flat \pcdm. (f) Nonflat \pcdm.}
\label{fig1}
\vspace{-50pt}
\end{figure*}

\begin{figure*}[htbp]
\centering
 \subfloat[]{%
    \includegraphics[width=0.35\textwidth,height=0.33\textwidth]{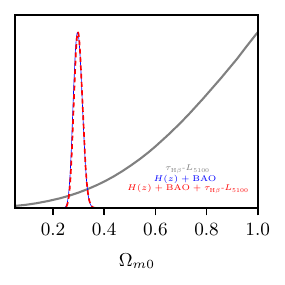}}
 \hspace{0.1\textwidth}
 \subfloat[]{%
    \includegraphics[width=0.35\textwidth,height=0.33\textwidth]{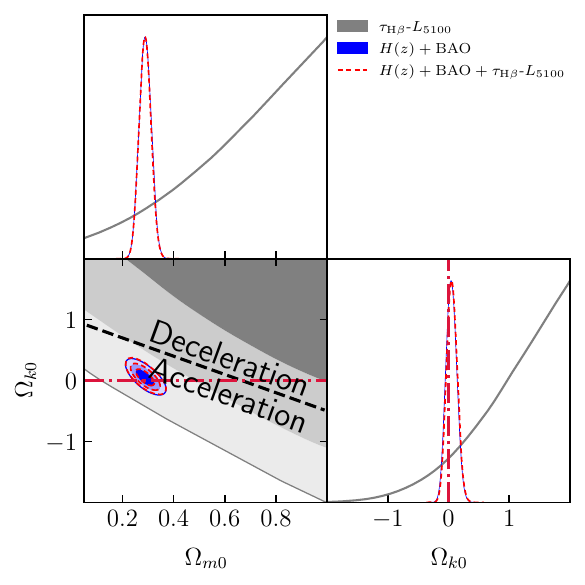}}\\
 \subfloat[]{%
    \includegraphics[width=0.35\textwidth,height=0.33\textwidth]{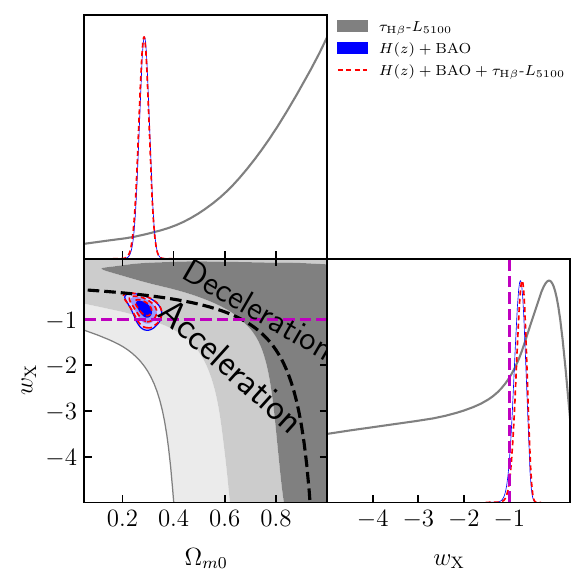}}
 \hspace{0.1\textwidth}
 \subfloat[]{%
    \includegraphics[width=0.35\textwidth,height=0.33\textwidth]{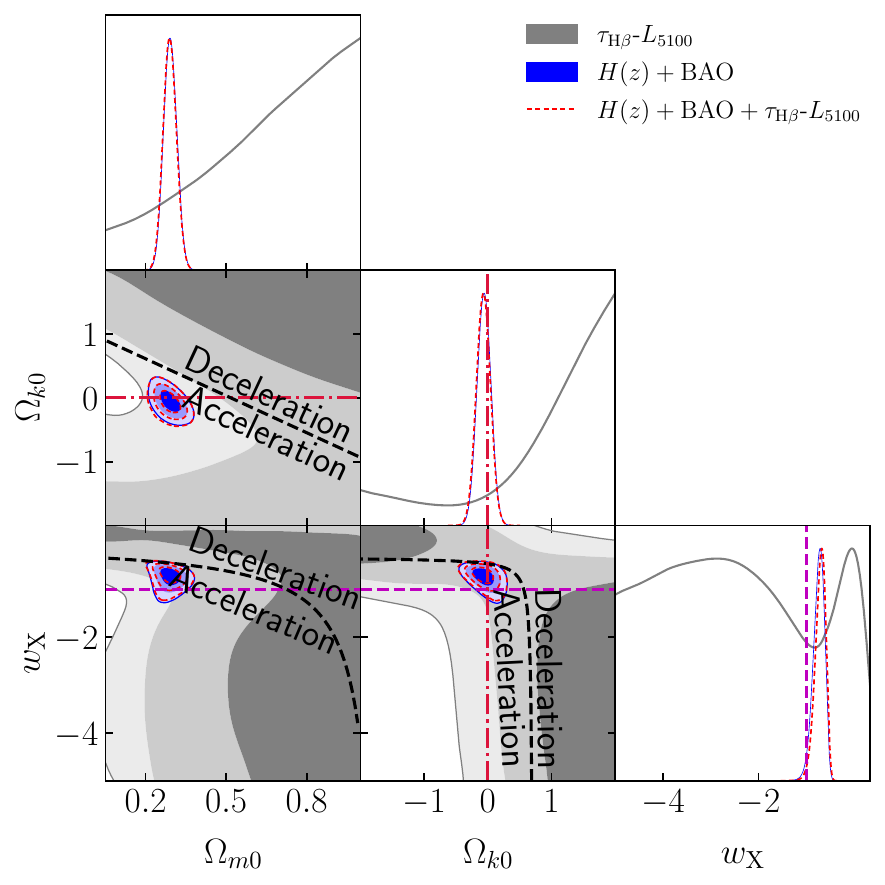}}\\
 \subfloat[]{%
    \includegraphics[width=0.35\textwidth,height=0.33\textwidth]{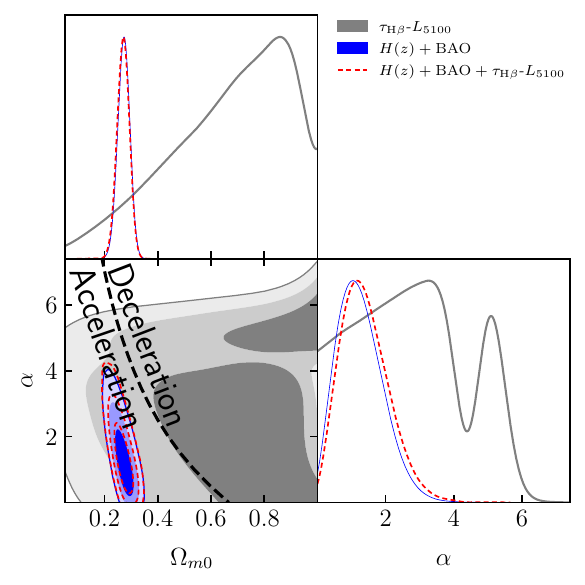}}
 \hspace{0.1\textwidth}
 \subfloat[]{%
    \includegraphics[width=0.35\textwidth,height=0.33\textwidth]{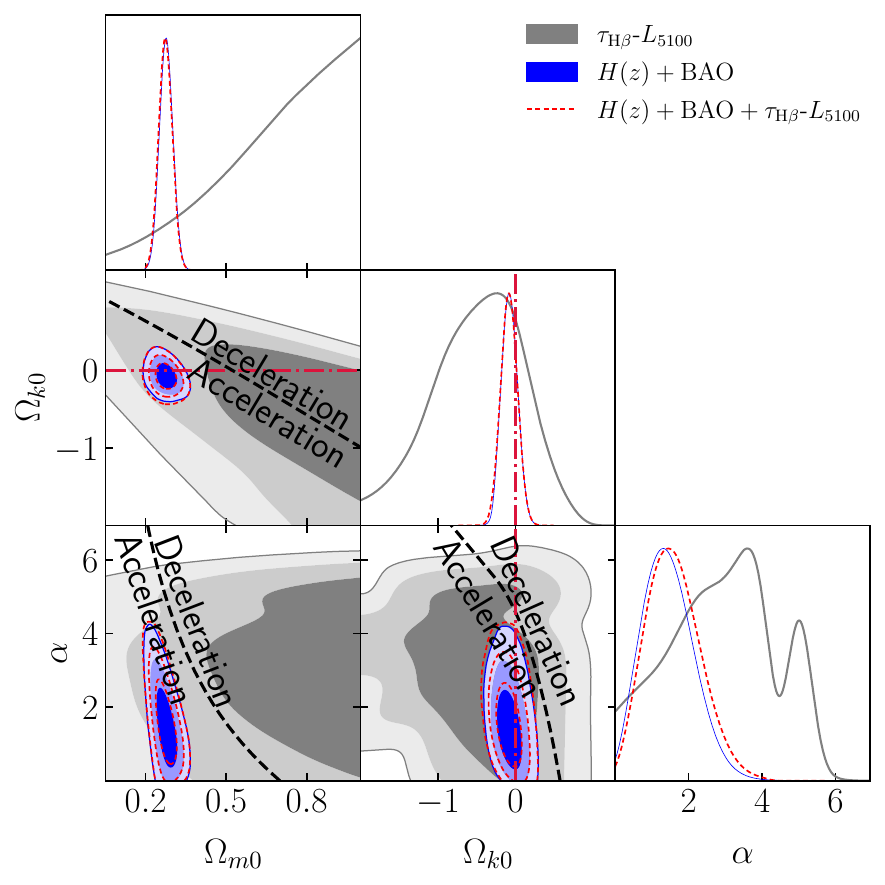}}\\
\caption{Same as Fig.\ \ref{fig1}, but for cosmological parameters only. (a) Flat \lcdm. (b) Nonflat \lcdm. (c) Flat XCDM. (d) Nonflat XCDM. (e) Flat \pcdm. (f) Nonflat \pcdm.}
\label{fig2}
\end{figure*}

\begin{figure*}[htbp]
\centering
 \subfloat[]{%
    \includegraphics[width=0.35\textwidth,height=0.33\textwidth]{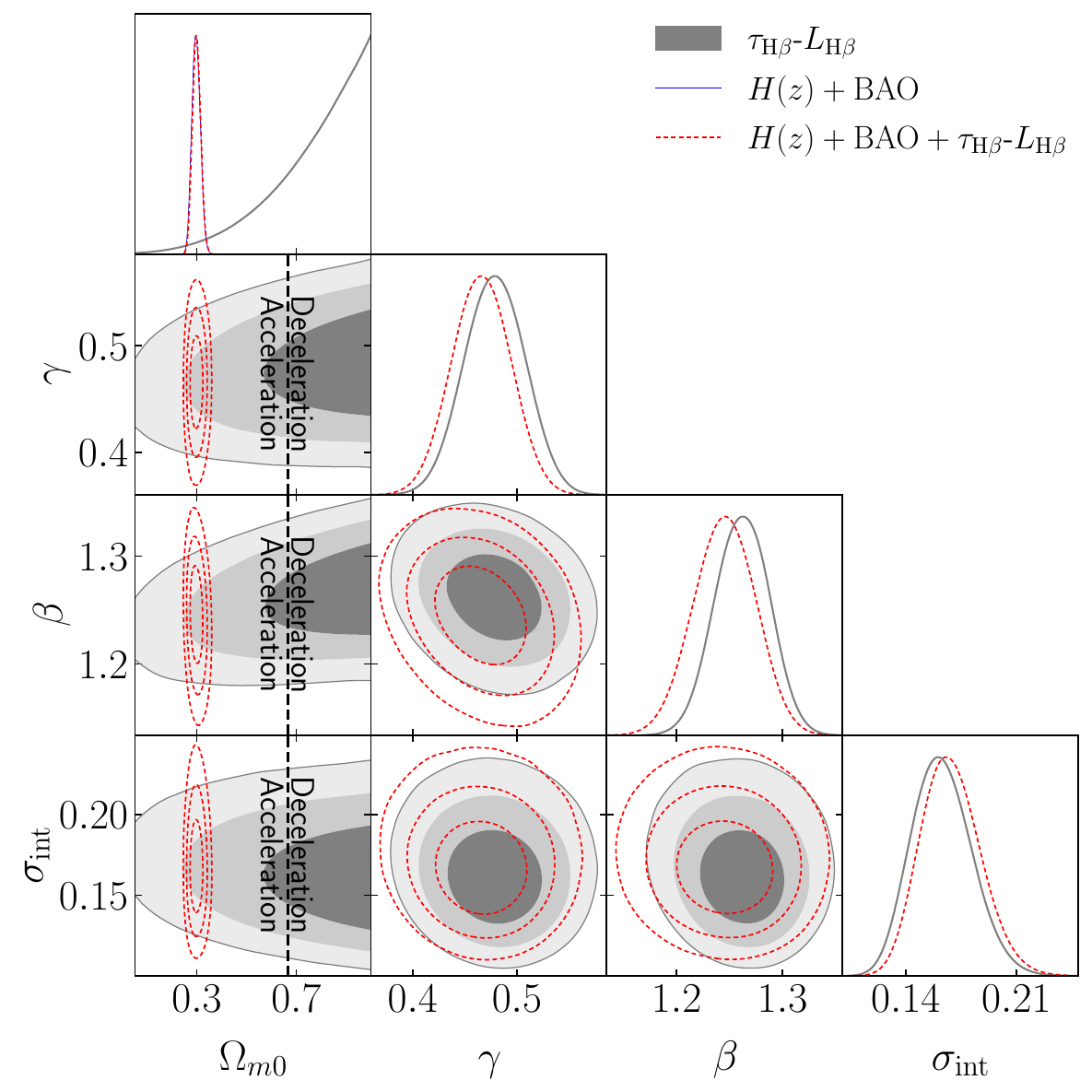}}
 \hspace{0.1\textwidth}
 \subfloat[]{%
    \includegraphics[width=0.35\textwidth,height=0.33\textwidth]{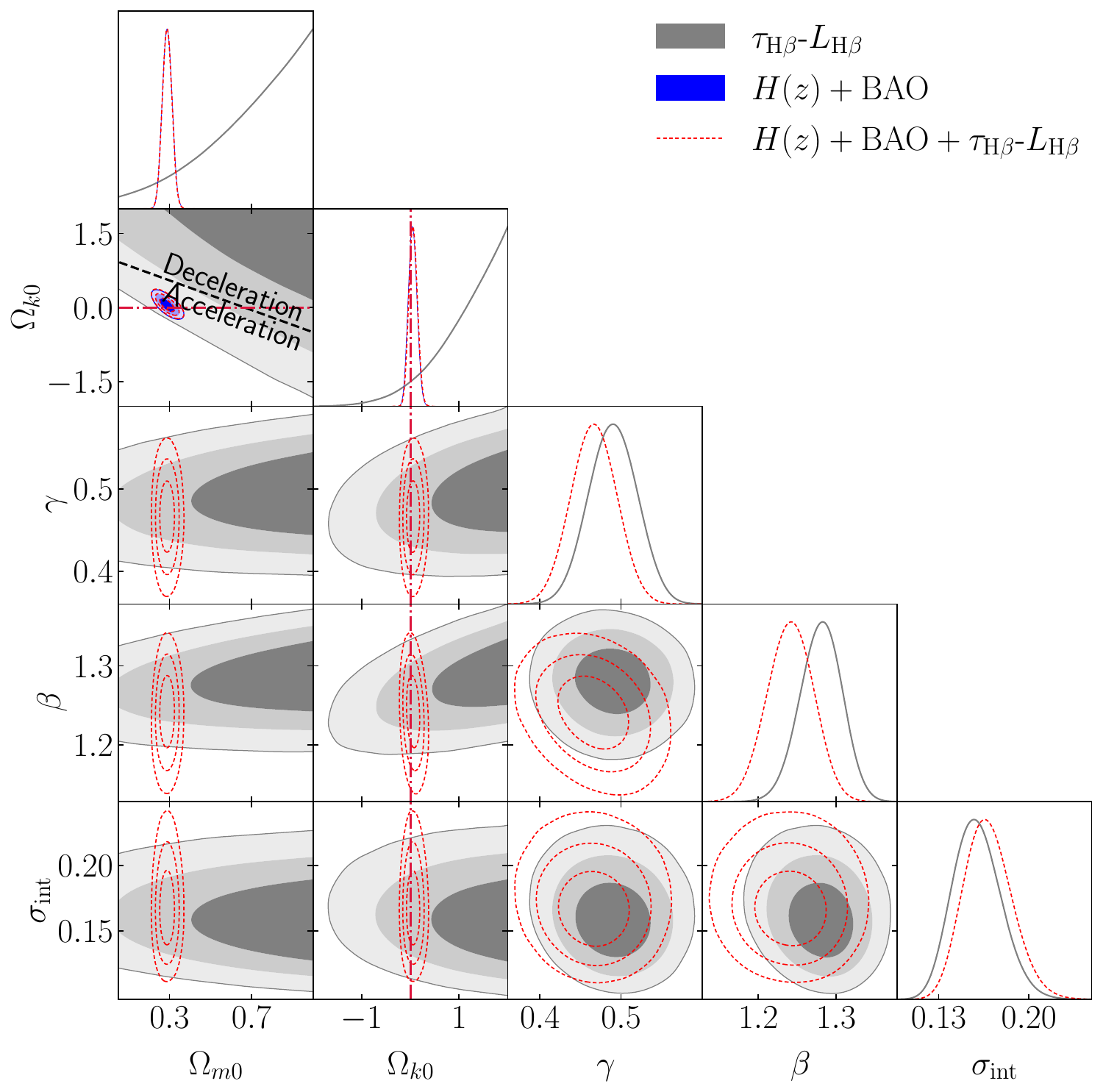}}\\
 \subfloat[]{%
    \includegraphics[width=0.35\textwidth,height=0.33\textwidth]{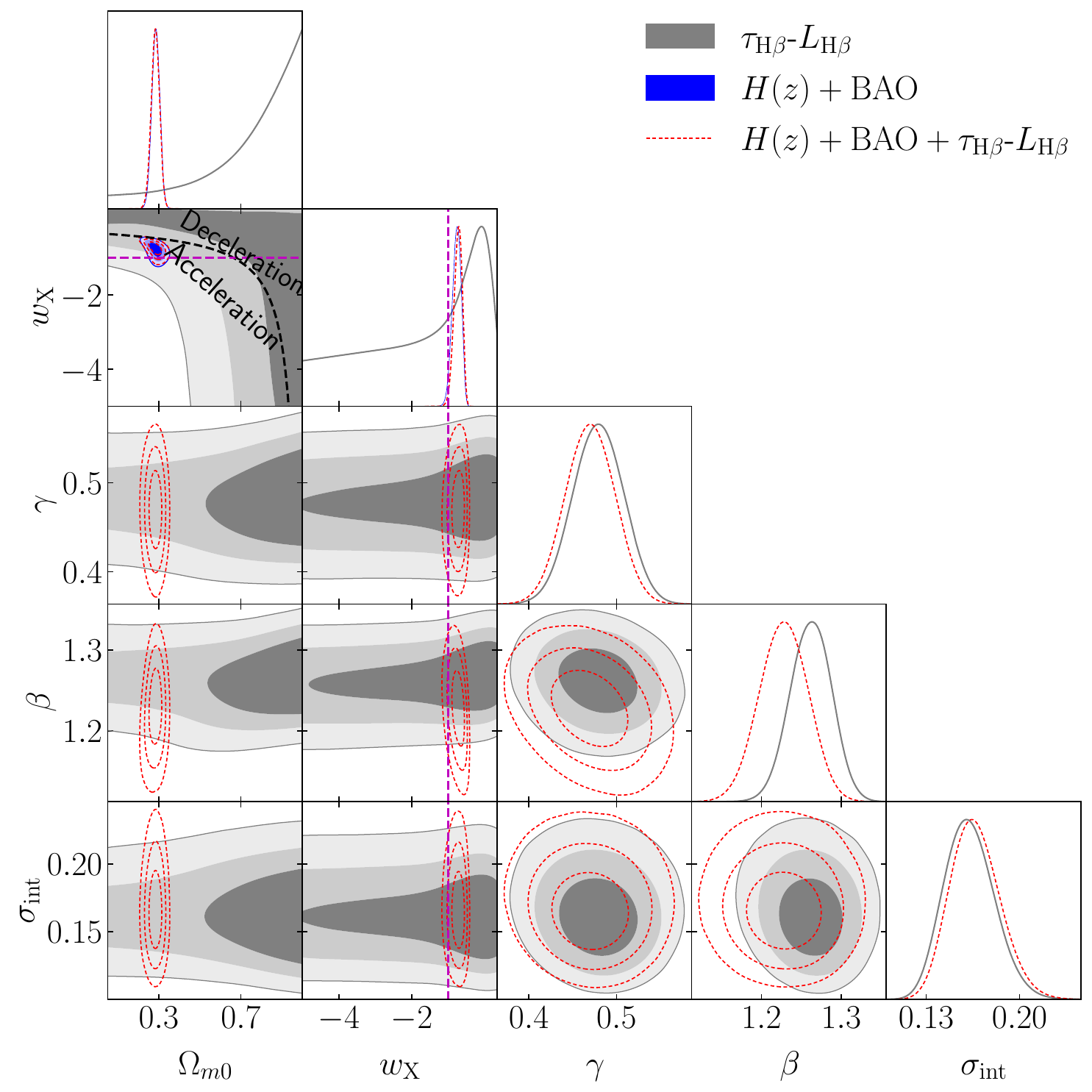}}
 \hspace{0.1\textwidth}
 \subfloat[]{%
    \includegraphics[width=0.35\textwidth,height=0.33\textwidth]{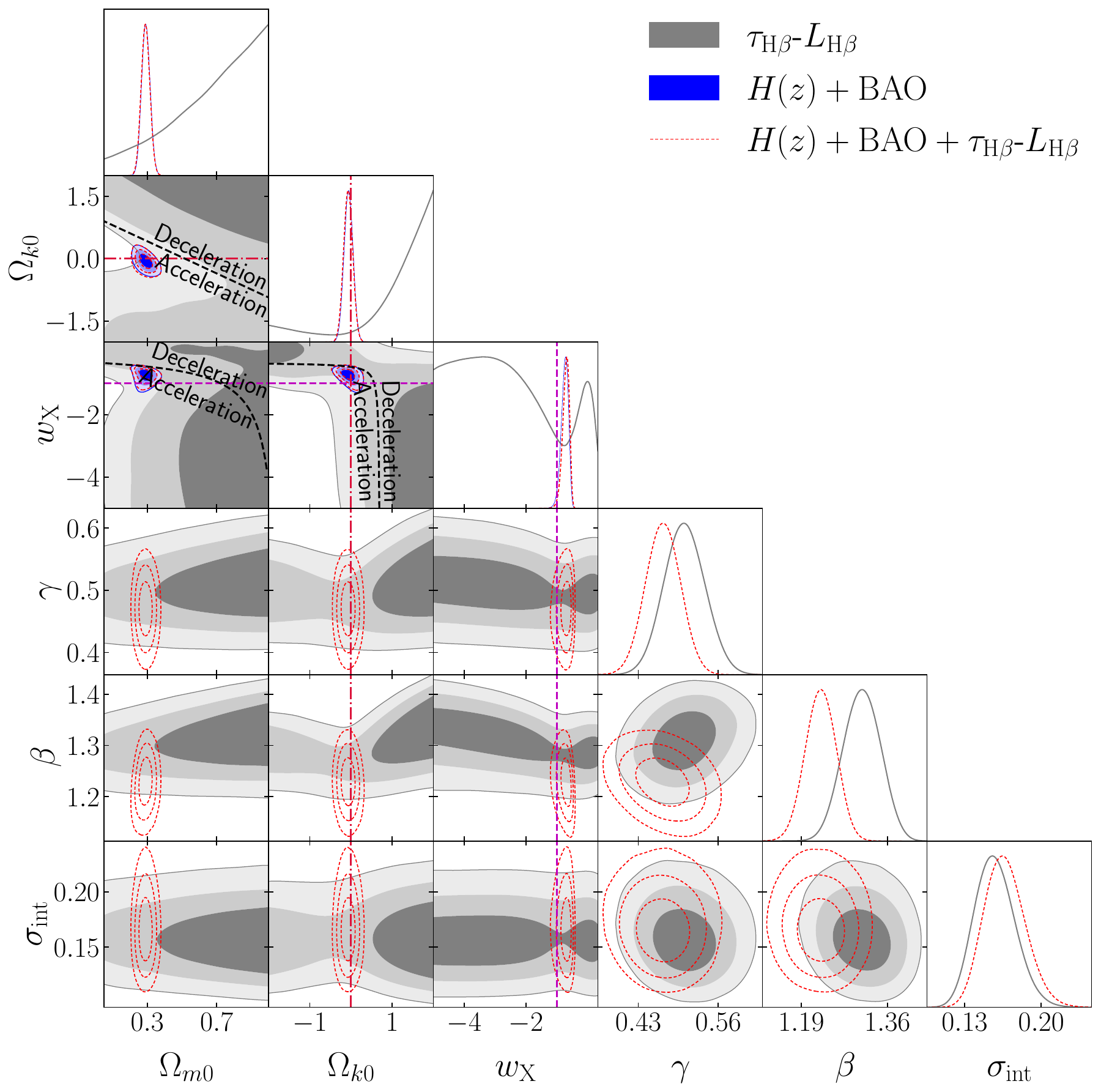}}\\
 \subfloat[]{%
    \includegraphics[width=0.35\textwidth,height=0.33\textwidth]{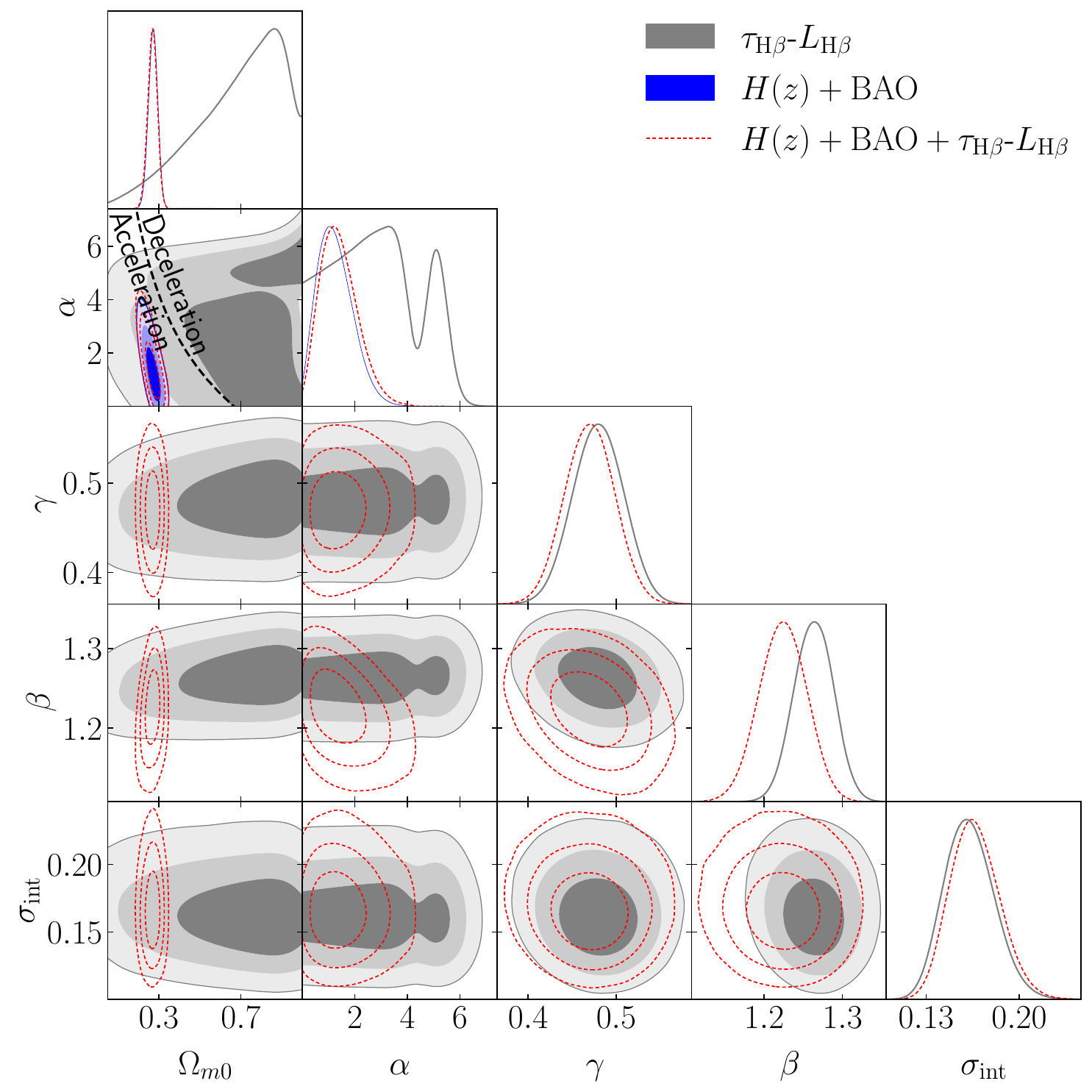}}
 \hspace{0.1\textwidth}
 \subfloat[]{%
    \includegraphics[width=0.35\textwidth,height=0.33\textwidth]{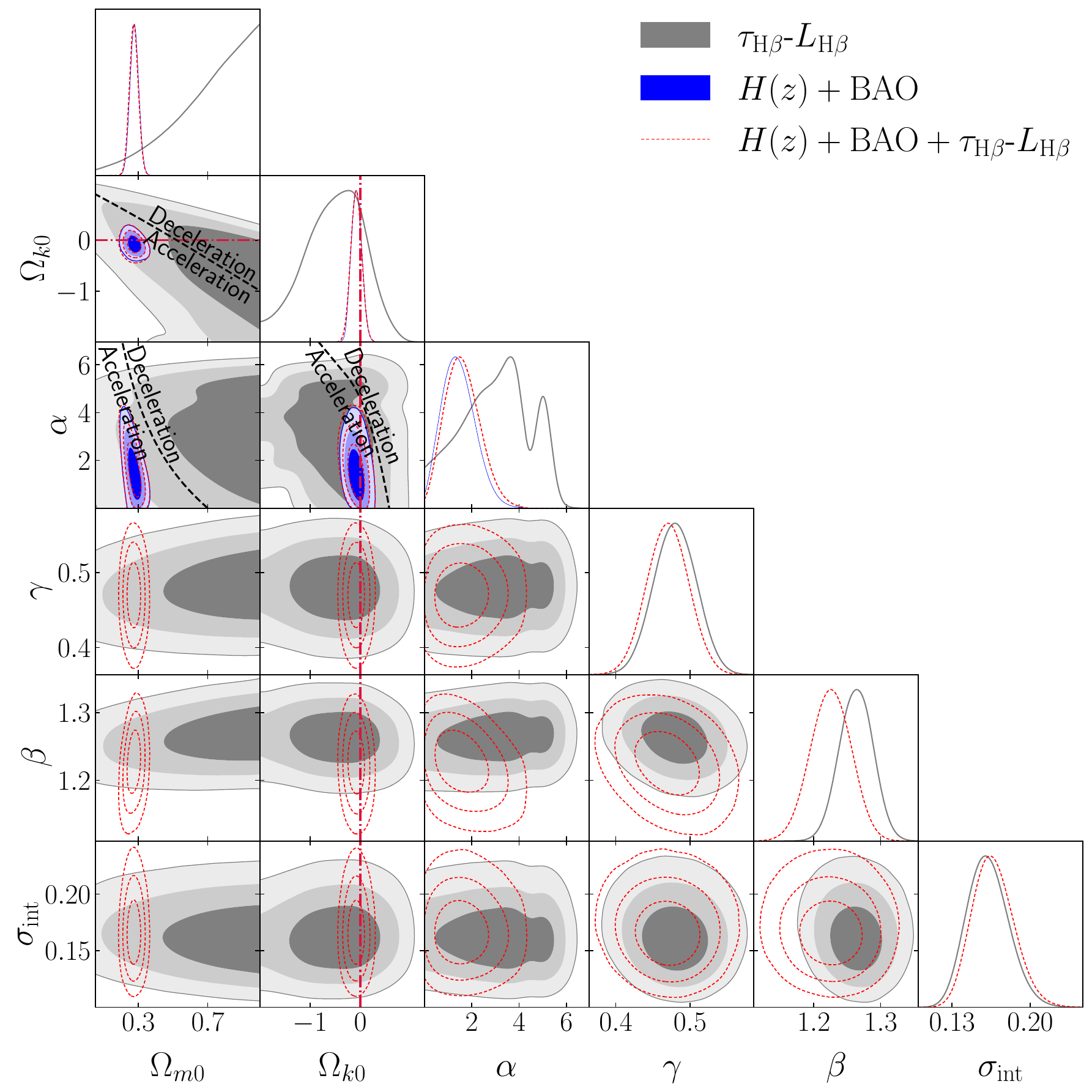}}\\
\caption{Same as Fig.~\ref{fig1}, but for $\tau_{\mathrm{H}\beta}\text{-}L_{\mathrm{H}\beta}$ (gray), $H(z)$ + BAO (blue), and $H(z)$ + BAO + $\tau_{\mathrm{H}\beta}\text{-}L_{\mathrm{H}\beta}$ (dashed red) data. (a) Flat \lcdm. (b) Nonflat \lcdm. (c) Flat XCDM. (d) Nonflat XCDM. (e) Flat \pcdm. (f) Nonflat \pcdm.}
\label{fig3}
\end{figure*}

\begin{figure*}
\centering
 \subfloat[]{%
    \includegraphics[width=0.35\textwidth,height=0.33\textwidth]{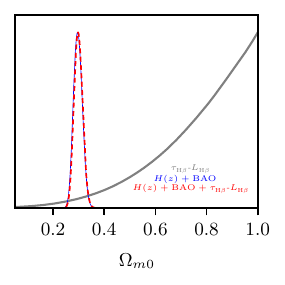}}
 \hspace{0.1\textwidth}
 \subfloat[]{%
    \includegraphics[width=0.35\textwidth,height=0.33\textwidth]{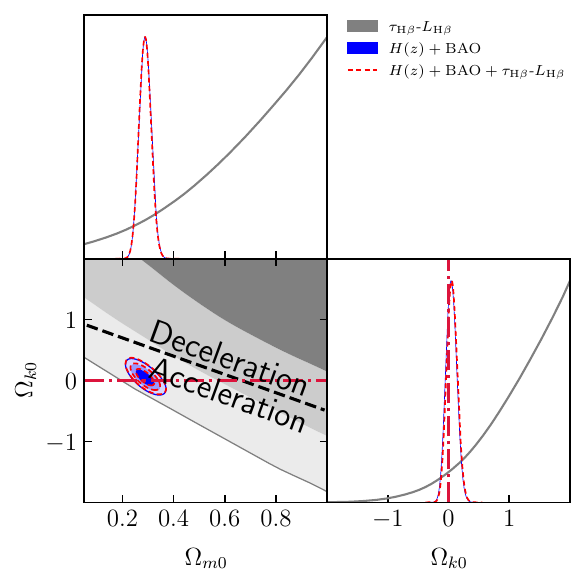}}\\
 \subfloat[]{%
    \includegraphics[width=0.35\textwidth,height=0.33\textwidth]{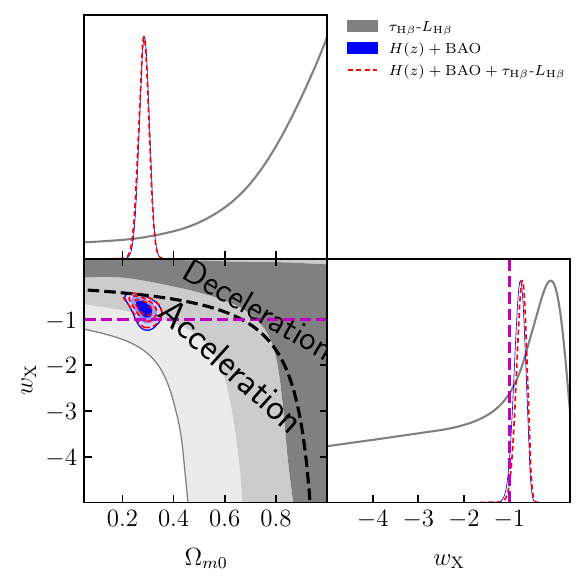}}
 \hspace{0.1\textwidth}
 \subfloat[]{%
    \includegraphics[width=0.35\textwidth,height=0.33\textwidth]{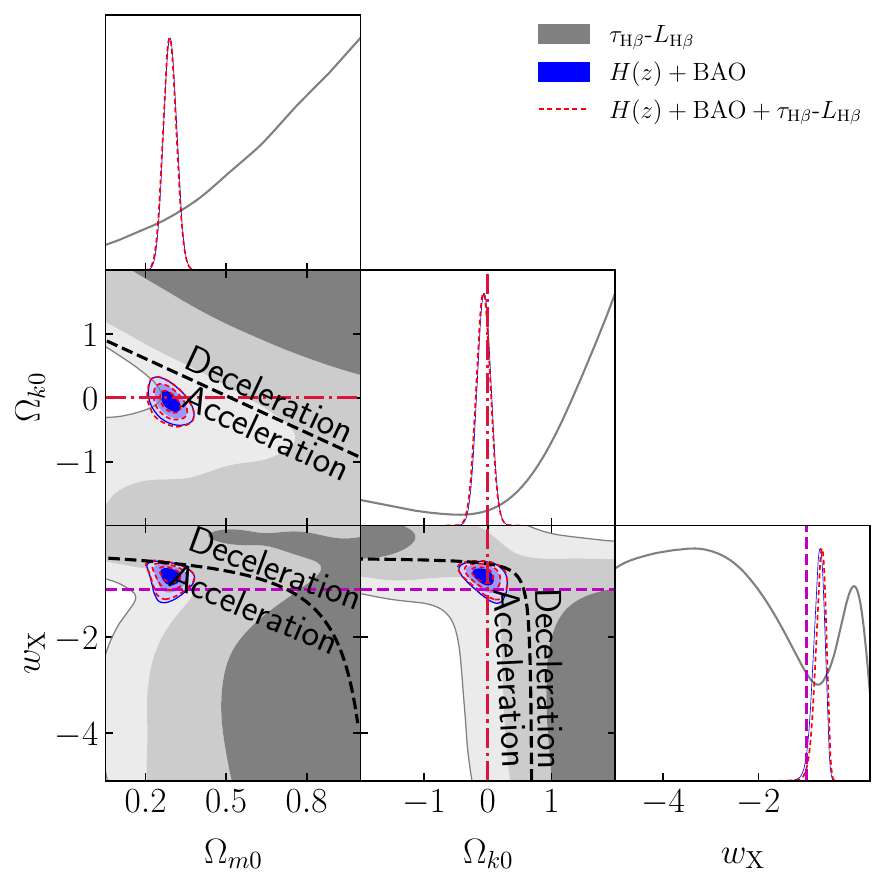}}\\
 \subfloat[]{%
    \includegraphics[width=0.35\textwidth,height=0.33\textwidth]{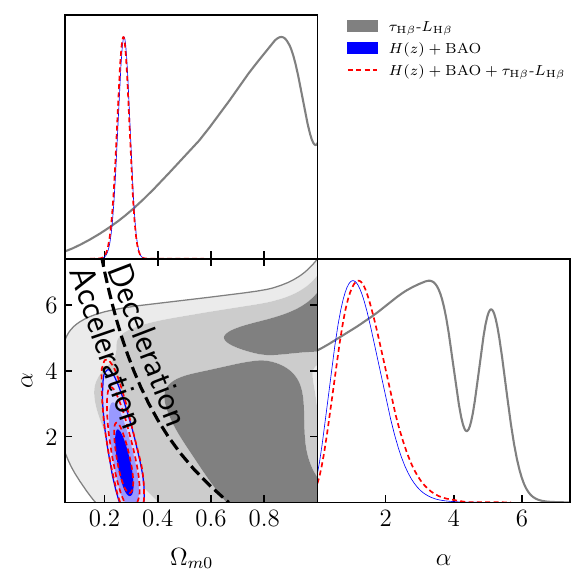}}
 \hspace{0.1\textwidth}
 \subfloat[]{%
    \includegraphics[width=0.35\textwidth,height=0.33\textwidth]{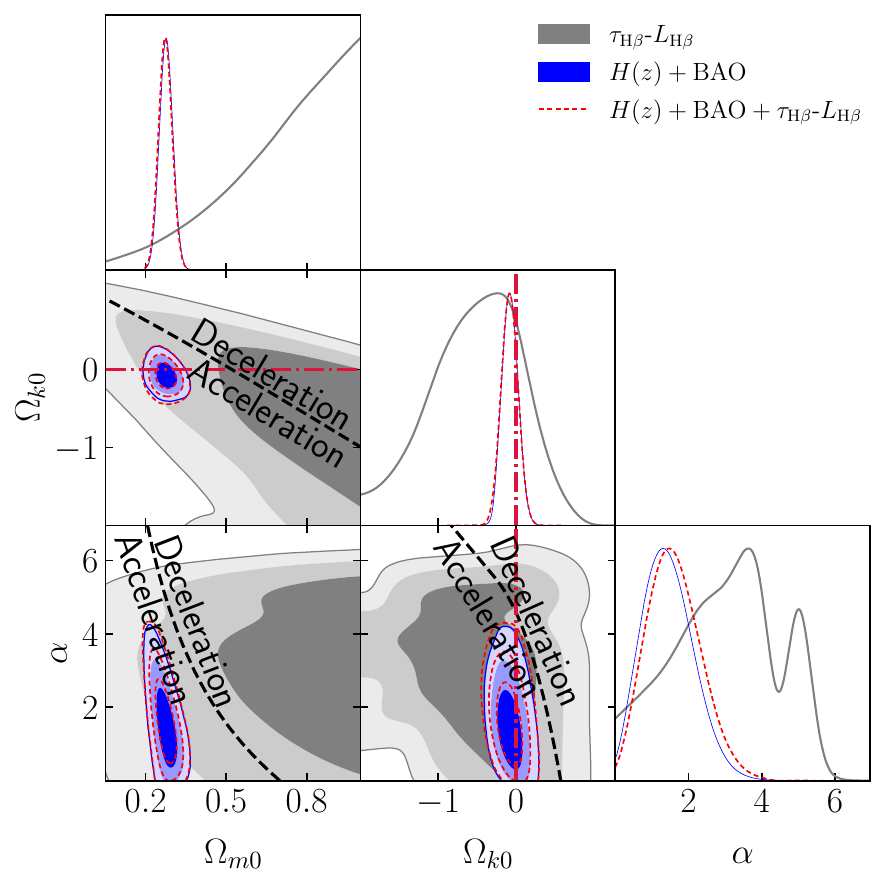}}\\
\caption{Same as Fig.\ \ref{fig3}, but for cosmological parameters only. (a) Flat \lcdm. (b) Nonflat \lcdm. (c) Flat XCDM. (d) Nonflat XCDM. (e) Flat \pcdm. (f) Nonflat \pcdm.}
\label{fig4}
\end{figure*}

\begin{figure*}[htbp]
\centering
 \subfloat[]{%
    \includegraphics[width=0.35\textwidth,height=0.33\textwidth]{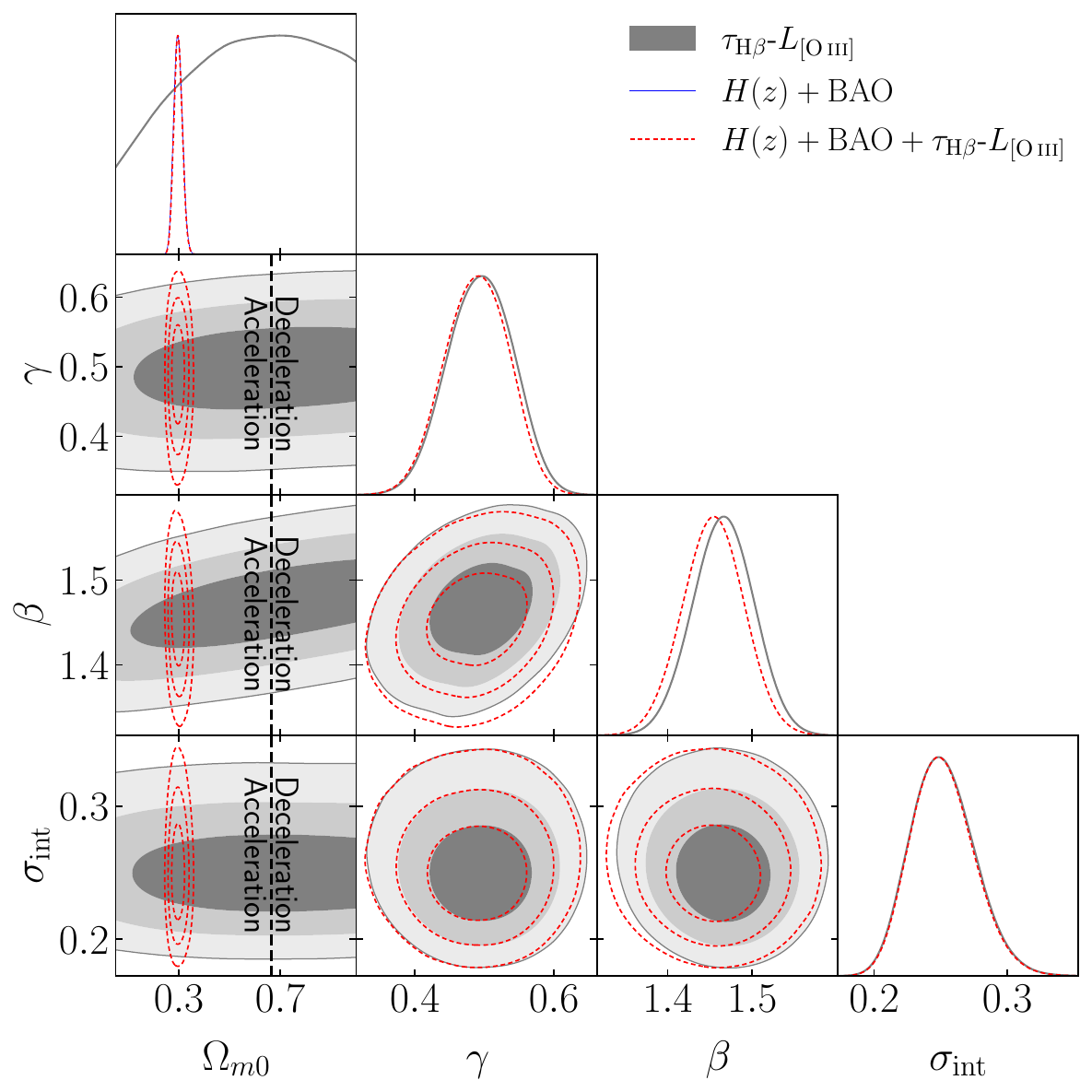}}
 \hspace{0.1\textwidth}
 \subfloat[]{%
    \includegraphics[width=0.35\textwidth,height=0.33\textwidth]{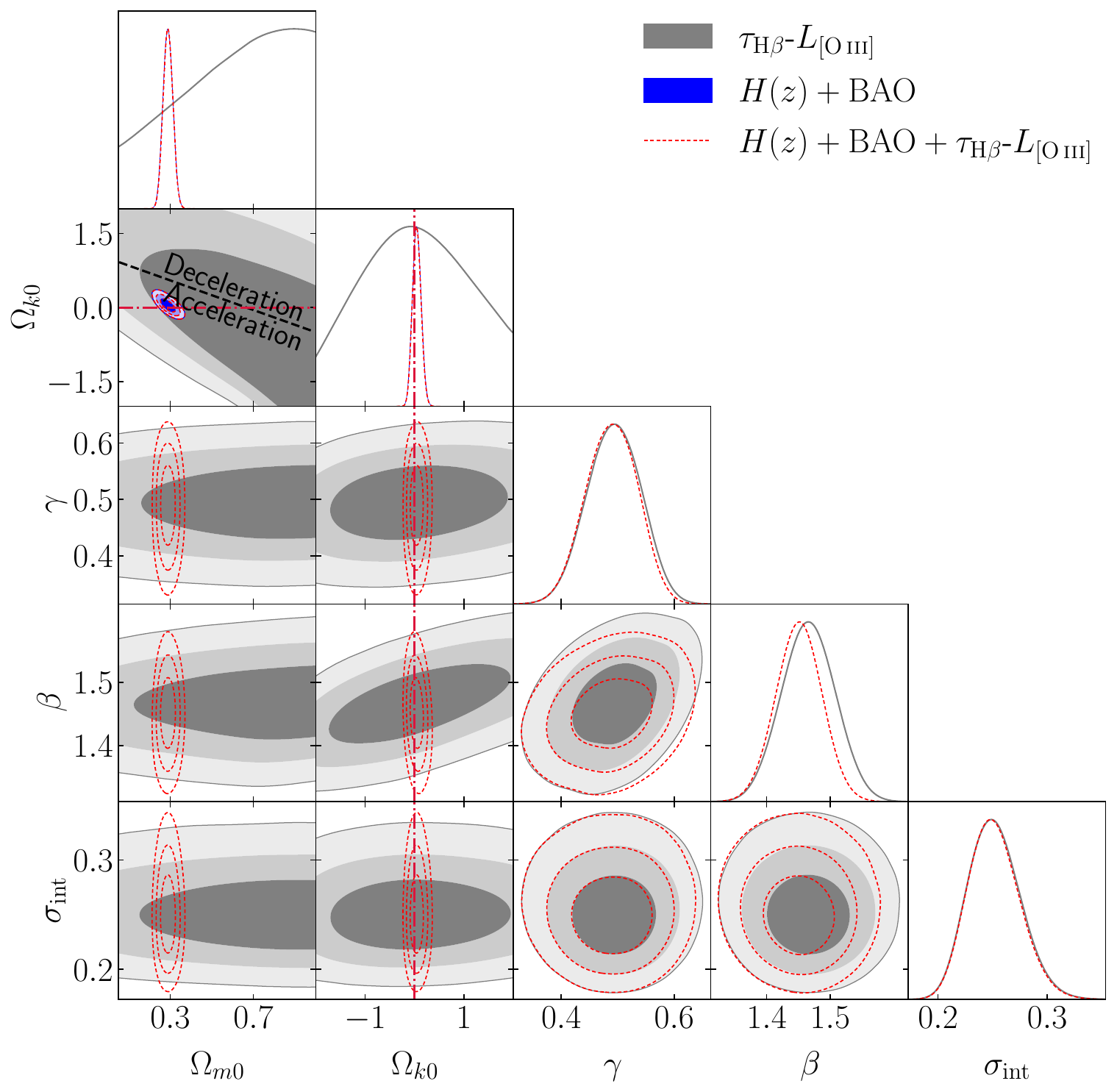}}\\
 \subfloat[]{%
    \includegraphics[width=0.35\textwidth,height=0.33\textwidth]{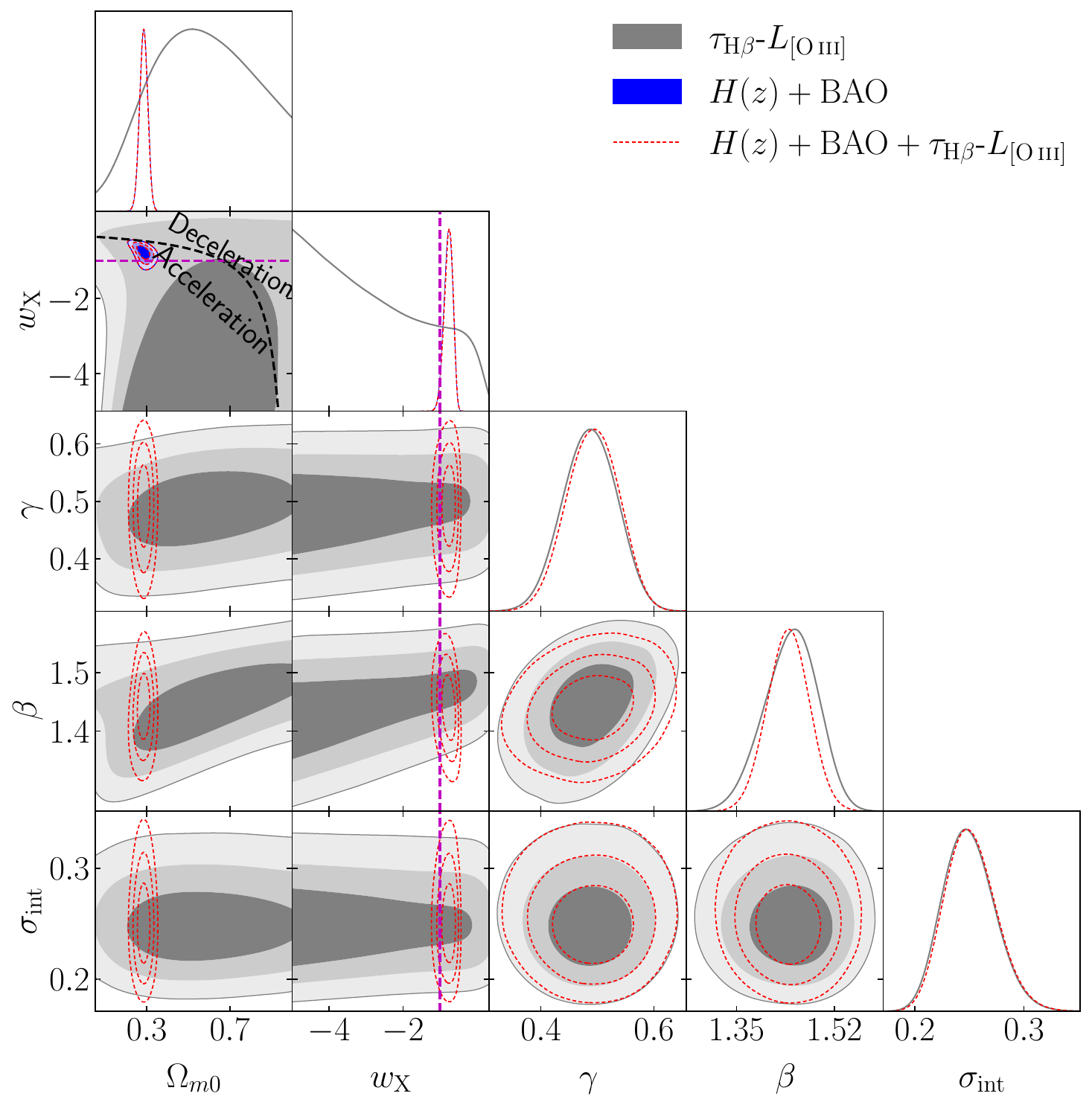}}
 \hspace{0.1\textwidth}
 \subfloat[]{%
    \includegraphics[width=0.35\textwidth,height=0.33\textwidth]{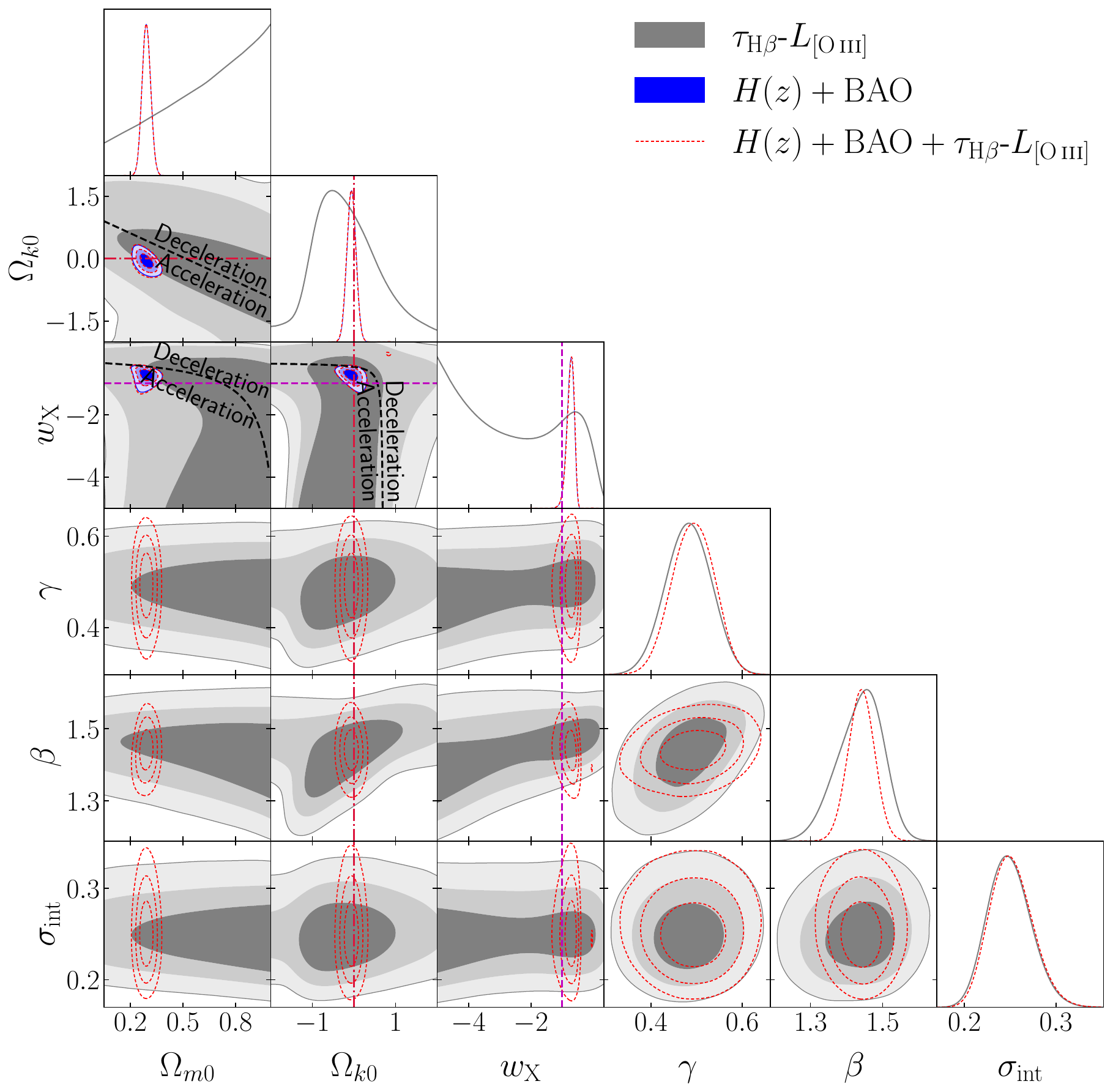}}\\
 \subfloat[]{%
    \includegraphics[width=0.35\textwidth,height=0.33\textwidth]{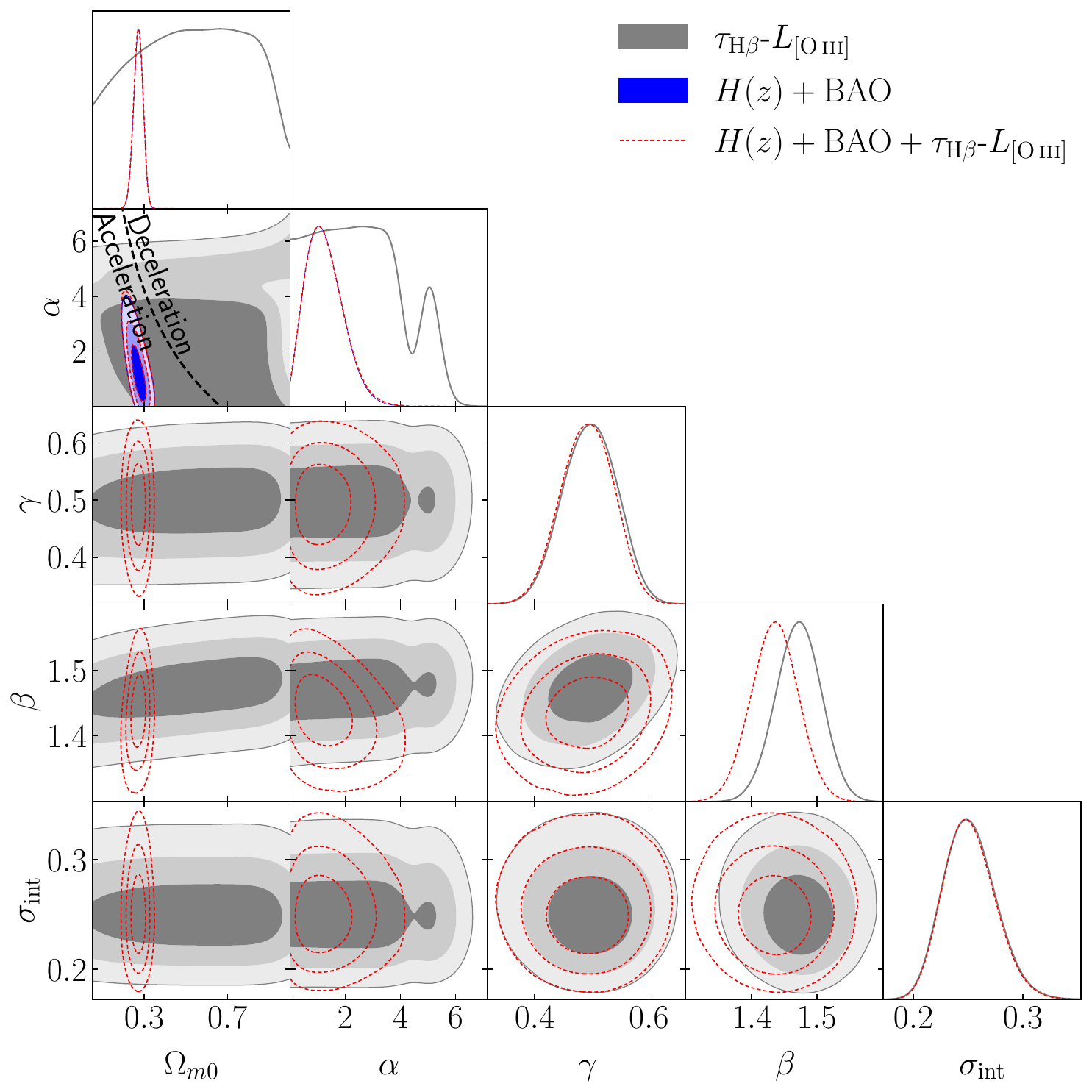}}
 \hspace{0.1\textwidth}
 \subfloat[]{%
    \includegraphics[width=0.35\textwidth,height=0.33\textwidth]{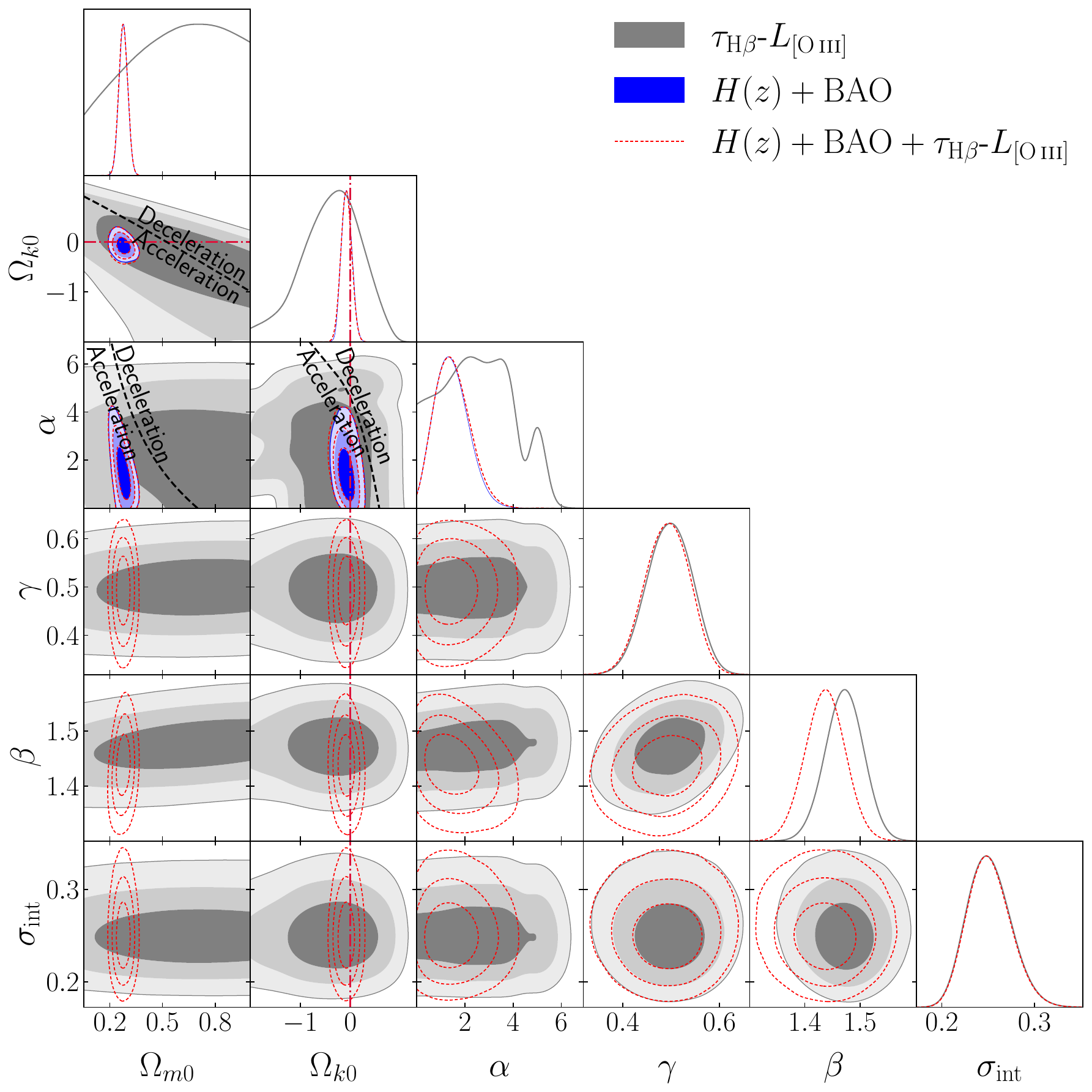}}\\
\caption{Same as Fig.~\ref{fig1}, but for $\tau_{\mathrm{H}\beta}\text{-}L_{\mathrm{{\OIII}}}$ (gray), $H(z)$ + BAO (blue), and $H(z)$ + BAO + $\tau_{\mathrm{H}\beta}\text{-}L_{\mathrm{{\OIII}}}$ (dashed red) data. (a) Flat \lcdm. (b) Nonflat \lcdm. (c) Flat XCDM. (d) Nonflat XCDM. (e) Flat \pcdm. (f) Nonflat \pcdm.}
\label{fig5}
\end{figure*}

\begin{figure*}[htbp]
\centering
 \subfloat[]{%
    \includegraphics[width=0.35\textwidth,height=0.33\textwidth]{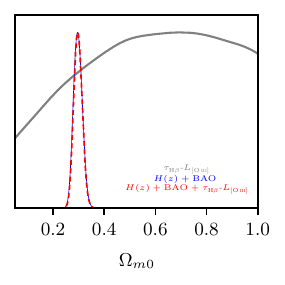}}
 \hspace{0.1\textwidth}
 \subfloat[]{%
    \includegraphics[width=0.35\textwidth,height=0.33\textwidth]{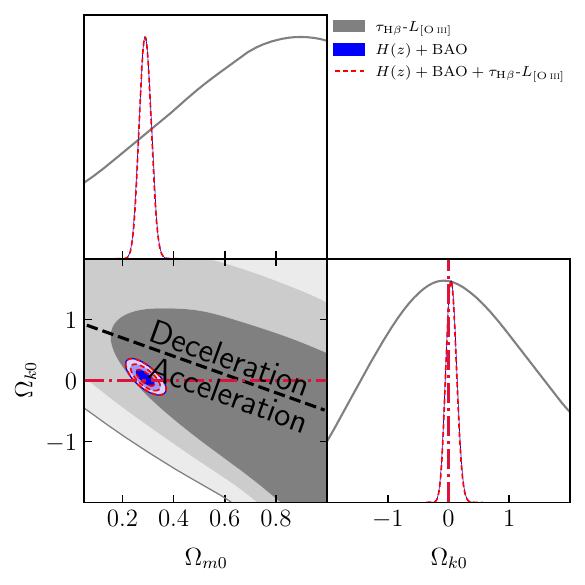}}\\
 \subfloat[]{%
    \includegraphics[width=0.35\textwidth,height=0.33\textwidth]{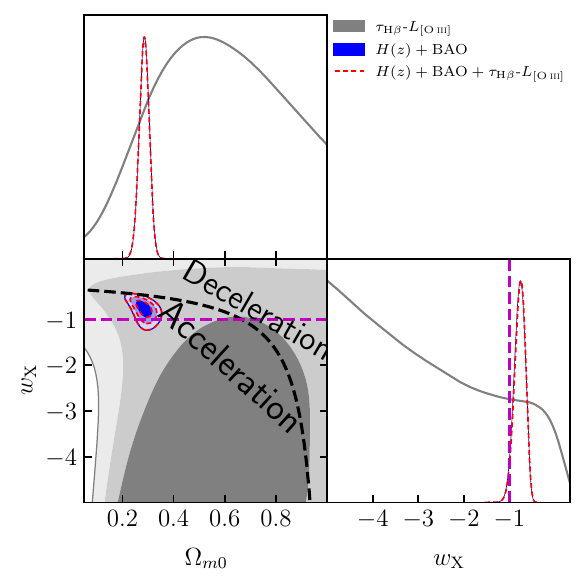}}
 \hspace{0.1\textwidth}
 \subfloat[]{%
    \includegraphics[width=0.35\textwidth,height=0.33\textwidth]{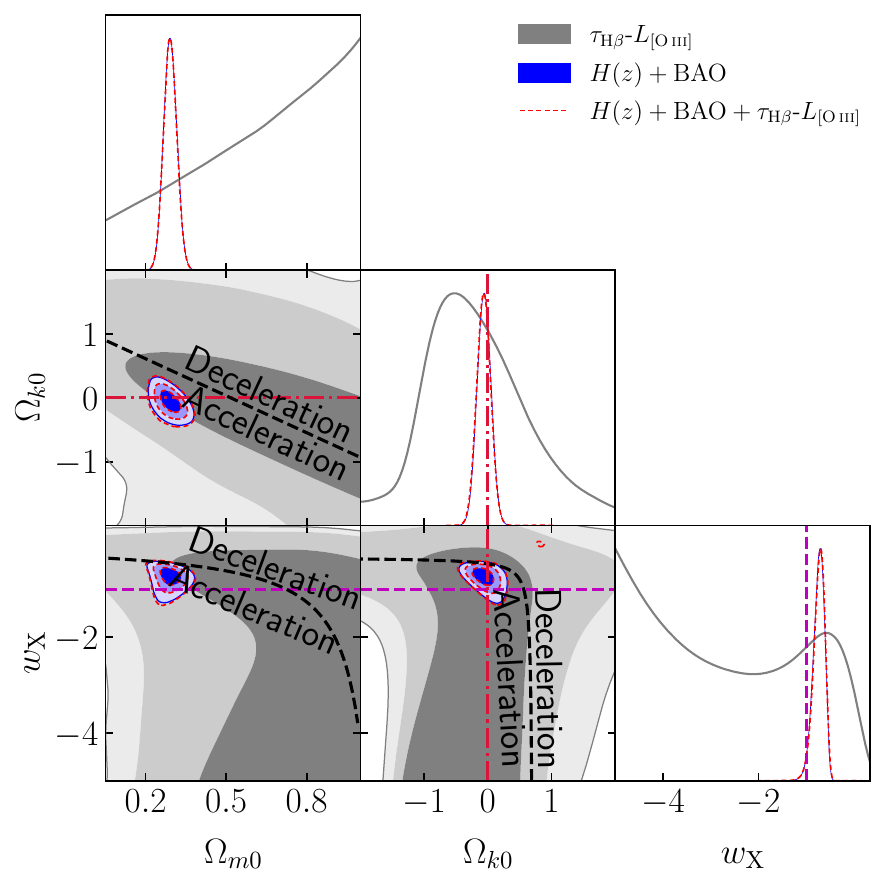}}\\
 \subfloat[]{%
    \includegraphics[width=0.35\textwidth,height=0.33\textwidth]{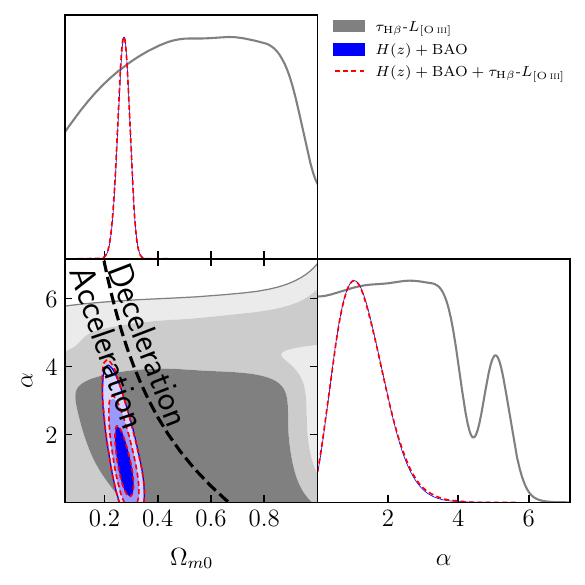}}
 \hspace{0.1\textwidth}
 \subfloat[]{%
    \includegraphics[width=0.35\textwidth,height=0.33\textwidth]{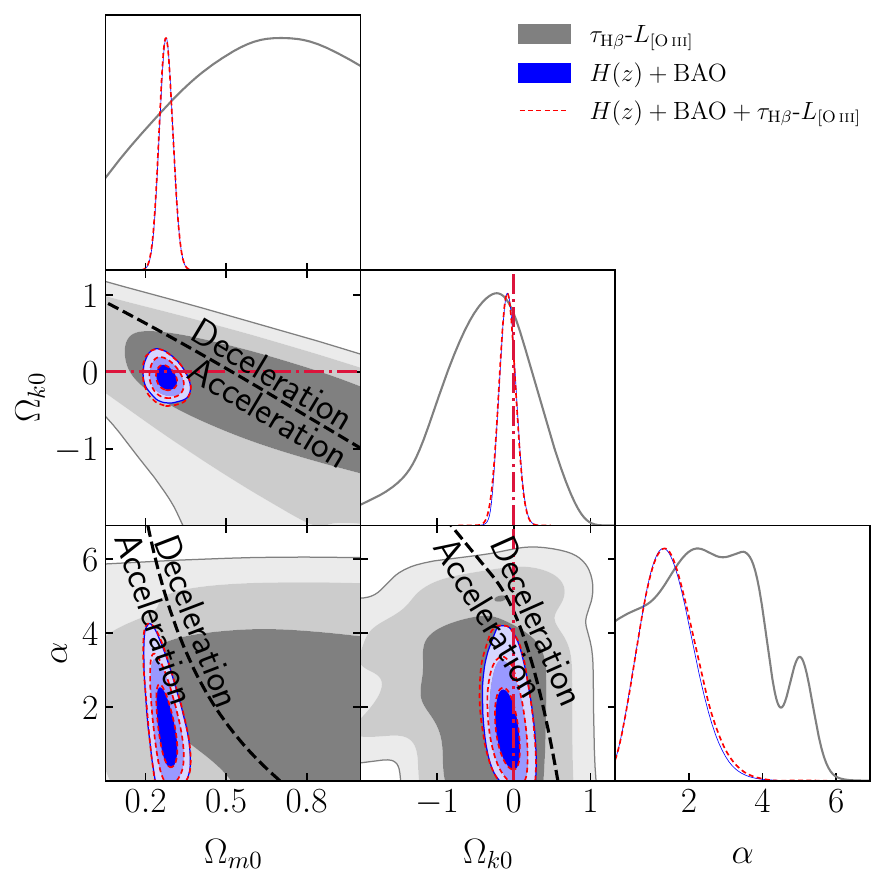}}\\
\caption{Same as Fig.\ \ref{fig5}, but for cosmological parameters only. (a) Flat \lcdm. (b) Nonflat \lcdm. (c) Flat XCDM. (d) Nonflat XCDM. (e) Flat \pcdm. (f) Nonflat \pcdm.}
\label{fig6}
\end{figure*}

Figures~\ref{fig1}--\ref{fig6} present the results of the posterior analyses for the six different cosmological models. The one-dimensional (1D) probability distributions (normalized) and 2D confidence regions for each parameter are illustrated, alongside the $R-L$ relation parameters and intrinsic scatter for $\tau_{\mathrm{H}\beta}\text{-}L_{5100}$, $\tau_{\mathrm{H}\beta}\text{-}L_{\mathrm{H}\beta}$, and $\tau_{\mathrm{H}\beta}\text{-}L_{\mathrm{{\OIII}}}$. Table~\ref{tab:BFP} summarizes the unmarginalized best-fitting values, the corresponding maximum likelihood estimates ($\mathcal{L}_{\rm max}$), and model comparison AIC, BIC, and DIC metrics, as well as their respective differences ($\Delta \mathrm{AIC}$, $\Delta \mathrm{BIC}$, and $\Delta \mathrm{DIC}$). The posterior mean values of the parameters, together with their $1\sigma$ uncertainties or $2\sigma$ limits, are listed in Table~\ref{tab:1d_BFP}.

Table~\ref{tab:diff1} presents the maximum variations in the $R-L$ relation parameters and the intrinsic scatter among the cosmological models considered (with or without nonflat \lcdm\ and XCDM). These findings suggest that the three RM AGN datasets can each be standardized through their corresponding $R-L$ relations. It is worth noting that omitting nonflat \lcdm\ and XCDM---which are observationally disfavored unless $\Ok$ is extremely small \citep{deCruzPerez:2024shj}---further reinforces this conclusion.

The slope parameter $\gamma$ for $\tau_{\mathrm{H}\beta}\text{-}L_{5100}$ ranges from $0.449\pm0.027$ in flat XCDM to $0.470\pm0.030$ in nonflat XCDM; for $\tau_{\mathrm{H}\beta}\text{-}L_{\mathrm{H}\beta}$ it ranges from $0.479\pm0.029$ in flat \lcdm\ to $0.505\pm0.033$ in nonflat XCDM; and for $\tau_{\mathrm{H}\beta}\text{-}L_{\mathrm{{\OIII}}}$ it ranges from $0.482\pm0.049$ in nonflat XCDM to $0.497\pm0.047$ in flat \pcdm.

These slope values are largely consistent with the slope of 0.5 predicted in the simple photoionization model. For $\tau_{\mathrm{H}\beta}\text{-}L_{5100}$ they deviate from 0.5 by $1.0\sigma$ to $1.9\sigma$, for $\tau_{\mathrm{H}\beta}\text{-}L_{\mathrm{H}\beta}$ they deviate from 0.5 by $0.15\sigma$ to $0.72\sigma$, and for $\tau_{\mathrm{H}\beta}\text{-}L_{\mathrm{{\OIII}}}$ they deviate from 0.5 by $0.065\sigma$ to $0.37\sigma$, depending on cosmological model. 

The intercept parameter $\beta$ for $\tau_{\mathrm{H}\beta}\text{-}L_{5100}$ ranges from $1.380\pm0.025$ in flat XCDM to $1.423^{+0.035}_{-0.039}$ in nonflat XCDM; for $\tau_{\mathrm{H}\beta}\text{-}L_{\mathrm{H}\beta}$ it ranges from $1.262\pm0.025$ in flat XCDM and flat \lcdm\ to $1.309^{+0.037}_{-0.036}$ in nonflat XCDM; and for $\tau_{\mathrm{H}\beta}\text{-}L_{\mathrm{{\OIII}}}$ it ranges from $1.437^{+0.070}_{-0.052}$ in nonflat XCDM to $1.472\pm0.034$ in flat and nonflat \pcdm.

The intrinsic scatter parameter $\sigma_{\rm int}$ for $\tau_{\mathrm{H}\beta}\text{-}L_{5100}$ ranges from $0.168^{+0.016}_{-0.019}$ in nonflat XCDM to $0.172^{+0.016}_{-0.020}$ in flat \lcdm\ and XCDM; for $\tau_{\mathrm{H}\beta}\text{-}L_{\mathrm{H}\beta}$ it ranges from $0.158^{+0.016}_{-0.020}$ in nonflat XCDM to $0.163^{+0.017}_{-0.020}$ in flat \lcdm; and for $\tau_{\mathrm{H}\beta}\text{-}L_{\mathrm{{\OIII}}}$ it ranges from $0.250^{+0.020}_{-0.026}$ in nonflat XCDM to $0.252^{+0.021}_{-0.026}$ in flat and nonflat \lcdm. Among these datasets, the $\tau_{\mathrm{H}\beta}\text{-}L_{\mathrm{H}\beta}$ case has the lowest intrinsic scatter, $0.4\sigma$ and $3\sigma$ smaller than the other cases. This outcome is reasonable, as H$\beta$ originates directly from hydrogen ionization in the BLR and is comparatively unaffected by host-galaxy correction uncertainties, making it a better proxy for the ionizing luminosity to constrain the $R-L$ relation. However, a dependence on the Eddington ratio is evident in the $\tau_{\mathrm{H}\beta}\text{-}L_{5100}$ and $\tau_{\mathrm{H}\beta}\text{-}L_{\mathrm{H}\beta}$ relations, whereas such a dependence is not apparent in the $\tau_{\mathrm{H}\beta}\text{-}L_{\mathrm{{\OIII}}}$ relation \citep{2024ApJS..275...13W}.

Among the six cosmological models analyzed, for $\tau_{\mathrm{H}\beta}\text{-}L_{5100}$ and $\tau_{\mathrm{H}\beta}\text{-}L_{\mathrm{H}\beta}$, only in the $\wX-\Om$ plane of the nonflat XCDM parametrization and in the $\alpha-\Ok$ plane of the nonflat \pcdm\ model do these RM AGN data favor currently accelerating cosmological expansion more. In contrast, for $\tau_{\mathrm{H}\beta}\text{-}L_{\mathrm{{\OIII}}}$, only in the flat \pcdm\ model and in the $\alpha-\Om$ plane of the nonflat \pcdm\ model do they more favor currently decelerating cosmological expansion. In this sense and based on the cosmological parameter constraints, the $\tau_{\mathrm{H}\beta}\text{-}L_{\mathrm{{\OIII}}}$ case is the most reliable.

In contrast to the relatively tight constraints obtained for the $R-L$ relation parameters, these RM AGN datasets yield only weak constraints on cosmological parameters. For $\tau_{\mathrm{H}\beta}\text{-}L_{5100}$ ($\tau_{\mathrm{H}\beta}\text{-}L_{\mathrm{H}\beta}$), the 1D marginalized $2\sigma$ lower limits on \om\ span from 0.210 (0.196) in nonflat \lcdm\ (nonflat XCDM) to 0.348 (0.400) in flat \lcdm, with only a $1\sigma$ lower limit available for the $\tau_{\mathrm{H}\beta}\text{-}L_{5100}$ nonflat XCDM case. In these two instances, the $2\sigma$ contours do not overlap with the constraints from $H(z)$ + BAO data, whereas for $\tau_{\mathrm{H}\beta}\text{-}L_{\mathrm{{\OIII}}}$ they do. For $\tau_{\mathrm{H}\beta}\text{-}L_{\mathrm{{\OIII}}}$, the 1D marginalized $1\sigma$ uncertainties range from $0.569^{+0.245}_{-0.259}$ in flat XCDM to $>0.511$ in nonflat XCDM.

The \ok\ constraints are all consistent with flat spatial hypersurfaces within $2\sigma$. For the $\tau_{\mathrm{H}\beta}\text{-}L_{5100}$ and $\tau_{\mathrm{H}\beta}\text{-}L_{\mathrm{H}\beta}$ relations, open hypersurfaces are preferred in nonflat \lcdm\ and XCDM, whereas closed hypersurfaces are favored in nonflat \pcdm. In contrast, for the $\tau_{\mathrm{H}\beta}\text{-}L_{\mathrm{{\OIII}}}$ relation, open hypersurfaces are mildly preferred in nonflat \lcdm, while closed hypersurfaces are mildly preferred in nonflat XCDM and \pcdm. 

More specifically, for $\tau_{\mathrm{H}\beta}\text{-}L_{5100}$, the 1D marginalized \ok\ constraints are
$>-0.359$ ($2\sigma$) in nonflat \lcdm, $>0.792$ ($1\sigma$) in nonflat XCDM, and $-0.492^{+0.653}_{-0.502}$ in nonflat \pcdm. For $\tau_{\mathrm{H}\beta}\text{-}L_{\mathrm{H}\beta}$, the constraints are $>-0.150$ ($2\sigma$) in nonflat \lcdm, $>1.015$ ($1\sigma$) in nonflat XCDM, and $-0.522^{+0.659}_{-0.498}$ in nonflat \pcdm. For $\tau_{\mathrm{H}\beta}\text{-}L_{\mathrm{{\OIII}}}$, the constraints are $0.028^{+1.125}_{-1.103}$ in nonflat \lcdm, $-0.169^{+0.589}_{-0.881}$ in nonflat XCDM, and $-0.396^{+0.655}_{-0.501}$ in nonflat \pcdm.

The dynamical dark energy parameter constraints from RM AGN are likewise weak. For $\tau_{\mathrm{H}\beta}\text{-}L_{5100}$, the 1D marginalized \wx\ constraints are $-1.784^{+1.943}_{-1.169}$ in the flat case and $<0.043$ ($2\sigma$) in the nonflat case. For $\tau_{\mathrm{H}\beta}\text{-}L_{\mathrm{H}\beta}$, the corresponding constraints are $-1.679^{+1.841}_{-1.189}$ and $<0.040$ ($2\sigma$). For $\tau_{\mathrm{H}\beta}\text{-}L_{\mathrm{{\OIII}}}$, the constraints are $<-0.187$ and $<-0.148$ ($2\sigma$) in flat and nonflat XCDM, respectively. For $\tau_{\mathrm{H}\beta}\text{-}L_{5100}$, the 1D marginalized $\alpha$ constraints are $2.747^{+2.571}_{-1.319}$ in flat \pcdm\ and $2.933^{+2.285}_{-0.916}$ in nonflat \pcdm. For $\tau_{\mathrm{H}\beta}\text{-}L_{\mathrm{H}\beta}$, the corresponding constraints are $2.779^{+2.562}_{-1.400}$ and $3.014^{+2.222}_{-0.890}$. For $\tau_{\mathrm{H}\beta}\text{-}L_{\mathrm{{\OIII}}}$, the constraints are $2.486^{+2.547}_{-1.767}$ and $2.537^{+2.442}_{-1.297}$ in flat and nonflat \pcdm, respectively.

For the $\tau_{\mathrm{H}\beta}\text{-}L_{5100}$ and $\tau_{\mathrm{H}\beta}\text{-}L_{\mathrm{H}\beta}$ relations, in the $\Om-\Ok$ subpanels the $2\sigma$ AGN data constraint contours do not overlap with the $2\sigma$ constraints from $H(z)$ + BAO data in nonflat \lcdm\ and XCDM. Although the marginalized 1D \om\ constraints in flat \lcdm\ are in $2\sigma$ tension with $H(z)$ + BAO, the two datasets remain broadly consistent when judged by the overlap in their marginalized 2D contours (e.g., in the $\gamma-\Om$ subpanel). In contrast, cosmological constraints derived from the $\tau_{\mathrm{H}\beta}\text{-}L_{\mathrm{{\OIII}}}$ RM AGN relation are fully consistent with those from $H(z)$ + BAO. 

When each RM AGN sample is combined with $H(z)$ + BAO, the joint constraints are dominated by $H(z)$ + BAO. Relative to the $H(z)$ + BAO constraints, the most notable shifts introduced by $\tau_{\mathrm{H}\beta}\text{-}L_{5100}$ are: $0.07\sigma$ in \ok\ for nonflat \lcdm, $0.20\sigma$ in \wx\ for flat XCDM, $0.14\sigma$ in $\alpha$ for nonflat \pcdm, and $0.14\sigma$ in $H_0$ for flat XCDM. For $\tau_{\mathrm{H}\beta}\text{-}L_{\mathrm{H}\beta}$, the corresponding shifts are: $0.08\sigma$ in \ok\ for nonflat \lcdm, $0.24\sigma$ in \wx\ for nonflat XCDM, $0.16\sigma$ in $\alpha$ for nonflat \pcdm, and $0.16\sigma$ in $H_0$ for flat XCDM. In contrast, for $\tau_{\mathrm{H}\beta}\text{-}L_{\mathrm{{\OIII}}}$, all shifts remain below $0.05\sigma$.

When jointly analyzed with $H(z)$ + BAO data, the parameter constraints for the $\tau_{\mathrm{H}\beta}\text{-}L_{5100}$ and $\tau_{\mathrm{H}\beta}\text{-}L_{\mathrm{H}\beta}$ relations remain broadly consistent with those obtained from the RM AGN datasets alone, except in the nonflat XCDM case. In contrast, the $\tau_{\mathrm{H}\beta}\text{-}L_{\mathrm{{\OIII}}}$ relation yields parameter constraints that are consistent across all six cosmological models.

For the slope parameter $\gamma$, the differences between the combined and AGN-only constraints range from $0.19\sigma$ (flat XCDM) to $0.72\sigma$ (nonflat XCDM) in the $\tau_{\mathrm{H}\beta}\text{-}L_{5100}$ relation, and from $0.23\sigma$ (flat XCDM) to $0.81\sigma$ (nonflat XCDM) in the $\tau_{\mathrm{H}\beta}\text{-}L_{\mathrm{H}\beta}$ relation. In contrast, for the $\tau_{\mathrm{H}\beta}\text{-}L_{\mathrm{{\OIII}}}$ relation the differences are only $0.15\sigma$ in nonflat XCDM and $0.08\sigma$ in all other models.

For the intercept parameter $\beta$, the differences in the $\tau_{\mathrm{H}\beta}\text{-}L_{5100}$ relation range from $0.45\sigma$ (flat \lcdm) to $1.54\sigma$ (nonflat XCDM) and to $1.04\sigma$ (flat \pcdm) after excluding nonflat \lcdm\ and XCDM. In the $\tau_{\mathrm{H}\beta}\text{-}L_{\mathrm{H}\beta}$ relation, the differences range from $0.44\sigma$ (flat XCDM) to $1.89\sigma$ (nonflat XCDM) and to $1.00\sigma$ (flat \pcdm) after excluding nonflat \lcdm\ and XCDM. In contrast, the $\tau_{\mathrm{H}\beta}\text{-}L_{\mathrm{{\OIII}}}$ relation shows differences from just $0.04\sigma$ (nonflat XCDM) to $0.76\sigma$ (flat \pcdm).

For the intrinsic scatter parameter $\sigma_{\rm int}$, the differences are generally modest; the largest deviations occur in the $\tau_{\mathrm{H}\beta}\text{-}L_{\mathrm{H}\beta}$ relation in nonflat \lcdm\ and XCDM, reaching $0.35\sigma$.

According to the more reliable DIC, the $\tau_{\mathrm{H}\beta}\text{-}L_{5100}$ and $\tau_{\mathrm{H}\beta}\text{-}L_{\mathrm{H}\beta}$ datasets individually favor nonflat XCDM the most, showing weak evidence against the remaining models in the former case and weak or positive evidence against the remaining models in the latter. When combined with $H(z)$ + BAO, however, the joint constraints favor flat \pcdm\ the most, with mildly strong evidence against nonflat \lcdm\ and weak or positive evidence against the other models. In contrast, the $\tau_{\mathrm{H}\beta}\text{-}L_{\mathrm{{\OIII}}}$ dataset alone favors flat \pcdm\ the most, with mildly strong evidence against nonflat XCDM and weak evidence against the remaining models. In combination with $H(z)$ + BAO, flat \pcdm\ is also favored the most, with only weak or positive evidence against the other models.

\section{Discussion}
\label{sec:discussion}

We earlier applied our method to a sample of 41 relatively low redshift AGN ($0.00415 \le z \le 0.474$) with both H$\alpha$ and H$\beta$ RM AGN measurements. Although these sources can be standardized, the resulting cosmological constraints were found to be weak \citep{cao_2025a}. This is primarily due to the limited dynamical range of the sample in luminosity ($10^{41.97} \, \mathrm{erg \, s^{-1}} < L_{5100} < 10^{45.94} \, \mathrm{erg \, s^{-1}}$) and Eddington ratio $\lambda_{\text{Edd}}$ ($-2.4 < \log \lambda_{\text{Edd}} < 0.4$), which also yields $R-L$ slopes that are steeper than those predicted by the simple photoionization model.

To address these limitations we next considered a substantially larger H$\beta$ RM sample of 157 AGN spanning a wider range in luminosity ($10^{41.73}  \, \mathrm{erg \, s^{-1}} < L_{5100} < 10^{45.9}  \, \mathrm{erg \, s^{-1}}$), Eddington ratio ($-2.92 < \log \lambda_{\text{Edd}} < 0.94$), and redshift ($0.00308 \le z \le 0.8429$) \citep{cao25}. This broader sample can also be standardized across different cosmological models, and the inferred cosmological parameters were found to be consistent within $2\sigma$ with those derived from better-established $H(z)$ + BAO data, except for the nonflat \lcdm\ and XCDM models, which are already disfavored by other observational data, \citep{deCruzPerez:2024shj}. These earlier studies further indicated that the comparatively shallower $R-L$ slopes arise from the inclusion of highly accreting AGN whose BLR lags are systematically shorter.

These results motivated us to search for improved proxies for the BLR-ionizing flux. Since AGN with different accretion rates have distinct UV/optical spectral energy distributions, the standard monochromatic luminosity $L_{5100}$ may not always reflect the fraction of ionizing photons responsible for setting the BLR radius. To explore this possibility, we incorporate two additional luminosity indicators: $L_{\mathrm{H}\beta}$ and $L_{\mathrm{{\OIII}}}$, alongside $L_{5100}$. We apply these tracers to our Averaged Scheme sample, consisting of 113 sources with $L_{5100}$ measurements and 100 sources with both $L_{\mathrm{H}\beta}$ and $L_{\mathrm{{\OIII}}}$ data, spanning a dynamical range comparable to that of \citet{cao25}.

Our results show that all three RM AGN datasets can be standardized through their respective $R-L$ relations, but each set yields only weak constraints on cosmological parameters. Among the three $R-L$ relations, $\tau_{\mathrm{H}\beta}\text{-}L_{\mathrm{H}\beta}$ shows the least scatter, possibly because both $\tau_{\mathrm{H}\beta}$ and $L_{\mathrm{H}\beta}$ originate within the BLR.  Notably, although the $\tau_{\mathrm{H}\beta}\text{-}L_{\mathrm{{\OIII}}}$ relation exhibits larger scatter, it is more consistent with currently accelerating cosmological expansion in all models except \pcdm\ (where it is still somewhat consistent with currently accelerating cosmological expansion), unlike the relations that use $L_{5100}$ or $L_{\mathrm{H}\beta}$. This improvement likely stems from the fact that both the $\tau_{\mathrm{H}\beta}\text{-}L_{5100}$ and $\tau_{\mathrm{H}\beta}\text{-}L_{\text{H}\beta}$ relations exhibit Eddington ratio–dependent deviations, possibly caused by self-shadowing effects in slim accretion discs \citep{2014ApJ...797...65W}.  In contrast, the $\tau_{\mathrm{H}\beta}\text{-}L_{\mathrm{{\OIII}}}$ relation does not display such a behavior. Because the response timescale of NLR {\OIII} emission is extremely long ($\gg 10^3$ years), $L_{\mathrm{{\OIII}}}$ reflects only the long-term average accretion state, making it insensitive to distinctions between super-Eddington and sub-Eddington phases. Moreover, the BLR lag is primarily governed by higher-energy photons, consistent with the higher ionization potential of narrow {\OIII} compared to hydrogen \cite[see, e.g.,][for more discussion]{2024ApJS..275...13W}. In addition, $L_{5100}$ measurements are strongly affected by host-galaxy corrections, which introduce further uncertainties, whereas such contamination is minimal for $L_{\mathrm{{\OIII}}}$.

In summary, our analysis demonstrates that RM AGN can be standardized using multiple luminosity tracers, but the reliability of the resulting cosmological constraints may depend  on the chosen indicator. Among the tested proxies analyzed in this work, $L_{\mathrm{{\OIII}}}$ provides the most robust and least accretion-dependent calibration. These findings highlight the importance of selecting ionization-sensitive luminosity measures for future RM-based cosmological studies.

\section{Conclusion}
\label{sec:conclusion}

Among the three RM AGN datasets, the $\tau_{\mathrm{H}\beta}\text{-}L_{\mathrm{{\OIII}}}$ relation delivers the most robust cosmological constraints. Specifically, although the $\tau_{\mathrm{H}\beta}\text{-}L_{\mathrm{H}\beta}$ relation exhibits the smallest intrinsic scatter parameter constraints, the constraints derived from $\tau_{\mathrm{H}\beta}\text{-}L_{\mathrm{{\OIII}}}$ are more aligned with those from better-established $H(z)$ + BAO data and largely favor currently accelerating cosmological expansion. This is a qualitative improvement with respect to previous low-redshift H$\beta$ AGN samples that employed photoionizing flux proxies more closely related to the accretion disc emission, specifically the monochromatic luminosity at 5100\,\AA\, and broad H$\beta$ and H$\alpha$ luminosities. These samples, both largely heterogeneous \cite{Khadkaetal2021c} and homogeneous \cite{cao_2025a}, favored currently decelerating cosmological expansion, hence being in tension with results from better-established cosmological probes.

Moreover, the slopes inferred from the $\tau_{\mathrm{H}\beta}\text{-}L_{\mathrm{{\OIII}}}$ relations across different cosmological models show a closer agreement with expectations from the simple photoionization model. This is related to the narrow-line \OIII\ emission tracing long-term AGN photoionizing activity ($\sim 10^2-10^4$ years), hence it is insensitive to the short-term accretion state or the current source Eddington ratio, unlike 5100\,\AA\, and broad H$\alpha$ and H$\beta$ emissions that vary on short timescales of weeks to months. In previous samples with a broad Eddington ratio distribution, the $\tau_{H\beta}-L$ relation flattening was caused by higher-accreting sources that exhibited shortened H$\beta$ time delays \cite{MartinezAldama2019}. 

Overall, this demonstrates the importance of identifying the proper proxy for the BLR photoionizing luminosity in addition to the application of a unified time-delay determination methodology across the whole sample. Monochromatic continuum luminosity does not capture consistently the photoionizing flux across the whole sample since the broad-band spectrum changes shape with the variations in the relative accretion rate and the SMBH mass \cite{2020ApJ...899...73F}. It appears that the narrow-line \OIII\ emission is more suitable for capturing the long-term photoionizing photon flux than proxies related to accretion-disc and BLR emission. The latter, apart from being sensitive to short-term effects, are also more likely to be contaminated by host emission and by circumnuclear extinction \citep{2004MNRAS.348L..54C,2023MNRAS.522.2869S,Zajaceketal2024}. This is also the first successful demonstration applying the NLR \OIII\, luminosity to AGN standardization and to cosmological parameter constraints.

More and better $\tau_{\mathrm{H}\beta}\text{-}L_{\mathrm{{\OIII}}}$ data from future surveys, such as the SDSS-V Black Hole Mapper \citep{SDSS_BH_Mapper_2017arXiv171103234K} and the Rubin Observatory Legacy Survey of Space and Time \citep{2019FrASS...6...75P,Kovacevic:2022msi,2023A&A...675A.163C}, will allow for better testing of the standardizability (and potential cosmological parameters constraining power) of such data.

\begin{acknowledgments}
A.K.M.\ acknowledges the support from the European Research Council (ERC) under the European Union’s Horizon 2020 research and innovation program (grant No. 951549). M.Z.\ acknowledges the Czech Science Foundation Junior Star grant no. GM24-10599M. The computations for this project were partially performed on the Beocat Research Cluster at Kansas State University, which is funded in part by NSF grants CNS-1006860, EPS-1006860, EPS-0919443, ACI-1440548, CHE-1726332, and NIH P20GM113109. 
\end{acknowledgments}



\bibliographystyle{apsrev4-2-author-truncate}
\bibliography{apssamp}

\end{document}